\documentclass[aps,twocolumn,showpacs,prl,superscriptaddress]{revtex4-2}
\usepackage{graphics}
\usepackage{times}
\graphicspath{ {images/} }
\usepackage{amsmath}
\usepackage{mathtools}
\usepackage{bm}
\usepackage{amssymb}
\usepackage{amsthm}
\usepackage{amsbsy}
\usepackage{dsfont}
\usepackage{bbm}
\usepackage{verbatim}
\usepackage{comment}
\usepackage{overpic}
\usepackage{graphicx}
\usepackage{epstopdf}
\usepackage{subfigure}
\usepackage{natbib}
\usepackage{epsfig}
\usepackage{amsfonts}
\usepackage{mathrsfs}
\usepackage{sidecap}
\usepackage{lipsum}
\usepackage{url}
\usepackage[toc,page,title,titletoc,header]{appendix}
\usepackage[colorlinks,linkcolor=blue,citecolor=blue,anchorcolor=blue,urlcolor=blue]{hyperref}
\usepackage{hyperref}
\usepackage{resizegather}
\usepackage{tikz}
\usetikzlibrary{arrows.meta}
\usepackage{float}
\usepackage{mathbbol}
\usepackage[normalem]{ulem}
\usepackage{cancel}
\usepackage{upgreek}
\usepackage{enumitem}
\usepackage{relsize}

\newcommand{\bl}{\begin{aligned}}
	\newcommand{\el}{\end{aligned}}

\def\be{\begin{equation}}
	\def\ee{\end{equation}}

\def\bi{\begin{itemize}}
	\def\ei{\end{itemize}}
\def\bn{\begin{enumerate}}
	\def\en{\end{enumerate}}
\def\bea{\begin{eqnarray}}
	\def\eea{\end{eqnarray}}
\def\no{\nonumber}
\def\ba{\begin{array}}
	\def\ea{\end{array}}
\def\bd{\begin{displaymath}}
	\def\ed{\end{displaymath}}

\usepackage{xr}            
\usepackage{hyperref}

\makeatletter
\newcommand*{\addFileDependency}[1]{%
	\typeout{(#1)}
	\@addtofilelist{#1}
	IfFileExists{#1}{}{\typeout{No file #1.}}
}
\makeatother

\begin{document}

	\title{Noise-Immune Attractors in Stochastically Driven Non-Hermitian Systems}
	
	\author{Hanieh Najafzadeh}
	\email{honeynjf@gmail.com}
	\affiliation{Department of Physics, Sharif University of Technology, Tehran 1458889694, Iran}
	
	\author{Radmehr Jafari}
	\email{raadmehr.jafari@gmail.com}
	\affiliation{Department of Physics, Institute for Advanced Studies in Basic Sciences (IASBS), Zanjan, Iran}
	\affiliation{School of Quantum Physics and Matter, Institute for Research in Fundamental Sciences (IPM), Tehran, Iran}
	
	\author{Abdollah Langari}
	\email{abdollah.langari@gmail.com}
	\affiliation{Department of Physics, Sharif University of Technology, Tehran 1458889694, Iran}
	
	\date{\today}
\begin{abstract}
Temporal stochastic fluctuations generically destroy quantum coherence and accelerate defect generation in driven quantum systems. Here, we demonstrate that non-Hermitian dynamics can fundamentally reverse this trend by stabilizing noise-resilient late-time states. By developing an exact noise master equation for non-Hermitian systems subject to Gaussian fluctuations, we uncover a projective noise-invariance mechanism governed by $\mathcal{T}$-symmetry-broken Liouvillian attractors. This mechanism produces robust defect-density plateaus and yields an exact geometric classification into noise-independent, fully polarized, and continuously tunable regimes. We substantiate these results in an imbalanced-pairing non-Hermitian Kitaev model—a prototypical pseudo-Hermitian system—and propose their experimental observation in coupled photonic waveguides. Our findings identify non-Hermitian attractor dynamics as a versatile route to controlling noise-induced excitations beyond conventional anti-Kibble–Zurek scaling.
\end{abstract}

\maketitle
\textit{Introduction---}Understanding the interplay between nonequilibrium driving and temporal fluctuations is a central challenge in modern quantum physics. When a quantum system is driven across a critical point, the formation of excitations is governed by the Kibble--Zurek mechanism (KZM), which predicts universal scaling relations between the defect density and the driving rate. The KZM has been extensively investigated, both theoretically~\cite{Kibble1976,Zurek1985,Dziarmaga2010,Polkovnikov2011,Sengupta2008,Duttabook,Sen2008,Shreyoshi2009,Das2017,Pellegrini2008,Polkovnikov2005,Kou2023,Damski2005,Zamora2020,Jafari2026a,Puebla2020b,Das2021,Dziarmaga2005,Jafari2026b,Zurek2005,Kou2022,Kou2025,Bermudez2009,Antunes2006,Minchul2015,Suzuki2010,Ladewig2020,Dario2008,Dario2009,Puebla2020,Balducci2023,Adolfo2012,Jafari2026bb,Grandi2010,Polkovnikov2008,Jara2026} and experimentally~\cite{Liu2018,Mark1994,Maniv2003,Corman2014,Navon2015,Sven2015,Keesling2019,Zhang2017,Bando2020,Higuera2022} and has found broad applications ranging from condensed matter systems~\cite{Kolodrubetz2012,Chesler2015,Silvi2016,Deng2026,Grandi2011,Chandran2012,Hwang2015,Adolfo2018,Nikoghosyan2016,Deffner2017,Polkovnikov2010} to quantum technologies~\cite{Albash2018}.
In realistic settings, however, temporal fluctuations in control parameters can disrupt coherent evolution and enhance excitation production. This competition gives rise to the anti-Kibble--Zurek (AKZ) regime, where defect production increases as the driving becomes slower~\cite{Dutta2016,Chenu2017,Sadeghizade2025,Singh2021,Gao2017,Singh2023,Yang2026,Gherardini2024}. An important open question is  \textit{whether this noise-induced proliferation is inevitable or can be fundamentally modified by an appropriate dynamical mechanism.}

\begin{figure}[t]
    \centering
    \vspace{-0.9cm}
    \begin{overpic}[width=\columnwidth, trim=-50bp 110bp 0bp 0bp, clip]{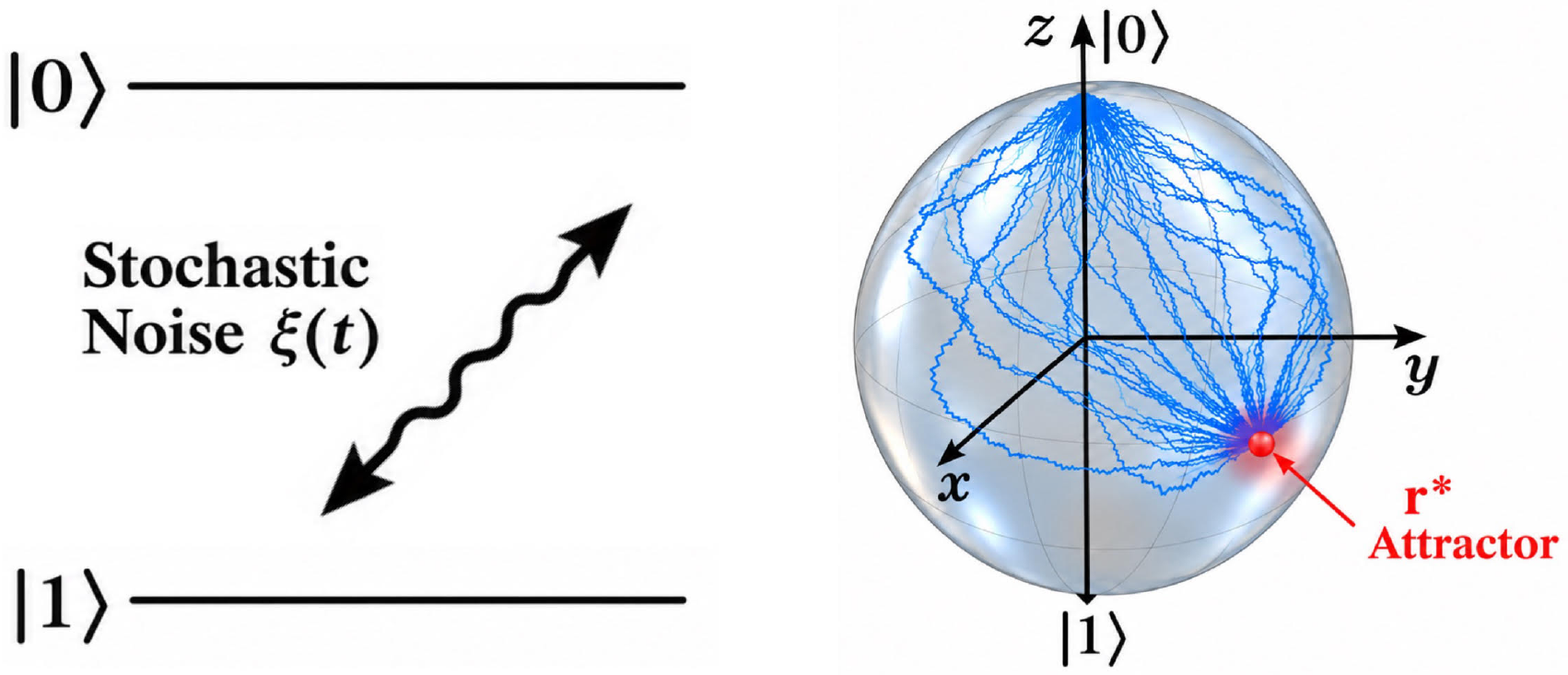}
        \put(1, 40){\textbf{(a)}}
        \put(50, 40){\textbf{(b)}}
    \end{overpic}
    \caption{(a) Effective two-level system subject to temporal stochastic noise $\xi(t)$. (b) Bloch-sphere dynamics illustrating the contraction of the normalized state toward a stable attractor $\mathbf r^*$ under non-Hermitian Liouvillian evolution.}
    \label{fig:bloch_attractor}
\end{figure}

Non-Hermitian systems offer a promising setting to address this question. Through effective gain and loss~\cite{Li2019,Ozawa2019}, they exhibit unconventional spectral and dynamical properties~\cite{Sim2023,Najafzadeh2026,Dora2019,Xiao2021,Deng2025,Agarwal2024,Naji2022}, including exceptional points (EPs)\cite{Doppler2016}, non-Hermitian skin modes, and state-selective amplification and suppression~\cite{Yao2018,Bergholtz2021,Ashida2020,Hatano1996,Yao2018,Rudner2009,Hu2011,Borgnia2020,Lee2019,Chen2020,Bender2007,Dvir2023}. These phenomena have attracted considerable interest in synthetic photonic waveguides, ultracold atoms, and acoustic systems~\cite{Bender1998,ElGanainy2018,Miri2019,Bergholtz2021,Ashida2020}. In particular, non-Hermitian evolution can selectively amplify or suppress state components, steering normalized states toward preferred configurations. This raises the possibility that non-Hermiticity may counteract noise-induced excitation production rather than merely modify its magnitude. Although recent work has shown that noise acting on the anti-Hermitian sector can induce anti-dephasing, stabilize lossy states, and facilitate state purification~\cite{Martinez2025}, a general understanding of driven non-Hermitian systems subject to simultaneous, multi-axis Hermitian and non-Hermitian fluctuations is still lacking. \textit{Whether such dynamics can suppress the conventional AKZ response, what determines the resulting stationary state remain open questions.}

While the Gorini-Kossakowski-Sudarshan-Lindblad (GKSL) approach~\cite{Minganti2019,Arkhipov2020,Cattaneo2020} provides a foundational description of open systems, obtaining exact, non-perturbative analytical solutions for multi-channel noise can be notoriously challenging, making stochastic Hamiltonian generators a powerful and complementary analytical tool.
In this letter, we formulate a general Liouvillian framework~\cite{Martinez2025,Gorini1976,Lindblad1976,Minganti2019,Arkhipov2020,Cattaneo2020} for two-level non-Hermitian Hamiltonian subject to simultaneous Hermitian and non-Hermitian Gaussian noise along arbitrary polarization axes. By unraveling the noise-averaged density matrix in an extended Pauli-Bloch representation, we obtain a closed algebraic expression that permits direct spectral analysis. We find that non-Hermitian dynamics can fundamentally reshape the effects of temporal noise through a mechanism of projective noise invariance. Specifically, we show that when the non-Hermitian drift enters an attractor regime, the long-time state space contracts onto a dominant Liouvillian eigenmode.


As illustrated in Fig.~\ref{fig:bloch_attractor}, while stochastic fluctuations induce diffusive transitions between bands [Fig.~\ref{fig:bloch_attractor}(a)], the non-Hermitian flow can dynamically purify the trajectories, potentially funneling them onto a localized attractor $\mathbf{r}^*$ on the generalized Bloch sphere [Fig.~\ref{fig:bloch_attractor}(b)]. Because normalized physical observables project out the overall stochastic growth, this mechanism can arrest conventional AKZ growth and produce robust late-time plateaus in the defect density.


We further derive an exact geometric compatibility condition that systematically classifies the resulting nonequilibrium plateau dynamics into three distinct regimes: 
(I) noise-immune plateaus, 
(II) fully polarized saturation limits, and 
(III) tunable noise-sensing plateaus. 
We apply our approach to the imbalanced-pairing Kitaev model and extend it to the non-Hermitian Su--Schrieffer--Heeger (SSH) model as paradigmatic topological platforms.
Finally, we propose a concrete experimental blueprint based on femtosecond laser-written coupled photonic waveguides, in which the predicted stationary attractor plateaus can be directly detected through the sublattice intensity imbalance.\\




\textit{General Two-level Non-Hermitian Liouvillian Dynamics---}We consider the time-dependent, two-level, non-Hermitian (NH) Hamiltonians with stochastic fluctuations in both the Hermitian (H) and anti-Hermitian sectors:
%
{\small
\begin{equation}
	H(t) = \big( \mathbf{d} + \eta_H(t)\mathbf{m}_H \big) \cdot \boldsymbol{\sigma} - i\big( \boldsymbol{\gamma} + \eta_{\text{NH}}(t)\mathbf{m}_{\text{NH}} \big) \cdot \boldsymbol{\sigma},
	\label{eq:generalized_H}
\end{equation}
}
%
where $H_0 = \mathbf{d}\cdot\boldsymbol{\sigma}$ and $\Gamma = \boldsymbol{\gamma}\cdot\boldsymbol{\sigma}$ denote deterministic operators, (and an overall trace shift is omitted). Independent Markovian white noises $\eta_{\nu}(t)$ ($\nu \in \{H, \text{NH}\}$) act along unit axes $\mathbf{m}_{\nu}$ with zero mean and variances $\langle \eta_{\nu}(t)\eta_{\nu}(t')\rangle = \xi_{\nu}^2 \delta(t-t')$.
Assuming that the system is initially prepared in the ground state of the right eigenvectors $|\psi^R_{g,\xi}(0)\rangle$, the time-evolved state $|\psi^R_{\xi}(t)\rangle$ is obtained by solving the time-dependent Schrödinger equation, with the corresponding density matrix given by $\rho_{\xi}(t) = |\psi^R_{\xi}(t)\rangle\langle\psi^R_{\xi}(t)|$. 
Averaging over all noise realizations via Novikov’s theorem~\cite{Novikov} yields the noise-averaged master equation for the non-Hermitian Hamiltonian~\cite{SuppMat}:
\begin{align}
	\frac{d\rho}{dt} &= -i[H_0, \rho] - \{\Gamma, \rho\} - \frac{\xi_H^2}{2} \Big[ (\mathbf{m}_H\cdot\boldsymbol{\sigma}), \big[ (\mathbf{m}_H\cdot\boldsymbol{\sigma}), \rho \big] \Big] \nonumber \\
	&\quad + \frac{\xi_{\text{NH}}^2}{2} \Big\{ (\mathbf{m}_{\text{NH}}\cdot\boldsymbol{\sigma}), \big\{ (\mathbf{m}_{\text{NH}}\cdot\boldsymbol{\sigma}), \rho \big\} \Big\},
	\label{eq:generalized_master}
\end{align}
%
where $\rho(t)=\langle \rho_{\xi}(t) \rangle_{\xi}$ is noise-averaged density matrix. 
In the noise master equation, the double commutator term describes standard dephasing, whereas the anticommutator arises from non-Hermitian fluctuations, which drive exponential population growth while preserving instantaneous $\mathcal{PT}$-symmetry.

By expanding the noise-averaged density matrix in the Pauli-Bloch basis, $\rho(t) = \frac{1}{2}\left(I_0 \mathds{1} + \mathbf{r}\cdot\boldsymbol{\sigma}\right)$ with the total population $I_0 = \mathrm{Tr}(\rho)$ and the coherence vector $\mathbf{r} = (x, y, z)^T$, the dynamics can be analyzed through a Liouvillian representation $d\mathbf{R}/dt = \mathcal{L} \mathbf{R}$, where $\mathbf{R} = (I_0, x, y, z)^T$ and
%
\begin{equation}
		\mathcal{L} = \begin{pmatrix}
		2\xi_{\mathrm{NH}}^2 & -2\gamma_x & -2\gamma_y & -2\gamma_z \\
		-2\gamma_x & \mathcal{M}_{xx} & -2d_z + \mathcal{S}_{xy} & 2d_y + \mathcal{S}_{xz} \\
		-2\gamma_y & 2d_z + \mathcal{S}_{xy} & \mathcal{M}_{yy} & -2d_x + \mathcal{S}_{yz} \\
		-2\gamma_z & -2d_y + \mathcal{S}_{xz} & 2d_x + \mathcal{S}_{yz} & \mathcal{M}_{zz}
	\end{pmatrix},
	\label{eq:L_gen_compact}
	\end{equation}
%
is Liouvillian matrix with diagonal rates $\mathcal{M}_{ii} = -2\xi_H^2 (1 - m_{H,i}^2) + 2\xi_{\mathrm{NH}}^2 m_{\mathrm{NH},i}^2$ and off-diagonal cross-couplings $\mathcal{S}_{ij} = 2\xi_H^2 m_{H,i}m_{H,j} + 2\xi_{\mathrm{NH}}^2 m_{\mathrm{NH},i}m_{\mathrm{NH},j}$ (for $i \neq j$).

\textit{Projective Noise Invariance---}An analysis of the Liouvillian generator $\mathcal{L}$ clarifies the conditions under which the system’s late-time dynamics attain an invariant plateau.
The dynamics of the system are fully governed by the corresponding eigenvalues ($\Lambda_{j}$) and eigenvectors ($\mathbf{V}_{j}$) of the Liouvillian generator. Let $\Lambda_+$ denote the dominant positive eigenvalue (defined as the one with the largest real part), and let $\mathbf{V}_+$ be its corresponding right eigenvector satisfying $\mathcal{L}\mathbf{V}_+ = \Lambda_+\mathbf{V}_+$. A projective steady state requires a positive, gapped leading eigenvalue:
\begin{equation}
	\operatorname{Re}\{\Lambda_{+}\} > \max\left(0,\, \max_{j\neq +}\operatorname{Re}\{\Lambda_{j}\}\right).
	\label{eq:attractor_condition}
\end{equation}
%
If Eq.~\eqref{eq:attractor_condition} holds, the state vector $\mathbf{R}(t) = \sum_j c_j e^{\Lambda_j t} \mathbf{V}_j$ is asymptotically dominated by the leading mode, $\mathbf{R}(t) \propto e^{\Lambda_+ t} \mathbf{V}_+$ with transient contributions decaying as $e^{(\Lambda_j - \Lambda_+)t} \to 0$. The rigorous validity of adiabatically tracking this instantaneous dominant mode and the suppression of non-adiabatic excitations are established in Ref.~\cite{SuppMat} for driven protocols. Violating this condition---e.g., when $\mathrm{Re}\{\Lambda_+\} \le 0$ or at exceptional points where the gap closes---yields an \textit{absence-of-plateau} regime governed by decaying states, anti-attractor dynamics, or limit cycles.

According to the criterion in Eq.~\eqref{eq:attractor_condition}, the mode-resolved projective observable converges to:
%
\begin{equation}
	\mathcal{O}_z \equiv \lim_{t\to\infty}\frac{\operatorname{Tr}(\sigma_z \rho(t))}{\operatorname{Tr}(\rho(t))} = \frac{z^*}{I_{0}^*},
	\label{eq:projective_observable}
\end{equation}
%
where $\mathbf{V}_{+} = (I_{0}^*, x^*, y^*, z^*)^T$ is the dominant right eigenvector.

In the secular problem, we define the detuned eigenvalues $\tilde{\Lambda}_0 \equiv \Lambda_+ - 2\xi_{\mathrm{NH}}^2$ and $\tilde{\Lambda}_z \equiv \Lambda_+ - \mathcal{M}_{zz} = \Lambda_+ + 2\xi_H^2(1 - m_{H,z}^2) - 2\xi_{\mathrm{NH}}^2 m_{\mathrm{NH},z}^2$, along with the transverse vectors $\mathbf{r}_\perp^* \equiv (x^*, y^*)^T$, $\boldsymbol{\gamma}_\perp \equiv (\gamma_x, \gamma_y)^T$, and the generalized transverse driving $\boldsymbol{\mathcal{D}}_\perp \equiv (-2d_y + \mathcal{S}_{xz}, 2d_x + \mathcal{S}_{yz})^T$. 
Accordingly, a stable dynamical plateau for the projective ratio $\mathcal{O}_z = z^* / I_0^*$, independent of transient transverse dynamics, requires $\boldsymbol{\gamma}_\perp$ and $\boldsymbol{\mathcal{D}}_\perp$ to be collinear, i.e., $\boldsymbol{\gamma}_\perp \times \boldsymbol{\mathcal{D}}_\perp = 0$~\cite{SuppMat}. This yields the geometric compatibility condition:
%
\begin{equation}
2(d_x \gamma_x + d_y \gamma_y) + (\gamma_x \mathcal{S}_{yz} - \gamma_y \mathcal{S}_{xz}) = 0.
	\label{eq:geometric_condition}
\end{equation}
%
%
Without loss of generality, by selecting an in-plane transverse frame (i.e., $\gamma_x = 0$), 
we establish the exact closed-form plateau value:
%
\begin{equation}
	\mathcal{O}_z = \frac{4\gamma_y \gamma_z + (2d_x + \mathcal{S}_{yz})\tilde{\Lambda}_0}{-2\gamma_y \tilde{\Lambda}_z - 2\gamma_z (2d_x + \mathcal{S}_{yz})}.
	\label{eq:plateau_formula}
\end{equation}
%
%


\textit{Classification of Dynamical Plateaus---}Provided that the attractor criterion in Eq.~\eqref{eq:attractor_condition} and the geometric compatibility condition in Eq.~\eqref{eq:geometric_condition} are satisfied (see~\cite{SuppMat} for explicit details), the asymptotic observable $\mathcal{O}_z$ [Eq.~\eqref{eq:plateau_formula}] settles into a robust stationary plateau. Depending on the drive orientations and noise tensors, the steady-state landscape partitions into three distinct regimes:

%
\textit{Case (I): Universal Noise-Independent Plateau---}For vanishing longitudinal dissipation ($\gamma_z = 0$) and noise preserving the longitudinal axis ($m_{H,z} = m_{\mathrm{NH},z} = 1$), it follows that $\mathcal{S}_{yz} = 0$ and $\tilde{\Lambda}_0 = \tilde{\Lambda}_z \equiv \tilde{\Lambda}$. Consequently, Eq.~\eqref{eq:plateau_formula} simplifies to
\begin{equation}
	\mathcal{O}_z = \frac{2d_x \tilde{\Lambda}}{-2\gamma_y \tilde{\Lambda}} = -\frac{d_x}{\gamma_y}.
	\label{eq:scenario_1}
\end{equation}
%
Since the stochastic shift $\tilde{\Lambda}$ cancels out, the late-time observable decouples entirely from the environment, being governed exclusively by the bare Hamiltonian and gain-loss parameters. This demonstrates topological bulk resilience against both Hermitian dephasing and non-Hermitian fluctuations, irrespective of their strengths. Importantly, this noise-independent plateau is not a mathematical artifact of state renormalization; rather, the non-Hermitian spectral gap induces genuine contracting dynamics on the Bloch sphere, exponentially compressing stochastic trajectories onto a zero-variance attractor manifold.

Algebraically, when the underlying deterministic system is pseudo-Hermitian, this noise-decoupled state is rooted in an exact Krein-charge conservation law~\cite{Gunther2005,Mostafazadeh2006,Gunther2008,Massarelli2022} that manifests as an invariant left zero-mode of the Liouvillian generator, $\mathbf{W}_\Theta^T \mathcal{L} = \mathbf{0}^T$. However, as established in Ref.~\cite{SuppMat}, static Krein conservation alone is insufficient to guarantee plateau formation: a strictly positive Liouvillian spectral gap $\delta \Lambda > 0$ is dynamically indispensable to purify the ensemble norm and erase initial-state memory, collapsing trajectories onto the attractor.

\textit{Case (II): Saturated Noise-Independent Plateau (Complete Polarization)---}Remarkably, noise independence also persists when $\gamma_z \neq 0$, provided the effective transverse coherent-stochastic field balances the non-Hermitian driving rate under longitudinal noise symmetry---namely,
$2d_x + \mathcal{S}_{yz} = \mp 2\gamma_y$ and $\tilde{\Lambda}_0 = \tilde{\Lambda}_z \equiv \tilde{\Lambda}$---which yields:
%
\begin{equation}
\mathcal{O}_z = \frac{4\gamma_y \gamma_z \mp 2\gamma_y \tilde{\Lambda}}{-2\gamma_y \tilde{\Lambda} \pm 4\gamma_y \gamma_z} = \pm 1.
	\label{eq:scenario_2}
\end{equation}
%
Here, the dynamical attractor collapses onto the poles of the generalized Bloch sphere ($z^* = \pm I_0^*$), rendering the fully saturated, maximally polarized steady state immune to fluctuations of $\xi_H$ and $\xi_{\mathrm{NH}}$.

\textit{Case (III): Noise-Tunable Plateau (Quantum Sensing Regime)---}If the assumptions underlying Cases~I and II are lifted (e.g., generic drive directions or noise anisotropy causing $\tilde{\Lambda}_0 \neq \tilde{\Lambda}_z$), the effective shifted eigenvalues no longer cancel. Consequently, while the plateau remains dynamically stable, its stationary magnitude acquires a direct dependence on the noise characteristics:
%
\begin{equation}
	\mathcal{O}_z = \mathcal{O}_z(\xi_H, \xi_{\mathrm{NH}}, \mathbf{m}_H, \mathbf{m}_{\mathrm{NH}}).
	\label{eq:scenario_3}
\end{equation}
%
Rather than destroying the attractor, the bath imprints itself onto the macroscopic steady state, establishing a versatile quantum sensing platform in which the robust plateau acts as a calibration-free probe of external noise strengths and directional anisotropies.


\textit{Driven Non-Hermitian Imbalanced-Pairing Kitaev Model---}For general translationally invariant systems that decouple into independent momentum modes, we define a momentum index $k$ within the underlying two-level model. Consequently, the dynamics of each $k$-mode are fully determined by the respective eigenvalues ($\Lambda_{j,k}$) and eigenvectors ($\mathbf{V}_{j,k}$) of the Liouvillian, i.e., $d\mathbf{R}_k/dt = \mathcal{L}(k) \mathbf{R}_k$.
To validate our universal framework, we investigate the driven non-Hermitian Kitaev model with imbalanced pairing in one and two dimensions. In the Nambu basis $\Psi_{\boldsymbol{k}} = (c_{\boldsymbol{k}}, c_{-\boldsymbol{k}}^\dagger)^T$ with anti-periodic boundary conditions, the Bogoliubov--de Gennes Hamiltonian for the $\boldsymbol{k}$-mode is expressed as~\cite{Jafari2026a,Jafari2026b,Li2018,Wang2017DKC,Li2006}:
\begin{figure*}[t]
	\centering
	\begin{minipage}[c]{0.30\textwidth}
		\centering
		\begin{overpic}[width=\linewidth]{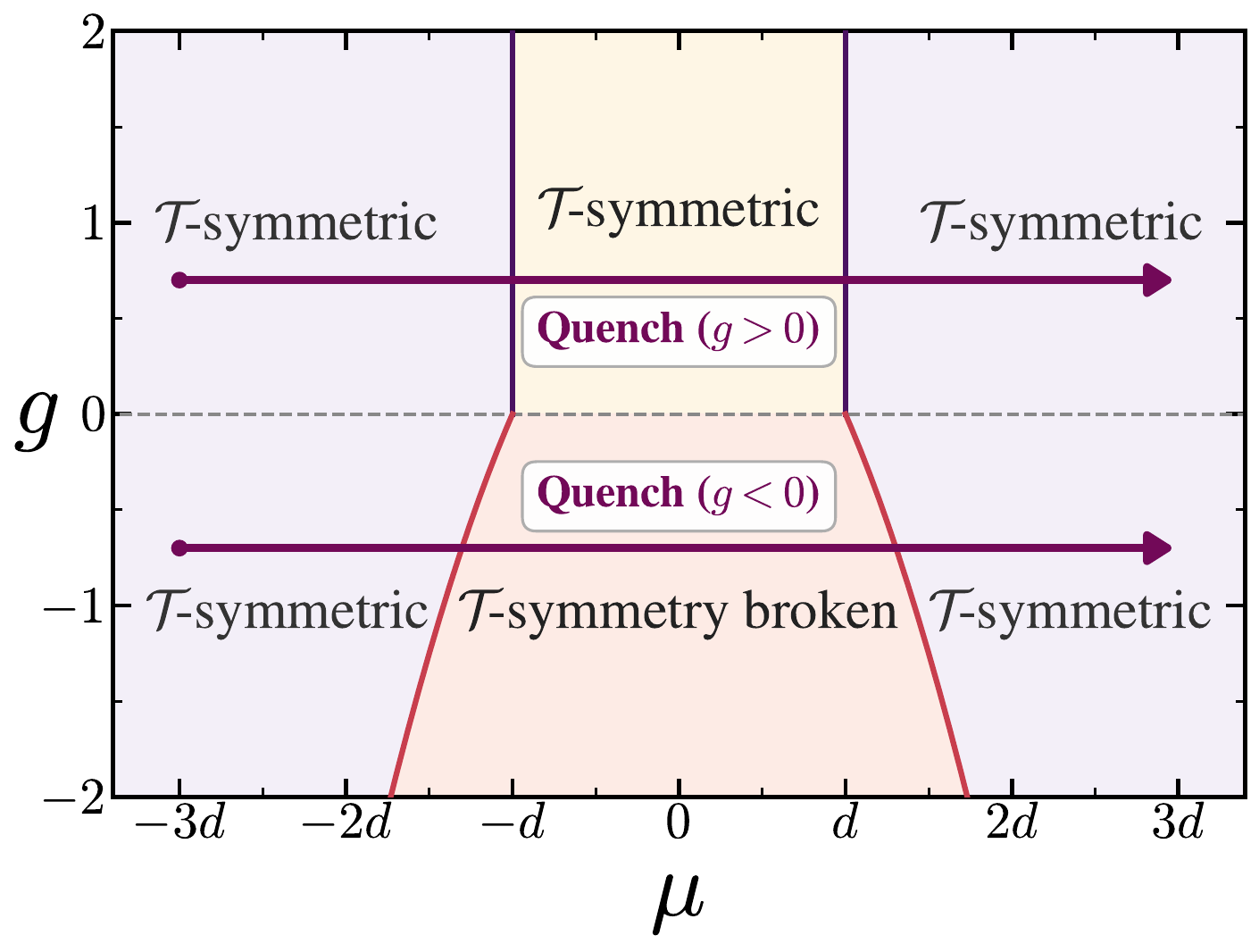}
			\put(5,78){\textbf{(a)}}
		\end{overpic}
	\end{minipage}\hfill
	\begin{minipage}[c]{0.32\textwidth}
		\centering
		\begin{overpic}[width=\linewidth]{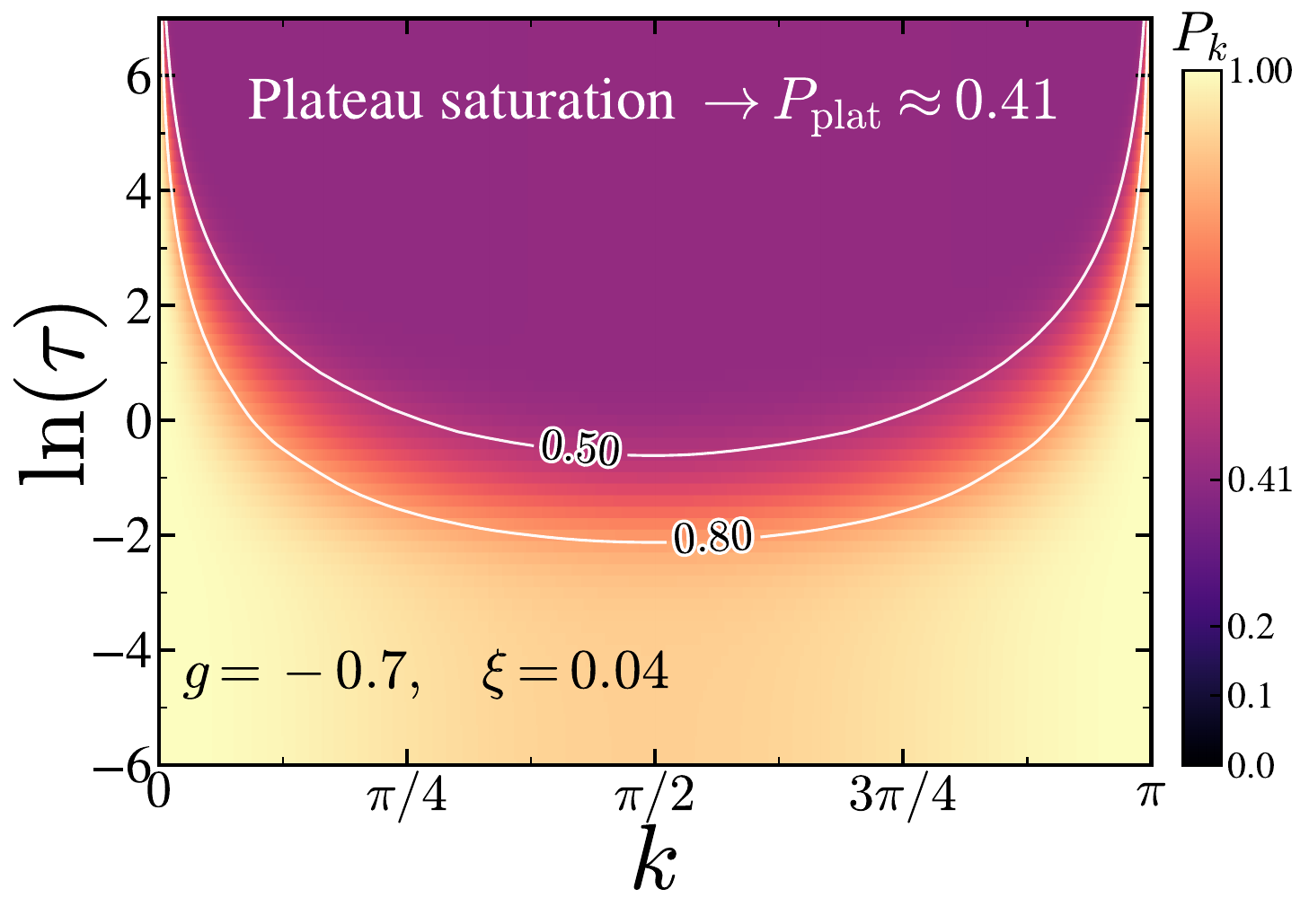}
			\put(5,75){\textbf{(b)}}
		\end{overpic}
	\end{minipage}\hfill
	\begin{minipage}[c]{0.32\textwidth}
		\centering
		\begin{overpic}[width=\linewidth]{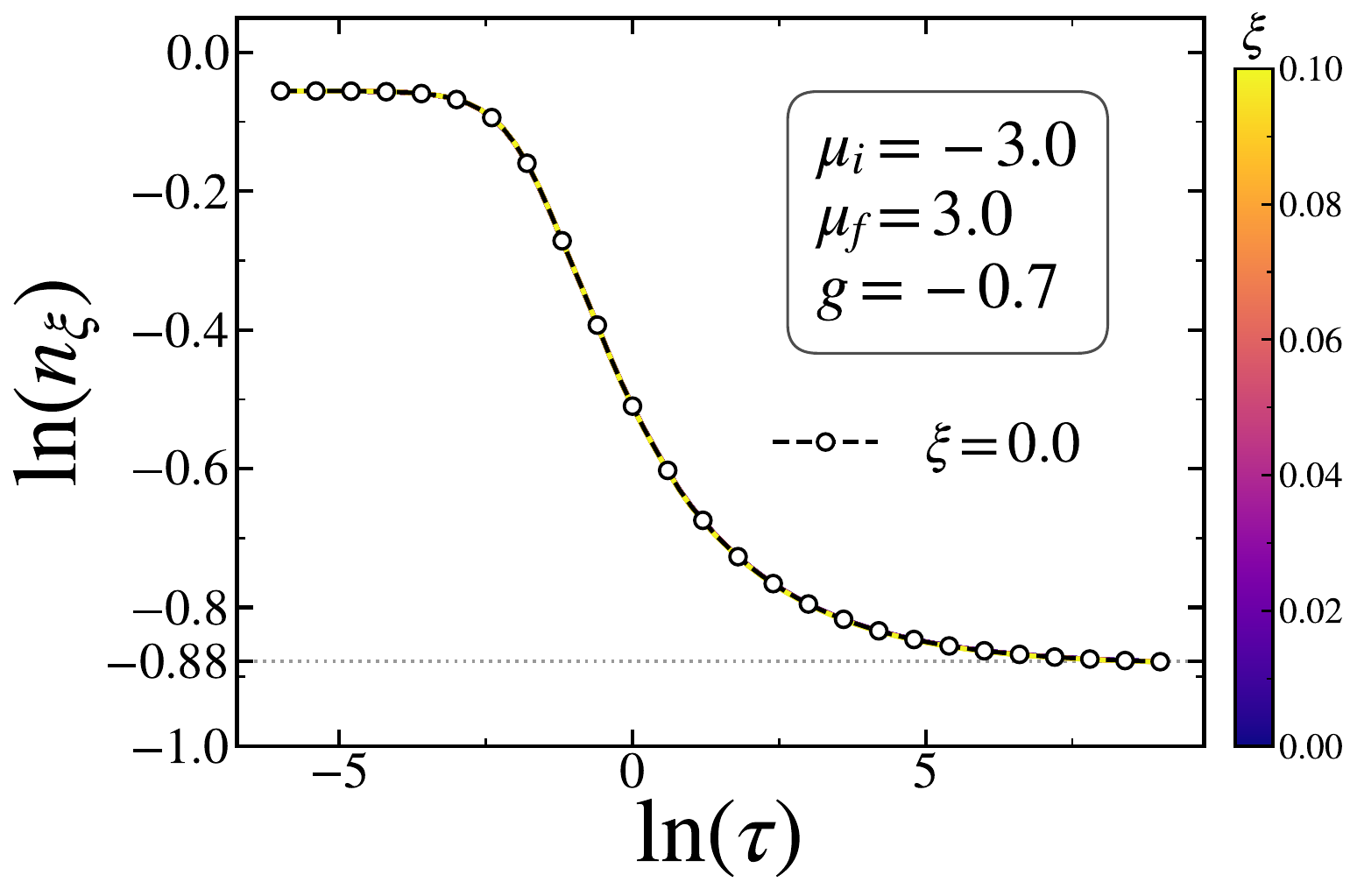}
			\put(5,75){\textbf{(c)}}
		\end{overpic}
	\end{minipage}
	\caption{\raggedright (a) Driven Kitaev phase diagram in the $(\mu, g)$ plane ($w = \Delta = 1$) along with quench paths. Quench through the $\mathcal{T}$-broken region at $g = -0.7 < 0$: (b) Momentum-resolved probability $P_k$ and the plateau saturation ($P_{\mathrm{plat}} \approx 0.41$) for long $\tau$ and (c) defect density locking onto $\ln(n_\xi) \approx -0.88$ across noise strength $\xi$. The horizontal dotted line in (c) corresponds to the analytical limit $\ln(P_{\text{plat}})$.
}
	\label{fig:main_results_three_panels_equal}
\end{figure*}
%
%
\begin{equation}
	H_{\boldsymbol{k}}(t) = 
	\begin{pmatrix}
		-\mu_{\boldsymbol{k}}(t) - \eta_H(t) & \Delta_{\boldsymbol{k}} \\
		g \Delta_{\boldsymbol{k}} & \mu_{\boldsymbol{k}}(t) + \eta_H(t)
	\end{pmatrix},
	\label{eq:BdG_Matrix}
\end{equation}
%
where the dispersion parameters are defined as
%
\begin{equation}
	\mu_{\boldsymbol{k}}(t) = \mu(t) + w \sum_{\alpha=1}^d \cos k_\alpha, \quad \Delta_{\boldsymbol{k}} = \Delta \sum_{\alpha=1}^d \sin k_\alpha,
	\label{eq:BdG_dispersion}
\end{equation}
%
with $\Delta_\pm = \frac{1\pm g}{2}\Delta_{\boldsymbol{k}}$. Here, $g \in \mathbb{R}$ parameterizes the pairing non-reciprocity (with $g=1$ recovering the Hermitian limit), and throughout this work we set the characteristic scales $w = \Delta = 1$. The chemical potential is swept linearly as $\mu(t) = t/\tau$ from $\mu_i$ at $t_i$ to $\mu_f$ at $t_f$ in the presence of stochastic longitudinal dephasing noise $\eta_H(t)$, which is characterized by the noise correlator $\langle \eta_H(t)\eta_H(t') \rangle = \xi^2 \delta(t-t')$.
Mapping the deterministic components of $H_{\boldsymbol{k}}(t)$ directly onto the generalized framework of Eq.~\eqref{eq:generalized_H}, the parameter vectors are identified as
%
\begin{equation}
	\mathbf{d}_{\boldsymbol{k}}(t) = (\Delta_+,\, 0,\, - \mu_{\boldsymbol{k}}(t)), \quad 
	\boldsymbol{\gamma}_{\boldsymbol{k}} = (0,\, -\Delta_-,\, 0).
	\label{eq:Kitaev_vector_mapping}
\end{equation}
%
In typical quench protocols, stochastic fluctuations in the control parameter predominantly couple along the driven axis [$\mu(t) \to \mu(t) + \eta_H(t)$]. This naturally gives rise to longitudinal dephasing noise ($\mathbf{m}_H = \hat{\mathbf{z}}$, $\xi_H = \xi$) in the absence of non-Hermitian fluctuations ($\xi_{\mathrm{NH}} = 0$). To demonstrate that the plateau emerges intrinsically from the microscopic dynamics, we independently integrate the full time-dependent master equation [Eq.~\eqref{eq:generalized_master}] for each $\boldsymbol{k}$-mode across the quench, initialized in the instantaneous right ground-state eigenvector $\rho_{\boldsymbol{k}}(t_i) = |\Phi_{g,\boldsymbol{k}}^R(t_i)\rangle\langle\Phi_{g,\boldsymbol{k}}^R(t_i)|$. 
Excitation dynamics are quantified via the normalized biorthogonal transition probability into the instantaneous left excited state $|\Phi_{e,\boldsymbol{k}}^L(t_f)\rangle$~\cite{Deng2025} at the ramp termination $t_f = \mu_f\tau$, given by 
$P_{\boldsymbol{k}}(t_f) = p_{e,\boldsymbol{k}}(t_f)/[p_{g,\boldsymbol{k}}(t_f)+p_{e,\boldsymbol{k}}(t_f)]$,
where $p_{\alpha,\boldsymbol{k}}(t_f)=\langle\Phi_{\alpha,\boldsymbol{k}}^L(t_f)|\rho_{\boldsymbol{k}}(t_f) |\Phi_{\alpha,\boldsymbol{k}}^L(t_f)\rangle$ ($\alpha=g, e$)~\cite{SuppMat}, yielding the total defect density $n_\xi = (2\pi)^{-d} \int_{[-\pi,\pi]^d} P_{\boldsymbol{k}}(t_f) \, d^d\boldsymbol{k}$.
As detailed in the phase diagram in Fig.~\ref{fig:main_results_three_panels_equal}(a), the phase boundaries are characterized by exceptional points at $\mu_c = \pm d\sqrt{1-g}$ for $g < 0$ and critical lines at $\mu_c = \pm d$ for $g > 0$.

%
\textit{a. Non-Hermitian Attractor Region ($g < 0$)---}For $g < 0$, the quench protocol traverses the $\mathcal{T}$-symmetry-broken phase [Fig.~\ref{fig:main_results_three_panels_equal}(a)], where the imaginary components of the eigenvalues trigger a positive exponential growth rate, $\Lambda_{+, \boldsymbol{k}} = 2\sqrt{|g|\Delta_{\boldsymbol{k}}^2 - \mu_{\boldsymbol{k}}^2} > 0$. Upon satisfying the conditions of \textit{Case~I} ($\gamma_z = 0$, $\mathcal{S}_{yz} = 0$, and $\boldsymbol{\gamma}_\perp \times \boldsymbol{\mathcal{D}}_\perp = \mathbf{0}$), the system dynamics undergo a contraction toward a noise-immune attractor:
%
\begin{equation}
	\mathcal{O}_z = -\frac{d_x}{\gamma_y} = \frac{1-|g|}{1+|g|}, \quad P_{\mathrm{plat}} = \frac{1-\mathcal{O}_z}{2} = \frac{|g|}{1+|g|}.
	\label{eq:attractor_plateau}
\end{equation}
%
As demonstrated in the momentum-resolved excitation landscape [Fig.~\ref{fig:main_results_three_panels_equal}(b)], the probability $P_{\boldsymbol{k}}$ for slow quenches ($\ln\tau \gtrsim 2$) saturates across the broad resonant momentum sector precisely onto the noise-shielded value $P_{\boldsymbol{k}} \to P_{\mathrm{plat}} \approx 0.41$ (for $g = -0.7$). 
Consequently, the macroscopic defect density asymptotically locks onto this single-mode value, $n_\xi \to P_{\mathrm{plat}}$ [corresponding to $\ln(P_{\mathrm{plat}}) \approx -0.887$]. 
This finding is corroborated by the numerical simulations in Fig.~\ref{fig:main_results_three_panels_equal}(c), which display an unambiguous locking at $\ln(n_\xi) \approx -0.88$ across diverse noise strengths $\xi$, collapsing entirely onto the noiseless limit ($\xi = 0$). 
Further numerical analysis confirms the robust persistence of this plateau saturation across a wide range of negative coupling strengths $g < 0$.

Furthermore, as detailed in Ref.~\cite{SuppMat}, this non-Hermitian attractor-induced noise shielding readily generalizes to two-dimensional systems. 
We note that a single-realization momentum-resolved probability $P_{\boldsymbol{k}}$ exhibits excellent agreement with the ensemble-averaged transition probability derived from the exact noise master equation. 
Moreover, for $g < 0$, a single-realization defect density $n_\xi$ coincides exactly with its noise-averaged counterpart, underscoring the robustly noise-immune nature of the imbalanced-pairing Kitaev model.
%

\begin{figure}[t!]
		\centering
\begin{minipage}{\linewidth}
		\begin{overpic}[width=0.49\columnwidth]{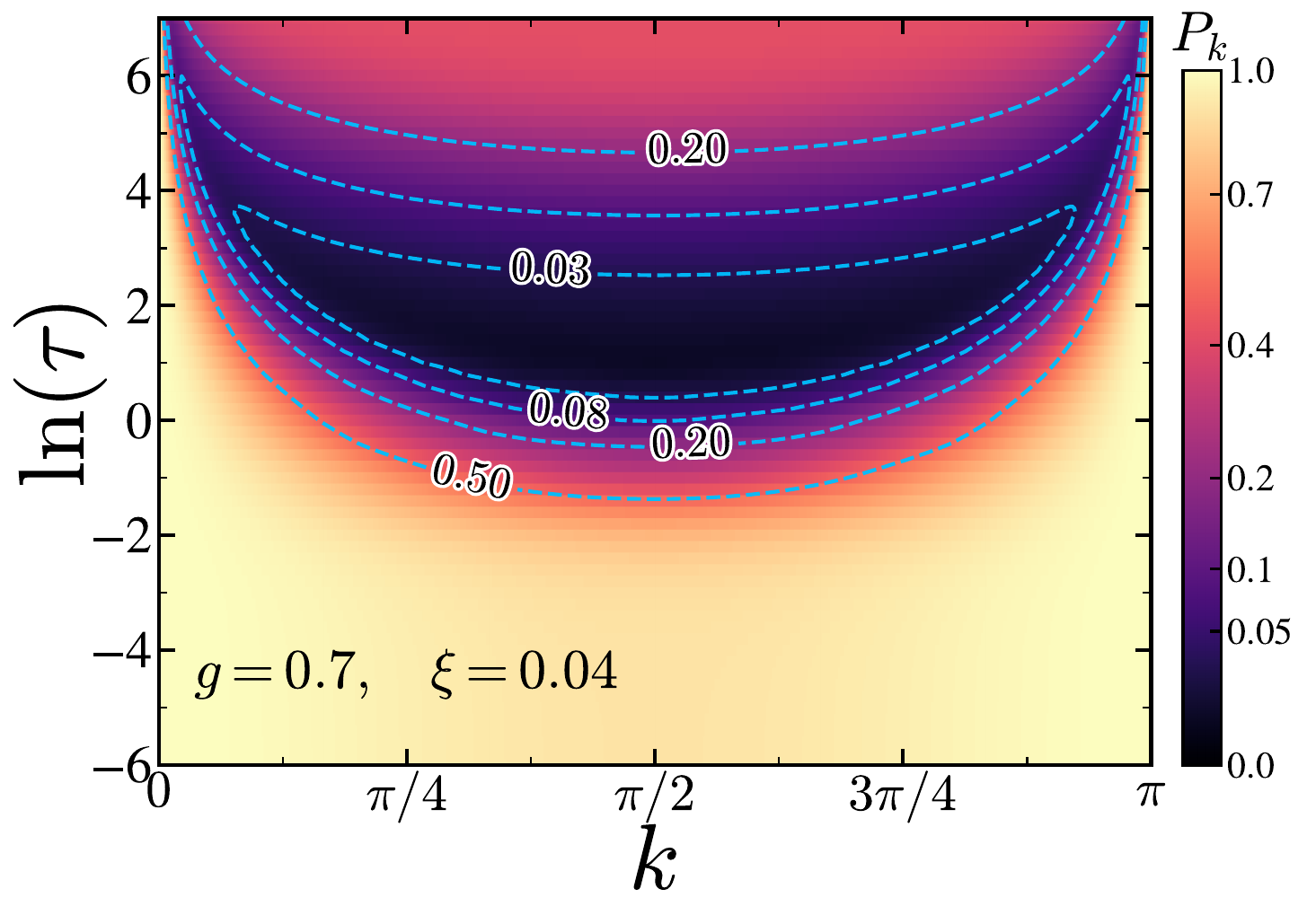}
			\put(5,78){\textbf{(a)}}
		\end{overpic}
		\centering
		\begin{overpic}[width=0.49\columnwidth]{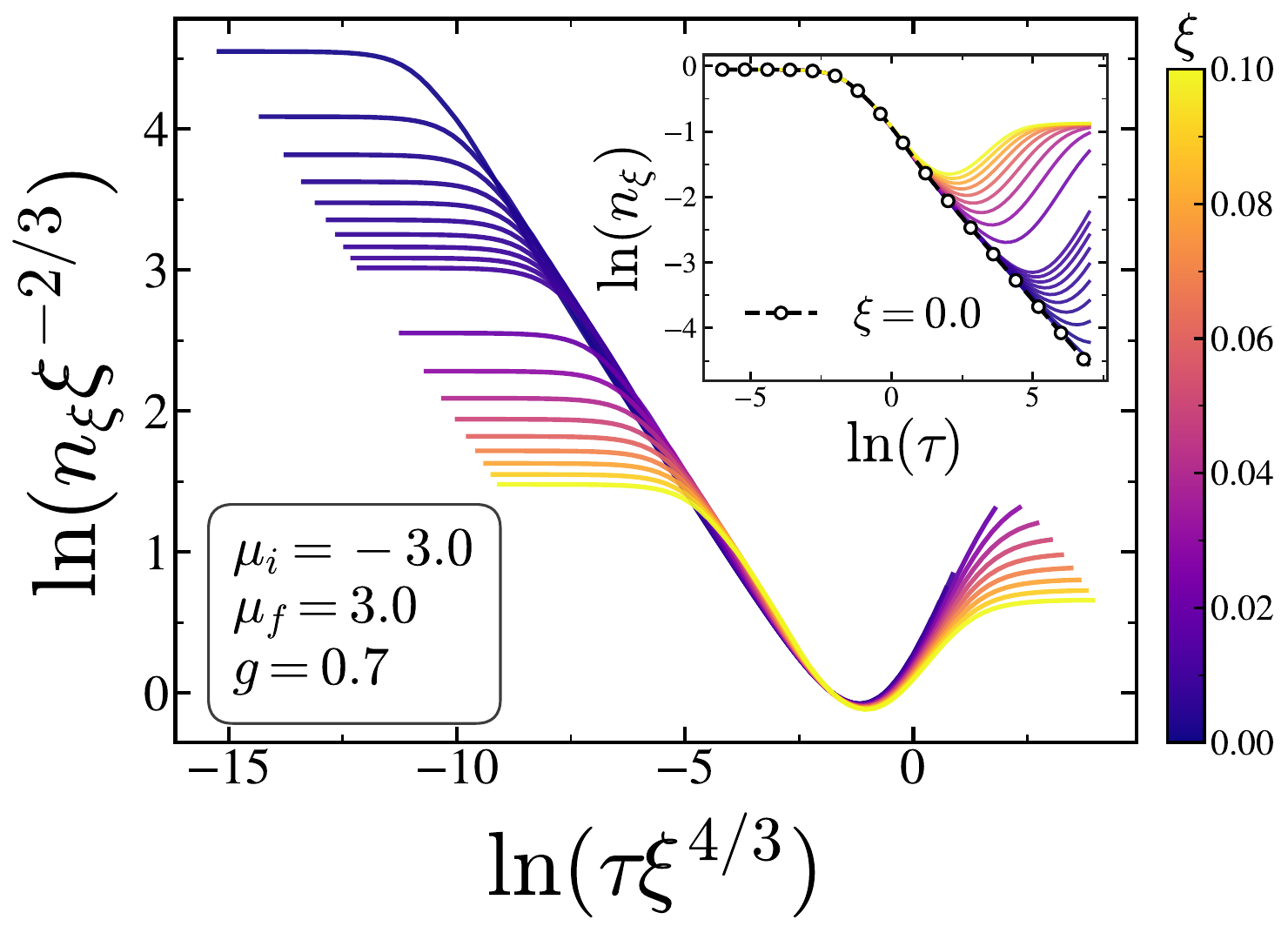}
			\put(5,75){\textbf{(b)}}
		\end{overpic}
\end{minipage}
\caption{(Color online) (a) Momentum-resolved probability $P_k$ for $g = 0.7 > 0$ ($\mathcal{PT}$-symmetric / AKZ). (b) Universal scaling collapse $n_\xi\xi^{-2/3}$ vs $\tau\xi^{4/3}$ (inset: raw defect density $\ln(n_\xi)$ vs $\ln\tau$ displaying non-monotonic AKZ behavior)}
	\label{fig:main_results_bottom}
\end{figure}

\textit{b. $\mathcal{T}$-symmetric region ($g > 0$)---}As the quench protocol traverses the $\mathcal{T}$-symmetric phase for $g > 0$ [Fig.~\ref{fig:main_results_three_panels_equal}(a)], the generator spectrum satisfies $\operatorname{Re}\{\Lambda_{+, \boldsymbol{k}}\} \le 0$ for all modes, thereby precluding the formation of a non-Hermitian attractor and preventing the convergence of $P_{\boldsymbol{k}}$ to a stable plateau.
Instead, the momentum-resolved excitation probability exhibits a non-monotonic landscape [Fig.~\ref{fig:main_results_bottom}(a)]: excitations reach a pronounced minimum at intermediate ramp rates $\tau$ before noise heating dominates at large $\tau$. 
Integrating over the Brillouin zone yields an anti-Kibble--Zurek (AKZ) crossover characterized by a non-monotonic defect density $n_\xi(\tau)$ [inset, Fig.~\ref{fig:main_results_bottom}(b)].

Our numerical simulations reveal that the optimal quench time scales as $\tau_{\mathrm{opt}} \propto \xi^{-4/3}$, while the minimum defect density scales as $n_{\min} \propto \xi^{2/3}$, yielding an excellent data collapse [Fig.~\ref{fig:main_results_bottom}(b)]. 
These scaling laws are consistent with the competition form $n_\xi(\tau) \approx A\tau^{-1/2} + B\xi^2\tau$ established for noisy Hermitian systems, confirming that universal AKZ scaling remains robust in this $\mathcal{T}$-symmetric driven system. 
Furthermore, benchmarking against individual trajectory unraveling via the stochastic Schrödinger equation (SSE) confirms excellent quantitative agreement across all parameter regimes~\cite{SuppMat}.

\textit{The Non-Hermitian SSH Model---}To illustrate \textit{Case III}, we analyze the Su--Schrieffer--Heeger (SSH) model subjected to an imaginary staggered potential with fluctuations~\cite{Lieu2018,SuppMat}:
%
\begin{equation}
	\no
	H_k(t) = (v + w \cos k) \sigma_x + w \sin k \sigma_y + i\big(u(t) + \eta(t)\big) \sigma_z,
\end{equation} 
where $v$ and $w$ denote the intra- and inter-cell hopping amplitudes, respectively, while $u(t)$ represents an imaginary staggered potential subjected to stochastic fluctuations $\eta(t)$.
Depending on the Liouvillian gap structure, these parameter trajectories exhibit qualitatively distinct nonequilibrium dynamics. 
For quenches into the globally $\mathcal{PT}$-symmetry-broken region (e.g., $u_i = 0 \to u_f = 2.5$) past a critical threshold $u_f > u_T \approx 2$, the $\mathcal{PT}$-symmetry-broken sector spans the entire Brillouin zone. 
In this regime, for sufficiently slow ramps, the leading Liouvillian eigenmode acquires a strictly positive real part across all momenta, generating a dominant dynamical attractor and a \textit{noise-tunable plateau} with quadratic defect scaling, $n_{\text{plat}} \propto \xi^2$. 
Conversely, trajectories terminating below this threshold ($u_f < u_T$) fail to establish an attractor spanning the entire Brillouin zone. 
Consequently, prolonged exposure to localized dissipative amplification during slow ramps yields anti-Kibble--Zurek (AKZ) scaling without an asymptotic plateau.

\textit{Experimental Feasibility---}The predicted phenomenology can be investigated in laser-written photonic waveguide arrays~\cite{Szameit2010,Cerjan2019,Slootman2024}, where the propagation coordinate $z$ emulates time. Alternating waveguide spacings implement dimerized SSH couplings, while an appropriate encoding of particle-hole degrees of freedom enables realization of the Kitaev model. Sublattice-dependent losses provide a route to engineering non-Hermitian dissipation, whereas controlled longitudinal modulations introduce effective stochastic fluctuations, with sufficiently short correlation lengths approximating white noise. Multiple samples with independent disorder realizations allow for the reconstruction of ensemble-averaged observables. For the SSH model, polarization is extracted from the normalized sublattice intensity imbalance $(I_A - I_B)/(I_A + I_B)$, with momentum-resolved measurements accessible via Fourier-plane imaging. These techniques enable the experimental distinction between the noise-insensitive Kitaev plateau and the noise-dependent SSH response, while testing the predicted quadratic scaling $n_{\mathrm{plat}} \propto \xi^2$ through systematic tuning of the fluctuation amplitude.

%
%
\textit{Conclusion and Discussion---}In summary, we have formulated a general Liouvillian framework that unifies and classifies nonequilibrium plateau dynamics in driven quantum systems via an exact geometric compatibility condition.
Crucially, the emergence of a Liouvillian attractor provides a transparent mechanism that elucidates defect freezing in non-Hermitian systems~\cite{Sim2023,Deng2025}. 
This framework naturally establishes distinct universal dynamical regimes: In the non-Hermitian Kitaev model [Case (I)], the attractor strictly pins the physical polarization, yielding an absolute, noise-independent plateau where individual trajectories coincide with the ensemble average~\cite{SuppMat}.
Conversely, in the non-Hermitian SSH model under a staggered-potential quench [Case (II)], exact Krein-charge conservation is insufficient to enforce full noise immunity; transverse fluctuations generate bounded deviations around a noise-dependent plateau ($n_{\mathrm{plat}} \propto \xi^2$). 
Our findings demonstrate that dynamical resilience against stochastic noise is fundamentally governed by the spectral gap and mode alignment of the Liouvillian generator rather than static conservation laws alone, opening new avenues for probing driven open quantum matter. 
Notably, although uncorrelated fluctuations are assumed for analytical tractability, the attractor dynamics remain robust against correlated (colored) noise with a finite correlation time $\tau_c$, provided $\tau_c \ll \tau_{\mathrm{relax}} \sim [\min_{j\neq+} \mathrm{Re}\{\Lambda_{+,\boldsymbol{k}} - \Lambda_{j,\boldsymbol{k}}\}]^{-1}$, thereby supporting broad physical applicability in realistic experimental settings.

\textit{Acknowledgments}---The authors would like to acknowledge financial support from Sharif University of Technology through Grant No.~QA050518.

\newpage
\appendix

\clearpage
\onecolumngrid  

\setcounter{equation}{0}
\setcounter{figure}{0}
\setcounter{table}{0}
\setcounter{page}{1}
\setcounter{section}{0}

\renewcommand{\theequation}{S\arabic{equation}}
\renewcommand{\thefigure}{S\arabic{figure}}
\renewcommand{\thetable}{S\arabic{table}}
\renewcommand{\thesection}{S\Roman{section}}
\renewcommand{\thepage}{S\arabic{page}}

\section*{Supplemental Material}
In this Supplemental Material, we provide additional technical details, extended mathematical derivations, and supplementary numerical results, which are intended to complement the findings reported in the main text \cite{main}.

\section{Derivation of the Master Equation}
\label{sec:master_equation_derivation}

In this section, we provide a detailed derivation of the master equation governing the noise-averaged density matrix $\rho(t) = \langle \rho_\xi(t) \rangle_\xi$,

Suppose that we can decompose the Hamiltonian into deterministic and stochastic parts, $H(t) = H_{\text{det}} + V(t)$, where the deterministic part is given by:
\begin{equation}
	H_{\text{det}} = H_0 - i\Gamma,
\end{equation}

and the stochastic perturbation reads:
\begin{equation}
	V(t) = \eta_H(t) M_H - i \eta_{\text{NH}}(t) M_{\text{NH}},
\end{equation}
with the defined Hermitian operators $M_H$ and $M_{\text{NH}}$. 
The noises $\eta_H(t)$ and $\eta_{\text{NH}}(t)$ are independent Gaussian white noises with zero mean and variance $\langle \eta_\nu(t)\eta_\nu(t') \rangle = \xi_\nu^2 \delta(t-t')$ for $\nu \in \{H, \text{NH}\}$.

For a specific noise realization $\xi$, the time evolution of the unnormalized state vector $|\psi^R_\xi(t)\rangle$ is governed by the time-dependent Schrödinger equation $d|\psi^R_\xi(t)\rangle/dt = -i H(t) |\psi^R_\xi(t)\rangle$ with the initial condition specified by the ground state of the right eigenvectors. The corresponding single-realization density matrix $\rho_\xi(t) = |\psi^R_\xi(t)\rangle \langle \psi^R_\xi(t)|$ obeys the Liouville-von Neumann equation:
\begin{equation}
	\frac{d\rho_\xi(t)}{dt} = -i H(t) \rho_\xi(t) + i \rho_\xi(t) H^\dagger(t).
	\label{eq:S_LvN}
\end{equation}
Substituting $H(t)$ and its Hermitian conjugate $H^\dagger(t) = H_0 + i\Gamma + \eta_H(t) M_H + i \eta_{\text{NH}}(t) M_{\text{NH}}$ into Eq.~\eqref{eq:S_LvN}, we obtain the stochastic differential equation:
\begin{equation}
	\frac{d\rho_\xi}{dt} = -i[H_0, \rho_\xi] - \{\Gamma, \rho_\xi\} - i \eta_H(t) [M_H, \rho_\xi] - \eta_{\text{NH}}(t) \{M_{\text{NH}}, \rho_\xi\}.
	\label{eq:S_rho_xi_dot}
\end{equation}

To obtain the physical dynamics, we take the ensemble average over all noise realizations, $\rho(t) = \langle \rho_\xi(t) \rangle_\xi$:
\begin{equation}
	\frac{d\rho}{dt} = -i[H_0, \rho] - \{\Gamma, \rho\} - i \langle \eta_H(t) [M_H, \rho_\xi] \rangle_\xi - \langle \eta_{\text{NH}}(t) \{M_{\text{NH}}, \rho_\xi\} \rangle_\xi.
	\label{eq:S_rho_avg_dot}
\end{equation}
To evaluate the noise-correlation terms, we employ Novikov's theorem~\cite{NovikovSM}, which states that for a Gaussian colored noise $\eta(t)$ and a functional $F[\eta]$, the correlation is given by $\langle \eta(t) F[\eta] \rangle = \int dt' \langle \eta(t)\eta(t') \rangle \langle \delta F[\eta] / \delta \eta(t') \rangle$. For the white noise limit $\langle \eta_\nu(t)\eta_\nu(t') \rangle = \xi_\nu^2 \delta(t-t')$, evaluated in the Stratonovich sense (which correctly describes physical broadband noise limits), the delta function integration yields a factor of $1/2$:
\begin{equation}
	\langle \eta_\nu(t) F[\rho_\xi(t)] \rangle_\xi = \frac{\xi_\nu^2}{2} \left\langle \frac{\delta F[\rho_\xi(t)]}{\delta \eta_\nu(t)} \right\rangle_\xi.
	\label{eq:S_Novikov}
\end{equation}
From the equation of motion \eqref{eq:S_rho_xi_dot}, the functional derivatives of the density matrix with respect to the noise fields at the same time $t$ are given by:
\begin{equation}
	\frac{\delta \rho_\xi(t)}{\delta \eta_H(t)} = -i [M_H, \rho_\xi(t)], \label{eq:S_func_H}
\end{equation}
and
\begin{equation}
	\frac{\delta \rho_\xi(t)}{\delta \eta_{\text{NH}}(t)} = - \{M_{\text{NH}}, \rho_\xi(t)\}. \label{eq:S_func_NH}
\end{equation}

Now, applying Eq.~\eqref{eq:S_Novikov} and Eq.~\eqref{eq:S_func_H} to the Hermitian noise term in Eq.~\eqref{eq:S_rho_avg_dot}:
\begin{equation}
	-i \langle \eta_H(t) [M_H, \rho_\xi] \rangle_\xi = -i \frac{\xi_H^2}{2} \left\langle \left[ M_H, \frac{\delta \rho_\xi}{\delta \eta_H(t)} \right] \right\rangle_\xi = -i \frac{\xi_H^2}{2} \langle [M_H, -i[M_H, \rho_\xi]] \rangle_\xi = - \frac{\xi_H^2}{2} [M_H, [M_H, \rho]].
	\label{eq:S_term_H}
\end{equation}
Similarly, applying Eq.~\eqref{eq:S_Novikov} and Eq.~\eqref{eq:S_func_NH} to the anti-Hermitian noise term yields:
\begin{equation}
	- \langle \eta_{\text{NH}}(t) \{M_{\text{NH}}, \rho_\xi\} \rangle_\xi = - \frac{\xi_{\text{NH}}^2}{2} \left\langle \left\{ M_{\text{NH}}, \frac{\delta \rho_\xi}{\delta \eta_{\text{NH}}(t)} \right\} \right\rangle_\xi = - \frac{\xi_{\text{NH}}^2}{2} \langle \{M_{\text{NH}}, -\{M_{\text{NH}}, \rho_\xi\}\} \rangle_\xi = + \frac{\xi_{\text{NH}}^2}{2} \{M_{\text{NH}}, \{M_{\text{NH}}, \rho\}\}.
	\label{eq:S_term_NH}
\end{equation}

Notice the crucial sign difference: the double commutator for Hermitian noise introduces a negative sign (dephasing), while the double anticommutator for anti-Hermitian fluctuations yields a positive sign, reflecting the amplification/attenuation inherent to non-Hermitian gain/loss dynamics.

Finally, substituting Eqs.~\eqref{eq:S_term_H} and \eqref{eq:S_term_NH} back into Eq.~\eqref{eq:S_rho_avg_dot} 
we recover the exact master equation:
\begin{equation}
	\frac{d\rho}{dt} = -i[H_0, \rho] - \{\Gamma, \rho\} - \frac{\xi_H^2}{2} \big[M_H , [M_h, \rho] \big] + \frac{\xi_{\text{NH}}^2}{2} \big\{ M_{NH}, \{M_{NH}, \rho\} \big\}.
\end{equation}

For two-level Hamiltonians defined in Eq. (1) in the main text $H_0 = \mathbf{d} \cdot \boldsymbol{\sigma}$, $\Gamma = \boldsymbol{\gamma} \cdot \boldsymbol{\sigma}$, $M_\nu = \mathbf{m}_\nu \cdot \boldsymbol{\sigma} \quad (\nu \in \{H, \text{NH}\})$,
the exact master equation reduces to: 
\begin{equation}
	\frac{d\rho}{dt} = -i[H_0, \rho] - \{\Gamma, \rho\} - \frac{\xi_H^2}{2} \big[ (\mathbf{m}_H \cdot \boldsymbol{\sigma}), [(\mathbf{m}_H \cdot \boldsymbol{\sigma}), \rho] \big] + \frac{\xi_{\text{NH}}^2}{2} \big\{ (\mathbf{m}_{\text{NH}} \cdot \boldsymbol{\sigma}), \{(\mathbf{m}_{\text{NH}} \cdot \boldsymbol{\sigma}), \rho\} \big\}.
\end{equation}
\section{Adiabatic Following of the Dominant Liouvillian Mode and Validity of the Dynamical Plateau}
\label{sec:adiabatic_following}

For the linear time dependent ramp, as used in the Kitaev and SSH models [e.g., $\mu(t) = t/\tau$], the underlying generator is intrinsically time-dependent:
\begin{equation}
	\dot{R}(t) = \mathcal{L}(t)R(t), \quad R(t) = \mathcal{T} \exp\left[\int_0^t \mathcal{L}(s) ds\right] R(0),
	\label{eq:time_evolution}
\end{equation}
where $\mathcal{T}$ denotes chronological time ordering. A crucial theoretical question is whether the instantaneous spectral dominance condition, $\text{Re}\, \Lambda_+(t) > \max_{j \neq +} \text{Re}\, \Lambda_j(t)$, rigorously guarantees that the physical density vector follows the instantaneous dominant eigenmode:
\begin{equation}
	R(t) \propto V_+^R(t).
\end{equation}

To address this, we consider the instantaneous biorthogonal eigensystem of $\mathcal{L}(t)$:
\begin{equation}
	\mathcal{L}(t)|V_j^R(t)\rangle = \Lambda_j(t)|V_j^R(t)\rangle, \quad \langle V_i^L(t)|\mathcal{L}(t) = \Lambda_i(t)\langle V_i^L(t)|,
\end{equation}
satisfying the biorthogonal normalization $\langle V_i^L(t)|V_j^R(t)\rangle = \delta_{ij}$ away from exceptional points. Expanding the instantaneous state vector in this instantaneous basis,
\begin{equation}
	|R(t)\rangle = \sum_j c_j(t) \exp\left[\int_0^t \Lambda_j(s) ds\right] |V_j^R(t)\rangle,
	\label{eq:instantaneous_basis_expansion}
\end{equation}
and substituting Eq.~\eqref{eq:instantaneous_basis_expansion} into Eq.~\eqref{eq:time_evolution}, projection onto $\langle V_i^L(t)|$ yields the exact equations of motion for the transition amplitudes:
\begin{equation}
	\dot{c}_i(t) = - \sum_{j \neq i} c_j(t) \langle V_i^L(t)| \partial_t V_j^R(t) \rangle \exp\left[ \int_0^t [\Lambda_j(s) - \Lambda_i(s)] ds \right].
\end{equation}

The off-diagonal non-adiabatic couplings can be rewritten via the Hellmann-Feynman relation:
\begin{equation}
	\langle V_i^L(t)| \partial_t V_j^R(t) \rangle = \frac{\langle V_i^L(t)| \dot{\mathcal{L}}(t) |V_j^R(t)\rangle}{\Lambda_j(t) - \Lambda_i(t)} \quad (i \neq j).
\end{equation}

Consequently, the condition for adiabatic mode following without inter-mode transitions requires that the geometric rotation rate remains negligible compared to the instantaneous complex Liouvillian spectral gap:
\begin{equation}
	\left| \frac{\langle V_i^L(t)|\dot{\mathcal{L}}(t)|V_j^R(t)\rangle}{[\Lambda_+(t) - \Lambda_j(t)]^2} \right| \ll 1 \quad \forall j \neq +.
\end{equation}

For a linear time driven protocol with rate $v \equiv \tau^{-1}$, the time derivative satisfies $\dot{\mathcal{L}}(t) \propto 1/\tau$. In the slow-driving (large $\tau$) limit, the non-adiabatic transition rate is suppressed as $\mathcal{O}(\tau^{-1})$. For non-linear quench protocols $\mu(t)$, the adiabatic condition is governed by the time-dependent driving rate $\dot{\mu}(t)$; while such protocols generally disrupt universal scaling laws, the mode purification mechanism remains robust provided the driving speed satisfies the local adiabatic criterion throughout the trajectory. Furthermore, since $\text{Re}[\Lambda_+(s) - \Lambda_j(s)] \equiv \Delta_j(s) > 0$ along the plateau manifold, any non-adiabatic excitation into transient modes $j \neq +$ is exponentially damped by the dynamical Liouvillian factor:
\begin{equation}
	\exp\left[ - \int_{t'}^t \Delta_j(s) ds \right] \to 0 \quad \text{for } t - t' \gg \Delta_j^{-1}.
\end{equation}
Thus, mode purification dynamically restores alignment with $V_+^R(t)$.

\textit{Structural Invariance in case (I).}— Importantly, we note that the emergence of the noise-independent plateau in case (I) does not strictly rely on instantaneous adiabatic tracking of the full eigenvector. In case (I), the target observable $O_z = \text{Tr}(\sigma_z \rho)/\text{Tr}(\rho) = z/I_0$ is governed by a persistent projective symmetry of the Liouvillian generator. Because the effective shifts act exclusively along complementary directions ($\mathcal{S}_{xz} = \mathcal{S}_{yz} = 0$), the orientation of the dominant projective subspace remains strictly invariant along the quench trajectory. Therefore, the noise-independent plateau is not an accidental asymptotic numerical artifact, and the exact structural property of the driven non-Hermitian master equation protected by the underlying Liouvillian geometry.

\section{Geometric Compatibility Condition and Classification of Plateau Regimes}

To establish the physical and mathematical origin of noise cancellation, we analyze the structure of the dominant Liouvillian eigenmode $V_+ = (I_0^*, x^*, y^*, z^*)^T$ governing the asymptotic polarization $\mathcal{O}_z \equiv z^*/I_0^*$.

The secular equation $\mathcal{L} V_+ = \Lambda_+ V_+$, with $\mathcal{L}$ defined in \eqref{eq:L_gen_compact}, implies that the coupling between transverse fluctuations $(x^*, y^*)$ and the longitudinal sector is mediated by an effective transverse drive vector, defined as:
\begin{equation}
	\mathbf{D}_\perp \equiv (-2d_y + \mathcal{S}_{xz}, \, 2d_x + \mathcal{S}_{yz})^T.
\end{equation}
When the deterministic Hamiltonian includes a non-vanishing transverse component $\boldsymbol{\gamma}_\perp \equiv (\gamma_x, \gamma_y)^T \neq \mathbf{0}$, steady-state noise immunity requires that the dissipative vector $\boldsymbol{\gamma}_\perp$ collimates with the effective drive $\mathbf{D}_\perp$. This condition, which ensures the Hamiltonian drift suppresses noise-induced dephasing, yields the cross-product compatibility condition:
\begin{equation}
	\boldsymbol{\gamma}_\perp \times \mathbf{D}_\perp \equiv \gamma_x (2d_x + \mathcal{S}_{yz}) - \gamma_y (-2d_y + \mathcal{S}_{xz}) = 0.
	\label{eq:geometric_compatibility}
\end{equation}

Starting from the coupled secular equations for the identity and longitudinal components derived from \eqref{eq:L_gen_compact}:
\begin{align}
	\tilde{\Lambda}_0 I_0^* &= -2\gamma_z z^* - 2(\gamma_x x^* + \gamma_y y^*), \nonumber \\
	\tilde{\Lambda}_z z^* &= -2\gamma_z I_0^* + (-2d_y + \mathcal{S}_{xz})x^* + (2d_x + \mathcal{S}_{yz})y^*,
\end{align}
we utilize the gauge freedom in the transverse plane to choose $\gamma_x = 0$ without loss of generality. Under this choice, the compatibility condition \eqref{eq:geometric_compatibility} imposes the constraint $-2d_y + \mathcal{S}_{xz} = 0$, which effectively decouples the $y^*$ fluctuations from the primary system. The equations simplify to:
\begin{align}
	\tilde{\Lambda}_0 I_0^* &= -2\gamma_z z^* - 2\gamma_y y^*, \nonumber \\
	\tilde{\Lambda}_z z^* &= -2\gamma_z I_0^* + (2d_x + \mathcal{S}_{yz})y^*.
\end{align}
To solve for $\mathcal{O}_z = z^*/I_0^*$, we isolate $y^*$ from the first equation, $y^* = -(\tilde{\Lambda}_0 I_0^* + 2\gamma_z z^*)/(2\gamma_y)$, and substitute it into the second:
\begin{equation}
	\tilde{\Lambda}_z z^* = -2\gamma_z I_0^* + (2d_x + \mathcal{S}_{yz}) \left[ -\frac{\tilde{\Lambda}_0 I_0^* + 2\gamma_z z^*}{2\gamma_y} \right].
\end{equation}
Rearranging terms to group $z^*$ and $I_0^*$ yields:
\begin{equation}
	z^* \left( \tilde{\Lambda}_z + \frac{2\gamma_z(2d_x + \mathcal{S}_{yz})}{2\gamma_y} \right) = -I_0^* \left( 2\gamma_z + \frac{\tilde{\Lambda}_0(2d_x + \mathcal{S}_{yz})}{2\gamma_y} \right).
\end{equation}
Multiplying by $2\gamma_y$ and dividing by $I_0^*$ results in the exact closed-form expression for the plateau value:
\begin{equation}
	\mathcal{O}_z = \frac{4\gamma_y \gamma_z + (2d_x + \mathcal{S}_{yz})\tilde{\Lambda}_0}{-2\gamma_y \tilde{\Lambda}_z - 2\gamma_z(2d_x + \mathcal{S}_{yz})}.
	\label{eq:plateau_formula}
\end{equation}
This derivation explicitly demonstrates that when the compatibility condition \eqref{eq:geometric_compatibility} is satisfied, dynamic mode purification locks the polarization $\mathcal{O}_z$ into this noise-independent value, ensuring the attractor's robustness against longitudinal stochastic dephasing. 

Whenever Eq. \eqref{eq:geometric_compatibility} holds with $\gamma_\perp \neq 0$, this process locks the ratio $z/I_0$ into a noise-independent value for both cases (I) and (II), since the effective shift $\tilde{\Lambda}$ factors out symmetrically from both the numerator and the denominator, yielding $\mathcal{O}_z = -d_x/\gamma_y$ and $\mathcal{O}_z = \pm 1$ at exceptional boundaries.

\paragraph{\textbf {Breakdown of Compatibility and Transition to Case~(III).}}
The protection mechanism leading to a noise-independent plateau breaks down under two distinct geometric configurations:
\begin{enumerate}
	\item{Explicit Geometric Misalignment:} The transverse fields are non-zero ($\boldsymbol{\gamma}_\perp \neq \mathbf{0}$), yet they fail to satisfy the geometric alignment condition $\boldsymbol{\gamma}_\perp \times \mathbf{D}_\perp \neq \mathbf{0}$.
	 In this case, transverse and longitudinal sectors remain dynamically coupled, preventing the factorization of noise rates and leading to a complete loss of protection.
	\item{Longitudinal Drift Singularities / Trivial Collinearity:} The non-Hermitian deterministic part of the Hamiltonian is strictly longitudinal ($\boldsymbol{\gamma}_\perp = \mathbf{0}$, $\gamma_z \neq 0$), as realized in the non-Hermitian SSH model where $\boldsymbol{\gamma} = (0, 0, -u(t))^T$. Although the algebraic cross-product vanishes identically ($\boldsymbol{\gamma}_\perp \times \mathbf{D}_\perp \equiv \mathbf{0}$), this trivial vanishing does \emph{not} enforce noise cancellation. Instead, the absence of the transverse non-Hermitian sector eliminates the denominator responsible for noise cancellation, and the longitudinal noise operator ($\mathbf{m}_{\mathrm{NH}} = \hat{\mathbf{z}}$) introduces an intrinsic eigenvalue anisotropy:
	\begin{equation}
		\tilde{\Lambda}_0 \neq \tilde{\Lambda}_z.
	\end{equation}
\end{enumerate}
Consequently, the effective shifts fail to cancel, leaving the dominant polarization explicitly parameterized by the noise variance:
\begin{equation}
	\mathcal{O}_z = \mathcal{O}_z(\xi_{\mathrm{NH}}).
\end{equation}
This operational breakdown establishes the foundational boundary for \textbf{\bfseries case~(III) (Noise-Tunable Plateau)}, transforming the system from a noise-immune topological simulator into an environmentally responsive quantum sensor.

\section{Numerical Benchmarking: Master Equation vs. Stochastic Schr\"odinger Equation (SSE)}
\label{sec:SSE}

To independently verify the validity of the ensemble-averaged master equation and thoroughly examine 
the physical reality of dynamical plateau behaviors at the single-shot level, we simulate the non-equilibrium 
quench dynamics via the stochastic Schr\"odinger equation (SSE) for individual unraveled trajectories. 
For each momentum mode $k$, the time evolution is governed by the It\^o-type stochastic differential equation:
\begin{equation}
	d|\psi_{k}(t)\rangle = -i H_{k}(t) |\psi_{k}(t)\rangle dt 
	- i \xi \sigma_z |\psi_{k}(t)\rangle dW(t) 
	- \frac{\xi^2}{2} |\psi_{k}(t)\rangle dt,
	\label{eq:SSE_dynamical}
\end{equation}
where $dW(t)$ is a Wiener increment satisfying $\mathbb{E}[dW(t)] = 0$ and $\mathbb{E}[dW(t)^2] = dt$, 
and $\xi$ parameterizes the longitudinal dephasing noise along $\sigma_z$.
\begin{figure*}[t]
	\centering
	\begin{minipage}[b]{0.49\textwidth}
		\centering
		\begin{overpic}[width=\textwidth]{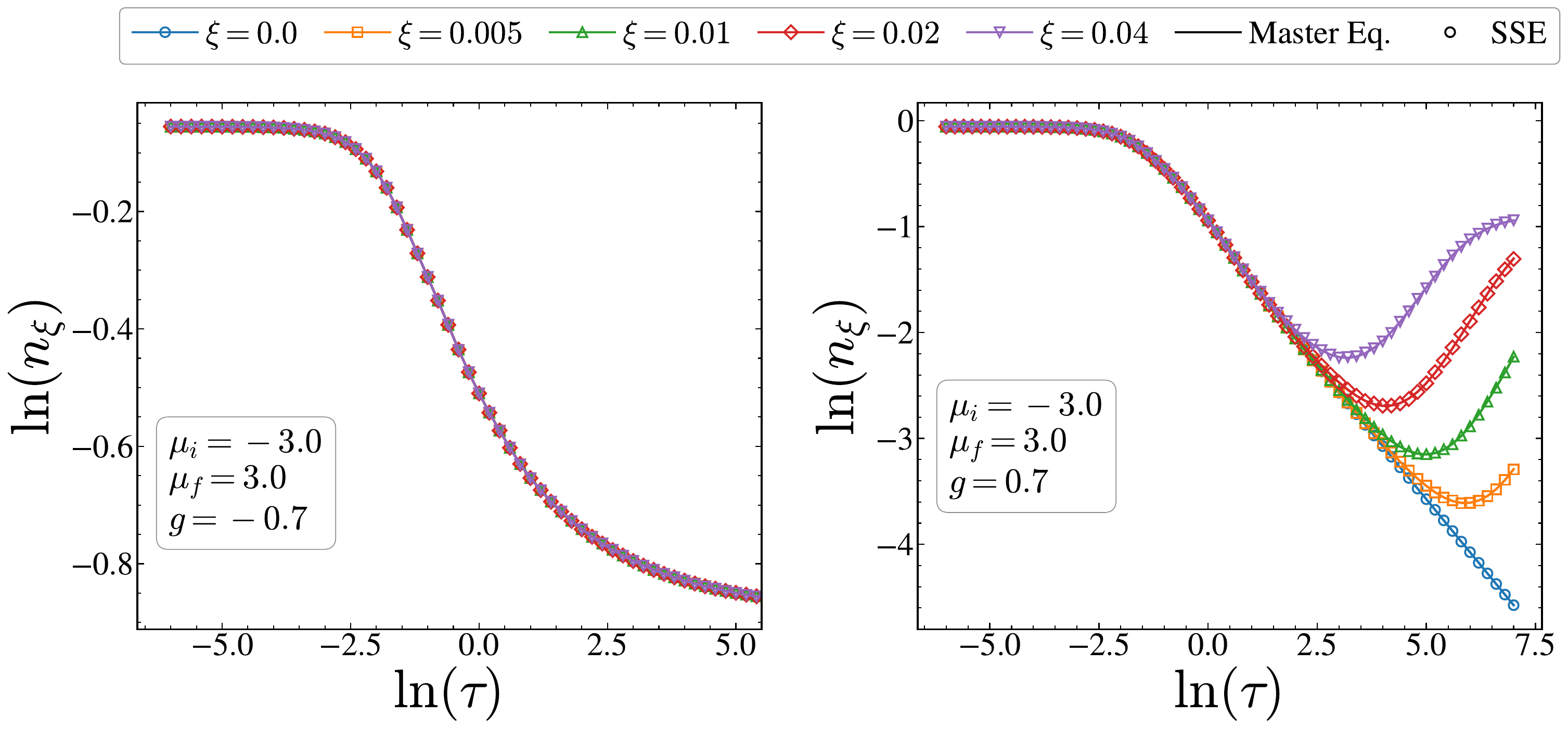}
			\put(1, 36){\textbf{(a)}}
			\put(51, 36){\textbf{(b)}}
		\end{overpic}
	\end{minipage}
	\hfill
	\begin{minipage}[b]{0.245\textwidth}
		\centering
		\begin{overpic}[width=\textwidth]{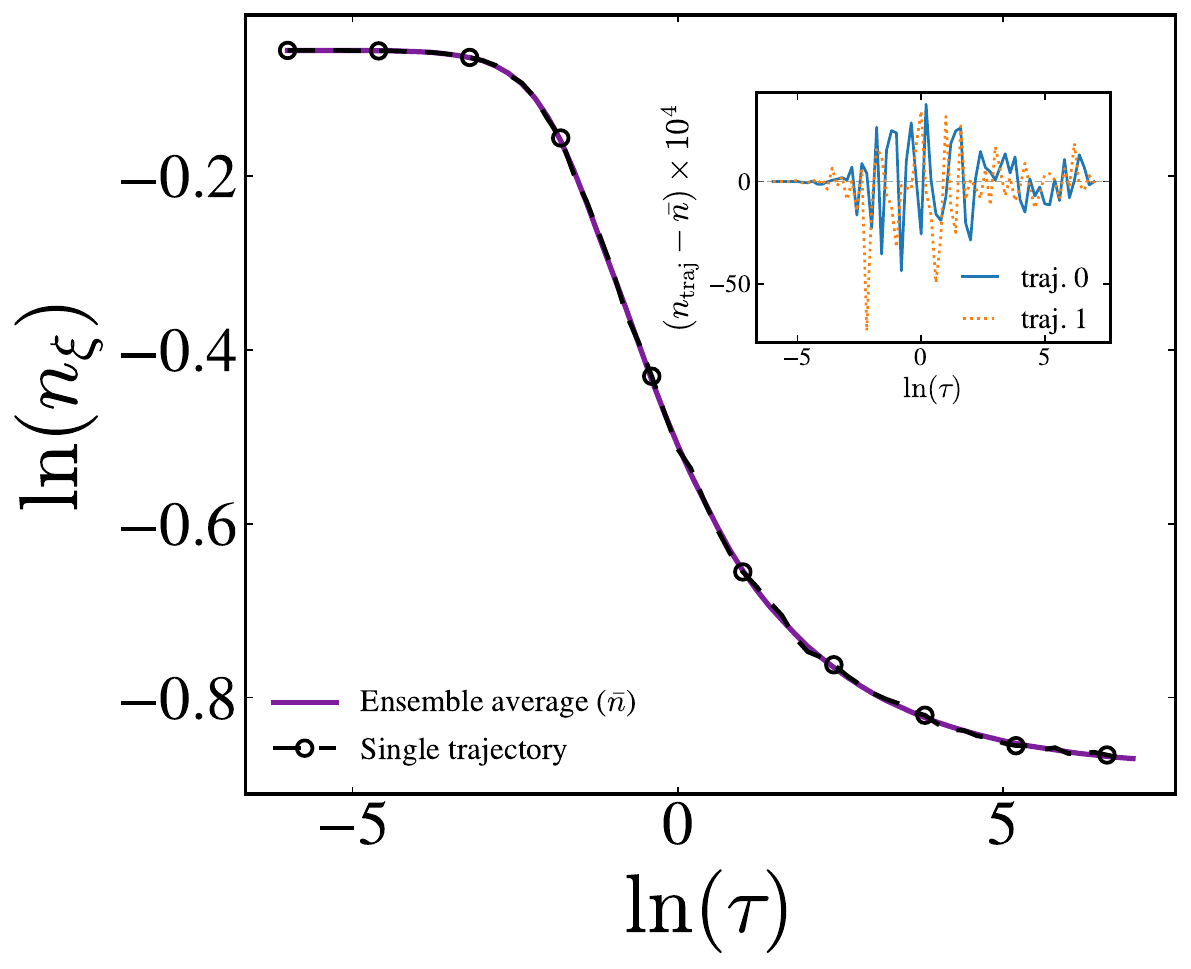}
			\put(2, 74){\textbf{(c)}}
		\end{overpic}
	\end{minipage}
	\hfill
	\begin{minipage}[b]{0.245\textwidth}
		\centering
		\begin{overpic}[width=\textwidth]{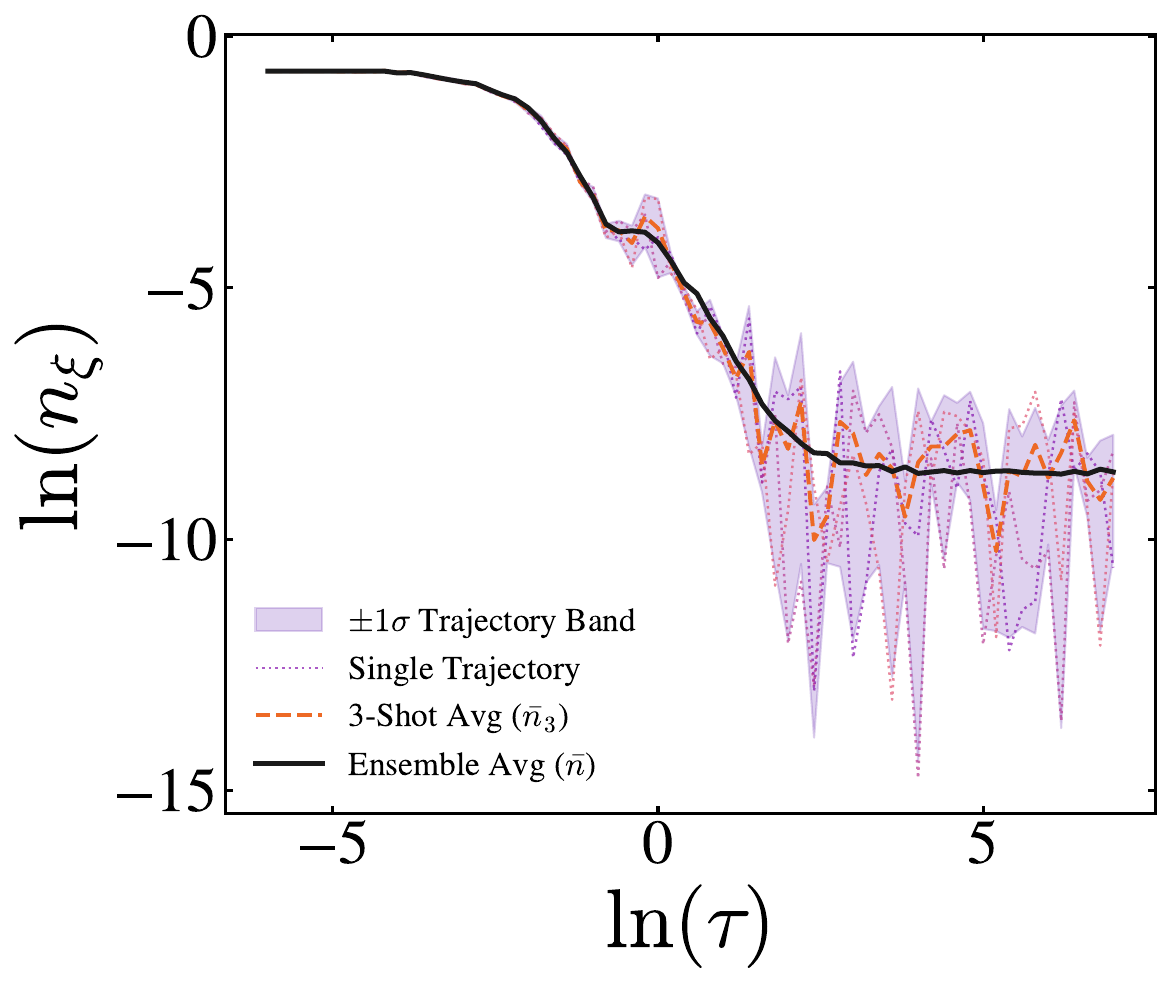}
			\put(2, 74){\textbf{(d)}}
		\end{overpic}
	\end{minipage}
	\caption{\justifying\textbf{Numerical benchmarking via the stochastic Schr\"odinger equation (SSE) and single-trajectory fluctuation analysis.} 
		\textbf{(a)}, \textbf{(b)} Validation of the ensemble-averaged master equation (solid curves) against SSE simulations ($N_{\mathrm{traj}} = 200$, open markers) in the driven non-Hermitian Kitaev model for (a) the attractor regime ($g = -0.7$) and (b) the anti-Kibble--Zurek regime ($g = 0.7$) across diverse longitudinal dephasing noise strengths $\xi \in [0.0, 0.04]$ ($\mu_i = -3.0, \mu_f = 3.0$). 
		\textbf{(c)} Single-trajectory dynamics in the Kitaev attractor regime ($g = -0.7, \xi = 0.04$): an unraveled single stochastic realization (dashed black curve with open circles) tracks the ensemble average $\bar{n}$ (purple curve) identically. Inset: residual fluctuations $(n_{\mathrm{traj}} - \bar{n}) \times 10^4$ for two independent trajectories, confirming tight confinement around the deterministic attractor. 
		\textbf{(d)} Unraveled dynamics in the non-Hermitian SSH model ($v = 1.0, w = 1.0, u_i = 0.0, u_f = 2.5$) under unprotected hopping noise ($\xi = 0.04$): while individual realizations (dotted purple curve) exhibit stochastic fluctuations within the $\pm 1\sigma$ band (shaded area), a modest 3-shot average $\bar{n}_3$ (dashed orange curve) rapidly reconstructs the macroscopic plateau floor ($\bar{n}$, solid black curve).}
	\label{fig:sse_combined_benchmark}
\end{figure*}
In our simulation, each stochastic trajectory is initialized at $t_i = \mu_i \tau$ in the ground state of the right eigenvectors, 
$|\psi_{k}(t_i)\rangle = |\Phi_{g,k}^{R}(t_i)\rangle$. The quench interval $[t_i, t_f]$ is discretized into 
uniform time steps $dt$, and the time evolution is executed via a two-step operator-splitting scheme:
\begin{equation}
	|\psi_{k}(t + dt)\rangle = \mathrm{diag}\left(e^{-i \xi dW}, \, e^{+i \xi dW}\right) U_0(t, dt) |\psi_{k}(t)\rangle,
	\label{eq:splitting_scheme}
\end{equation}
where $dW = \sqrt{dt} \, \eta$ with $\eta \sim \mathcal{N}(0, 1)$ independently sampled per realization. 
Here, the deterministic propagator $U_0(t, dt) = \exp[-i H_{k}(t) dt]$ is evaluated analytically as:
\begin{equation}
	U_0(t, dt) = \cos(\varepsilon_{k} dt)\mathbb{I} - i \frac{\sin(\varepsilon_{k} dt)}{\varepsilon_{k}} H_{k}(t),
\end{equation}
with $\varepsilon_{k}(t) = \sqrt{\mu_{k}(t)^2 + g \Delta_{k}^2}$ denoting the instantaneous complex eigenenergy.

For each of $N_{\mathrm{traj}} = 200$ single realizations, the state vector $|\psi_{k}^{(j)}(t)\rangle$ is integrated 
without intermediate normalization. Because non-Hermitian time evolution dynamically alters the state norm, 
normalizing individual trajectories at each time step would induce severe statistical distortions via Jensen's inequality, 
$\mathbb{E}[A/B] \neq \mathbb{E}[A]/\mathbb{E}[B]$. At the final quench time $t_f = \mu_f \tau$, the ensemble-averaged 
biorthogonal projection weights onto the left eigenstates $\langle \Phi_{g/e, k}^{L}(t_f)|$ are computed directly as:
\begin{equation}
	\mathcal{W}_{k}^{g,e} = \frac{1}{N_{\mathrm{traj}}} \sum_{j=1}^{N_{\mathrm{traj}}} \frac{|\langle \Phi_{g/e, k}^{L}(t_f) | \psi_{k}^{(j)}(t_f) \rangle|^2}{\|\langle \Phi_{g/e, k}^{L}(t_f)|\|^2}.
	\label{eq:SSE_weights}
\end{equation}
The single-mode excitation probability is given by $P_{k} = \mathcal{W}_{k}^{e} / (\mathcal{W}_{k}^{g} + \mathcal{W}_{k}^{e})$. 

For a lattice of size $N = 1000$ with discrete modes $k_p = \frac{2\pi(p - 1/2)}{N}$ ($p = 1, \dots, N/2$), the total 
defect density is evaluated through the discrete summation:
\begin{equation}
	n_\xi = \frac{2}{N} \sum_{p=1}^{N/2} P_{k_p}.
	\label{eq:defect_sum}
\end{equation}

As demonstrated in Figs.~\ref{fig:sse_combined_benchmark}(a) and (b), the defect density $n_\xi$ obtained from SSE simulations 
(open markers) perfectly tracks the numerical integration of the Lindblad master equation (solid curves) across all quench durations 
$\tau$ and noise strengths $\xi$. This absolute agreement holds both in the negative coupling regime ($g = -0.7$, non-Hermitian attractor) 
and in the positive coupling regime ($g = 0.7$, unbroken $\mathcal{PT}$ / anti-Kibble--Zurek regime), confirming the rigorous mathematical 
and numerical equivalence of both formalisms.

\subsection{Single-Trajectory Dynamics and Fluctuation Phenomenology}
A fundamental question regarding noisy quantum systems is whether the macroscopic defect plateaus 
survive at the single-shot level or represent an artifact of ensemble averaging. To elucidate this, we compute the 
single-realization defect density $n_{\mathrm{traj}}^{(j)}$ evaluated directly from individual unraveled trajectories:
\begin{equation}
	n_{\mathrm{traj}}^{(j)} = \frac{2}{N} \sum_{p=1}^{N/2} \frac{|\langle \Phi_{e, k_p}^{L}(t_f) | \psi_{k_p}^{(j)}(t_f) \rangle|^2}{|\langle \Phi_{g, k_p}^{L}(t_f) | \psi_{k_p}^{(j)}(t_f) \rangle|^2 + |\langle \Phi_{e, k_p}^{L}(t_f) | \psi_{k_p}^{(j)}(t_f) \rangle|^2}.
\end{equation}

In the non-Hermitian Kitaev chain within the attractor regime ($g = -0.7$), individual trajectories exhibit 
exceptional dynamical stability. As depicted in Fig.~\ref{fig:sse_combined_benchmark}(c), a single stochastic realization 
(dashed black line with circles) tracks the ensemble-averaged curve $\bar{n}$ (solid purple line) virtually without deviation. 
The inset displays the residual fluctuation $\Delta n = n_{\mathrm{traj}} - \bar{n}$ for two independent trajectories, 
demonstrating that stochastic noise is tightly bounded to within $\sim 10^{-3}$ and vanishes completely in the adiabatic/slow quench limit 
($\tau \gg 1$). This confirms that the noise-independent attractor is genuinely self-averaging: the non-Hermitian drift deterministically 
steers every individual wavepacket toward the dominant Liouvillian eigenmode, rendering defect saturation immune to stochastic phase slips 
even in single-shot experimental protocols.

Conversely, Fig.~\ref{fig:sse_combined_benchmark}(d) illustrates the stochastic unraveling in the non-Hermitian SSH model 
subject to unprotected hopping noise. Here, while the ensemble average $\bar{n}$ (solid black curve) saturates into a noise-tunable floor, 
individual realizations (dotted purple curve) exhibit pronounced stochastic variance, captured by the shaded $\pm 1\sigma$ band. 
Importantly, these fluctuations remain strictly bounded: an experimental average over as few as three shots ($N_{\mathrm{traj}} = 3$, 
dashed orange curve) rapidly suppresses the variance and recovers the plateau profile. This confirms that the noise-tunable plateau 
is a well-defined physical state that is accessible in finite-sampling quantum simulators.


\section{Biorthogonal Transition Probability Framework}
\label{sec:biorthogonal_framework}

In this section, we present the general formulation of transition probabilities for driven non-Hermitian systems and subsequently specialize it to the non-reciprocal topological superconductor model.

\subsection{General Formulation for Non-Hermitian Hamiltonians}
For a general non-Hermitian Hamiltonian $H(t)$, Hermiticity is broken ($H \neq H^\dagger$), and the conventional orthogonal eigenbasis is superseded by a biorthogonal dual set of right ($R$) and left ($L$) instantaneous eigenstates:
\begin{align}
	H(t) |\Phi_n^R(t)\rangle &= \varepsilon_n(t) |\Phi_n^R(t)\rangle, \label{eq:SM_right_eigen} \\
	\langle \Phi_n^L(t)| H(t) &= \varepsilon_n(t) \langle \Phi_n^L(t)| \Longleftrightarrow H^\dagger(t) |\Phi_n^L(t)\rangle = \varepsilon_n^*(t) |\Phi_n^L(t)\rangle, \label{eq:SM_left_eigen}
\end{align}
where $n$ labels the energy levels. Away from exceptional points, the dual sets satisfy the biorthogonal orthonormality and completeness relations:
\begin{equation}
	\langle \Phi_m^L(t) | \Phi_n^R(t) \rangle = \delta_{mn}, \quad \sum_n |\Phi_n^R(t)\rangle \langle \Phi_n^L(t)| = \mathbb{I}.
	\label{eq:SM_biorthogonal_completeness}
\end{equation}
Any evolving state vector $|\psi(t)\rangle$ admits the dual expansions:
\begin{equation}
	|\psi(t)\rangle = \sum_n c_n(t) |\Phi_n^R(t)\rangle, \, |\tilde{\psi}(t)\rangle = \sum_n c_n(t) \frac{|\Phi_n^L(t)\rangle}{\langle \Phi_n^L(t) | \Phi_n^L(t)\rangle},
	\label{eq:SM_state_expansions}
\end{equation}
where the projection amplitudes are gauge-invariantly extracted via $c_n(t) = \langle \Phi_n^L(t) | \psi(t)\rangle$. 

Because standard probability conservation is modified by non-unitary evolution, the normalized physical transition probability into the instantaneous eigenmode $n$ is defined by the biorthogonal projection of the state~\cite{Deng2025SM,Jafari2026aSM,Jafari2026bSM}:
\begin{equation}
	p_n(t) = \frac{|c_n(t)|^2 / \langle \Phi_n^L(t)|\Phi_n^L(t)\rangle}{\sum_m |c_m(t)|^2 / \langle \Phi_m^L(t)|\Phi_m^L(t)\rangle}.
	\label{eq:SM_prob_pure}
\end{equation}
For an open or stochastic quantum system described by an ensemble-averaged density matrix $\bar{\rho}(t)$, substituting $\bar{\rho}(t) = \mathbb{E}[|\psi(t)\rangle\langle\psi(t)|]$ and defining the projection operators $\Pi_n^L(t) \equiv |\Phi_n^L(t)\rangle\langle \Phi_n^L(t)|$ expresses the normalized transition probability directly in terms of $\bar{\rho}(t)$:
\begin{equation}
	p_n(t) = \frac{\operatorname{Tr}[\Pi_n^L(t) \bar{\rho}(t)] \,/\, \langle \Phi_n^L(t)|\Phi_n^L(t)\rangle}{\sum_m \operatorname{Tr}[\Pi_m^L(t) \bar{\rho}(t)] \,/\, \langle \Phi_m^L(t)|\Phi_m^L(t)\rangle}.
	\label{eq:SM_prob_density_matrix}
\end{equation}
Equation~\eqref{eq:SM_prob_density_matrix} is universally applicable to any driven non-Hermitian two-level or multi-band system.


\section{Two-Dimensional Generalization: Noise-Immune Attractor and Anti-Kibble--Zurek Scaling}
\label{sec:2D_generalization}

To demonstrate that the findings in the main text are independent of spatial dimensionality, we examine the driven non-Hermitian imbalanced-pairing superconductor in two dimensions ($2\mathrm{D}$). The structural requirements of \textbf{case~(I)} ($\gamma_z = 0$, $\mathcal{S}_{yz} = 0$, and $\boldsymbol{\gamma}_\perp \times \boldsymbol{\mathcal{D}}_\perp = 0$) remain rigorously fulfilled across the entire two-dimensional Brillouin zone.

\begin{figure*}[t]
	\centering
	\begin{minipage}[t]{0.32\textwidth}
		\raggedright \textbf{(a)} \\
		\includegraphics[width=\linewidth]{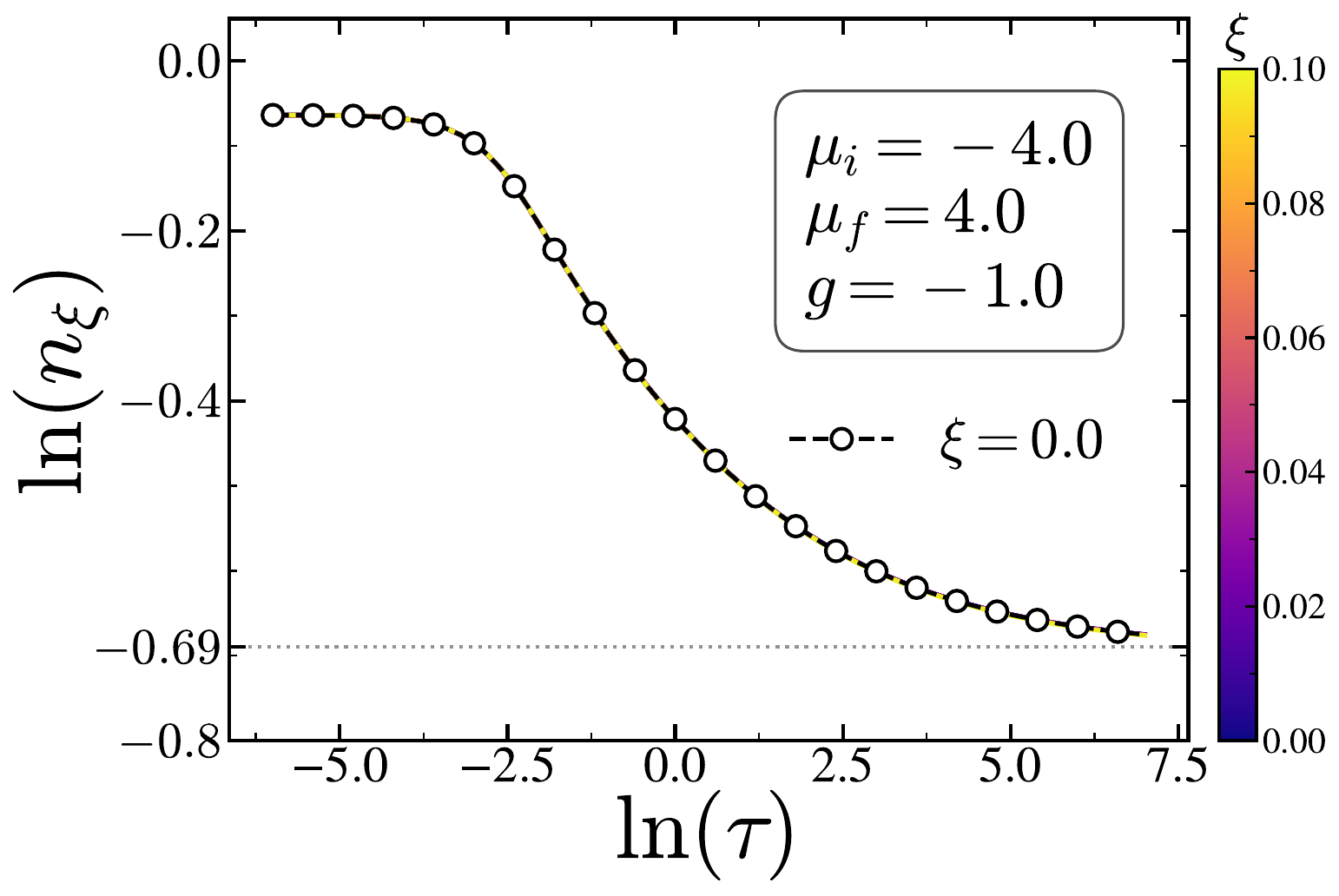}
	\end{minipage}\hfill
	\begin{minipage}[t]{0.32\textwidth}
		\raggedright \textbf{(b)} \\
		\includegraphics[width=\linewidth]{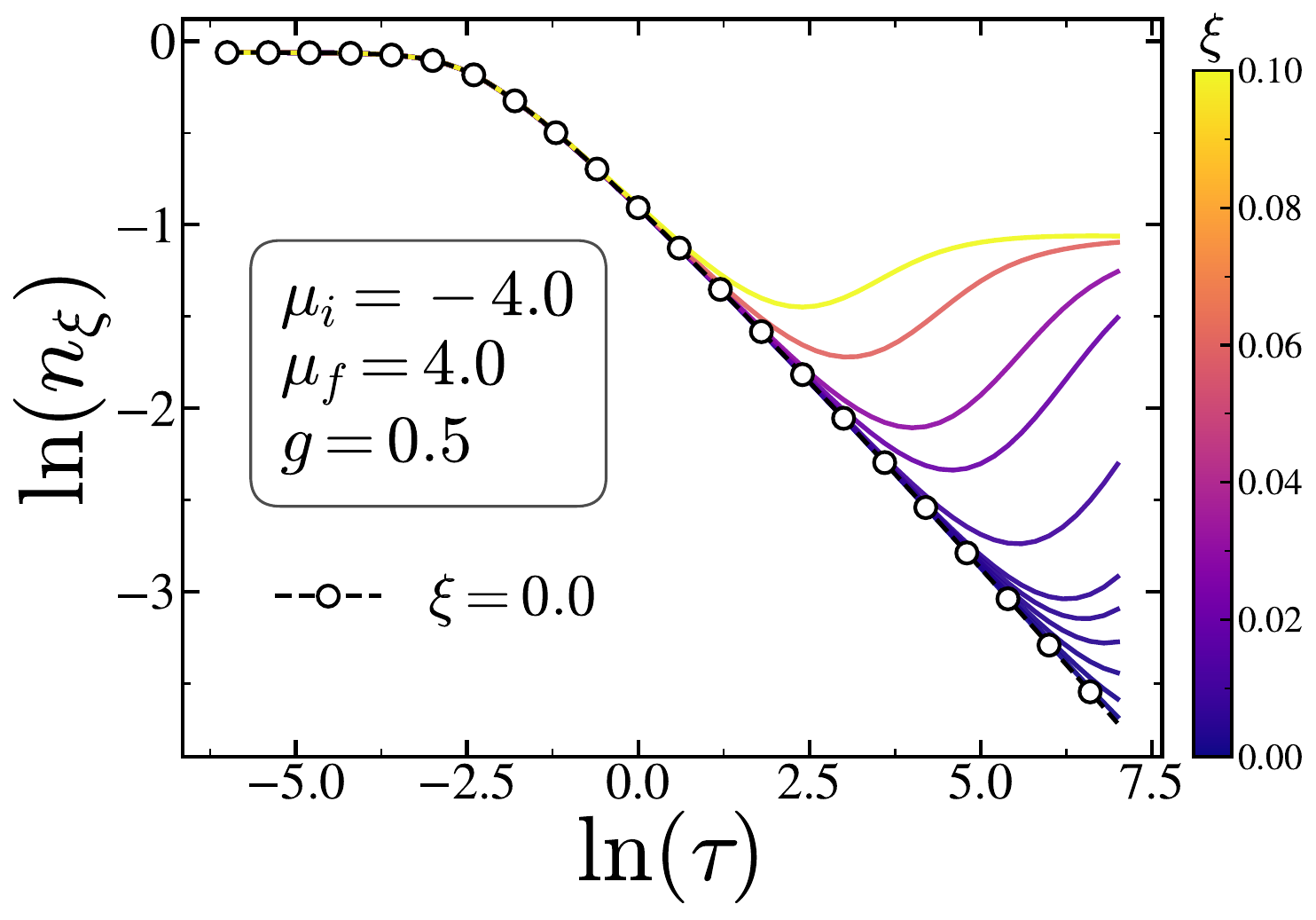}
	\end{minipage}\hfill
	\begin{minipage}[t]{0.28\textwidth}
		\raggedright \textbf{(c)} \\
		\includegraphics[width=\linewidth]{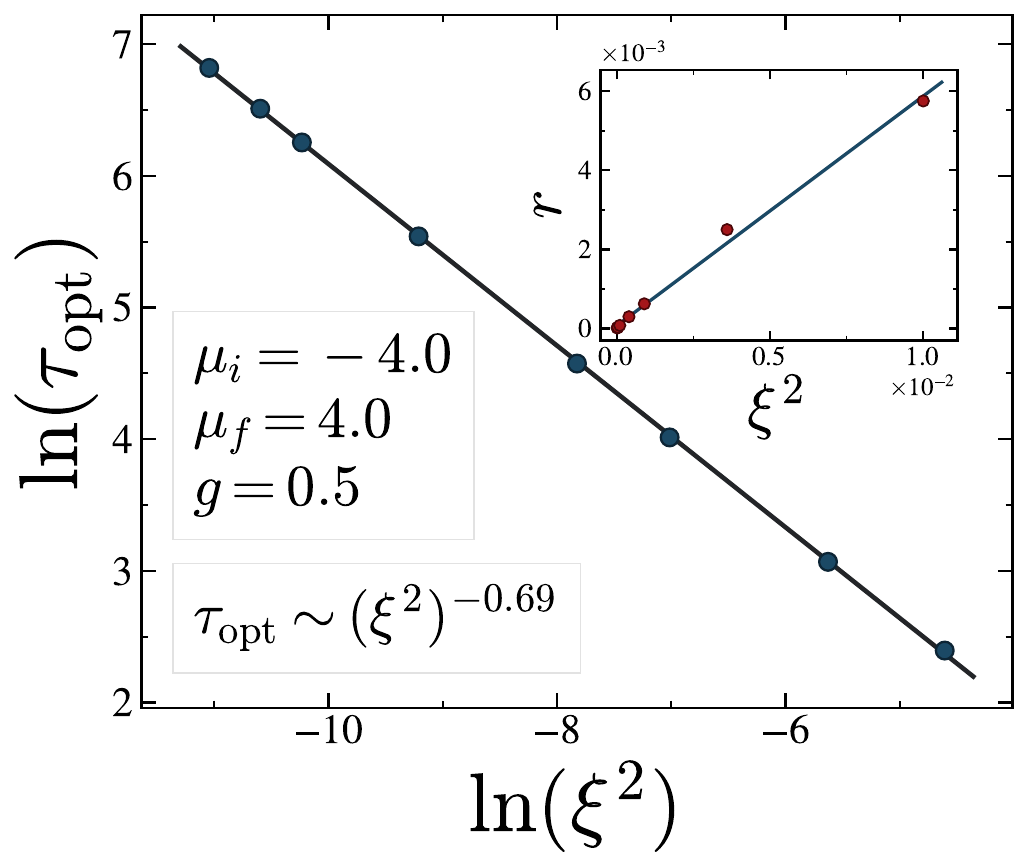}
	\end{minipage}
	\caption{\justifying \textbf{2D non-Hermitian dynamics and scaling ($\mu_i = -4 \to \mu_f = 4$).} 
		\textbf{(a)} Attractor regime ($g = -1.0$): defect density $\ln(n_\xi)$ collapses onto the noiseless baseline ($\xi = 0$), saturating at $\ln P_{\mathrm{plat}} = -\ln 2 \approx -0.69$ (dotted line). 
		\textbf{(b)} $\mathcal{PT}$ symmetric regime ($g = 0.5$): dephasing-induced heating yields non-monotonic AKZ profiles with minima $\tau_{\mathrm{opt}}$.
		\textbf{(c)} Optimal sweep time $\tau_{\mathrm{opt}} \sim (\xi^2)^{-0.69}$, in agreement with $\tau_{\mathrm{opt}} \sim (\xi^2)^{-2/3}$. Inset: heating rate $r \propto \xi^2$.}
	\label{fig:2D_summary}
\end{figure*}

In the non-Hermitian attractor regime ($g < 0$), resonant modes are projectively drawn into the noise-shielded fixed point $\mathcal{O}_z = (1 - |g|)/(1 + |g|)$. For $g = -1.0$, this dictates an asymptotic single-mode plateau $P_{\mathrm{plat}} = |g|/(1 + |g|) = 0.5$, setting a universal defect density floor given by $\lim_{\tau \to \infty} \ln(n_\xi) = \ln(P_{\mathrm{plat}}) = -\ln 2 \approx -0.693$. As shown in Fig.~\ref{fig:2D_summary}(a), full numerical simulations collapse identically onto the noiseless baseline ($\xi = 0$), precisely locking onto $\ln(n_\xi) \approx -0.69$ at long sweep durations, confirming that the non-Hermitian attractor shields quantum observables in higher dimensions.

In the $\mathcal{PT}$ symmetric ($g > 0$), continuous stochastic phase diffusion generates excess excitations at prolonged quench durations. This noise-induced defect generation is governed by a heating rate $r(\xi) \propto \xi^2$:

\begin{equation}
	\delta n_\xi(\tau) = n_\xi(\tau) - n_0(\tau) \approx r(\xi)\tau, \quad \text{with} \quad r(\xi) \propto \xi^2.
\end{equation}
As shown in Fig.~\ref{fig:2D_summary}(b), this yields the characteristic Anti-Kibble--Zurek (AKZ) non-monotonic turnaround, where stronger noise triggers heating at earlier quench times.

We systematically confirm the quadratic noise dependence $r \propto \xi^2$ in the inset of Fig.~\ref{fig:2D_summary}(c). Balancing this against the quench-induced decay yields the optimal sweep time $\tau_{\mathrm{opt}}$, which obeys the robust power-law scaling $\tau_{\mathrm{opt}} \sim (\xi^2)^{-0.69 \pm 0.02}$ as demonstrated in Fig.~\ref{fig:2D_summary}(c). This is in remarkable quantitative agreement with the analytical prediction $\tau_{\mathrm{opt}} \sim (\xi^2)^{-2/3}$, confirming that both the noise-immune attractor and AKZ mechanisms operate universally across dimensions.
\section{Application to the Non-Hermitian SSH Model}
\label{sec:SSH_model}
\subsection{Emergence of a Noise-Tunable Plateau}

We now apply our generalized Liouvillian formalism to the non-Hermitian Su--Schrieffer--Heeger (SSH) model subjected to gain/loss fluctuations~\cite{Deng2025SM,Lieu2018SM}. Following the standard two-level Bloch representation, the momentum-space Hamiltonian is explicitly written as:
\begin{equation}
	H_k(t) = h_x \sigma_x + h_y \sigma_y + i\big(u(t) + \eta(t)\big) \sigma_z,
	\label{eq:SSH_Hamiltonian}
\end{equation}
where $h_x = v + w \cos k$ and $h_y = w \sin k$ where $v$ and $w$ are the intracell and intercell hopping amplitudes, respectively, and $u(t)$ represents the deterministic non-Hermitian $\mathcal{PT}$-symmetric potential. The stochastic term $\eta(t)$ models a white Gaussian noise in the gain/loss amplitudes with zero mean and variance $\xi^2$, such that $\langle \eta(t)\eta(t') \rangle = \xi^2 \delta(t - t')$.

By mapping Eq.~\eqref{eq:SSH_Hamiltonian} onto our generalized framework, we identify the coherent driving vector $\mathbf{d} = (h_x, h_y, 0)$ and the non-Hermitian drift vector $\boldsymbol{\gamma} = (0, 0, -u(t))$. The system is driven exclusively by anti-Hermitian noise along the $\hat{z}$-axis ($\mathbf{m}_{\mathrm{NH}} = \hat{\mathbf{z}}$, $\xi_{\mathrm{NH}}^2 = \xi^2$), while purely Hermitian dephasing is structurally absent ($\xi_H^2 = 0$).

Substituting these vector fields into the generalized master equation allows us to track the dynamics of $\bar{\rho}_k(t)$ starting from the instantaneous ground state $|\Phi_{g,\boldsymbol{k}}^R(t_i)\rangle$. The resulting defect density $n_\xi$ at ramp termination is then evaluated in full analogy with the aforementioned biorthogonal projection scheme.

\begin{figure*}[t]
	\centering
	\begin{minipage}[c]{0.31\textwidth}
		\centering
		\setlength{\unitlength}{\textwidth}
		\begin{picture}(1, 0)
			\put(0.02, 0.02){\textbf{(a)}}
		\end{picture}
		\includegraphics[width=\textwidth]{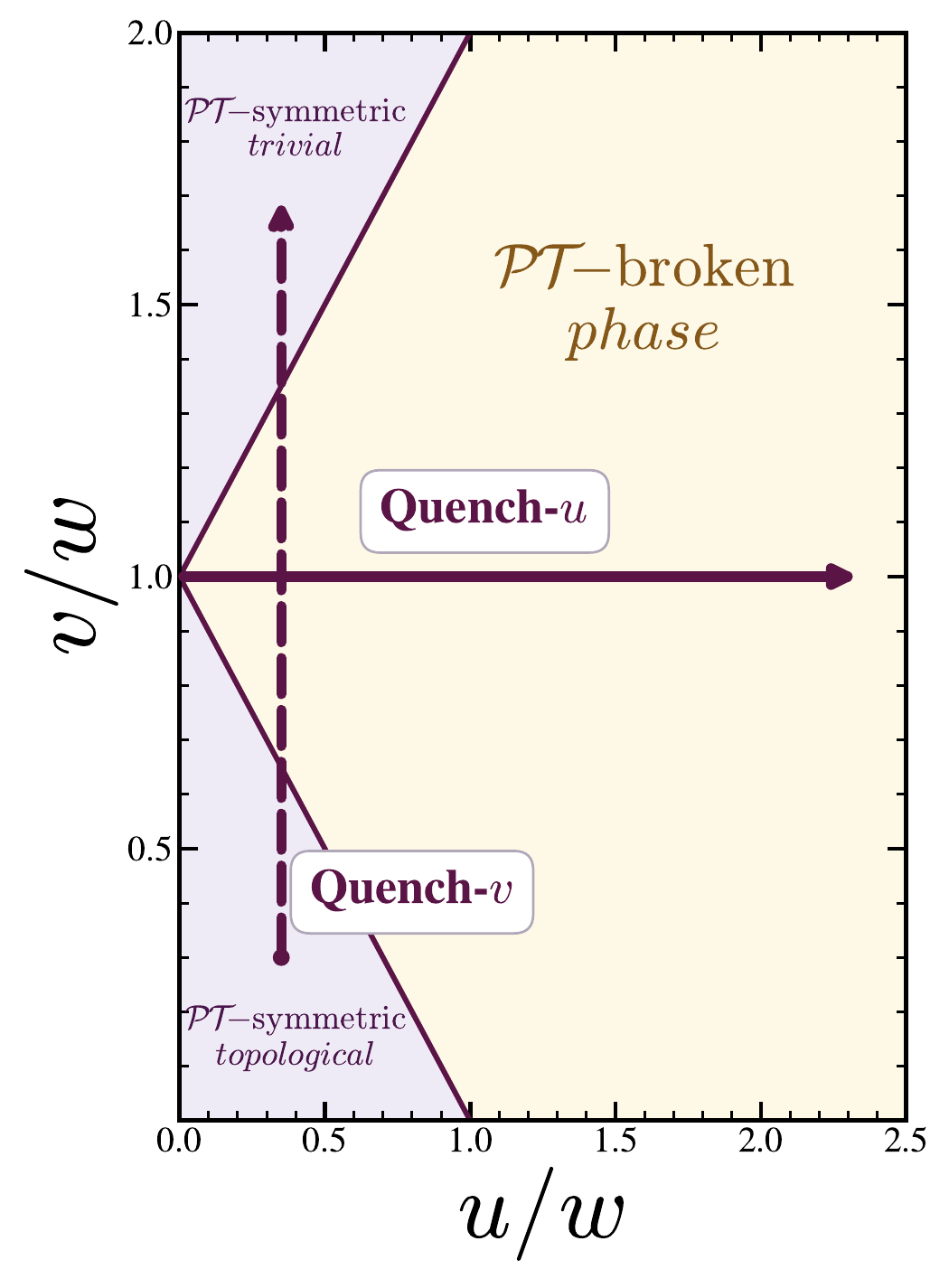}
		\label{fig:SSH_phase_diagram}
	\end{minipage}\hfill
	\begin{minipage}[c]{0.67\textwidth}
		\centering
		\begin{minipage}[t]{0.485\textwidth}
			\centering
			\setlength{\unitlength}{\textwidth}
			\begin{picture}(1, 0)
				\put(0.02, 0.02){\textbf{(b-1)}}
			\end{picture}
			\includegraphics[width=\textwidth]{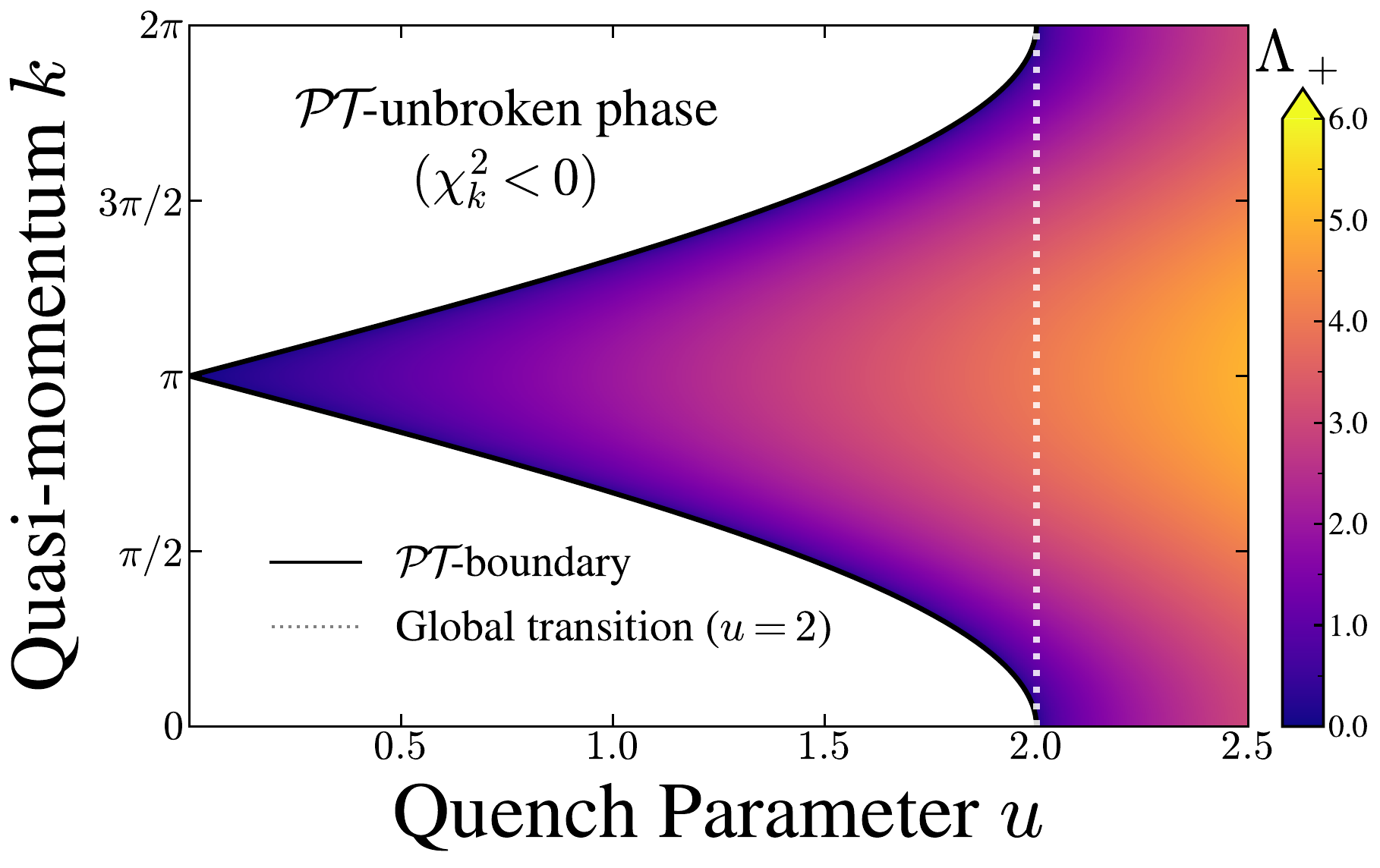}
			\label{fig:EV_quench_u}
		\end{minipage}\hfill
		\begin{minipage}[t]{0.485\textwidth}
			\centering
			\setlength{\unitlength}{\textwidth}
			\begin{picture}(1, 0)
				\put(0.02, 0.02){\textbf{(b-2)}}
			\end{picture}
			\includegraphics[width=\textwidth]{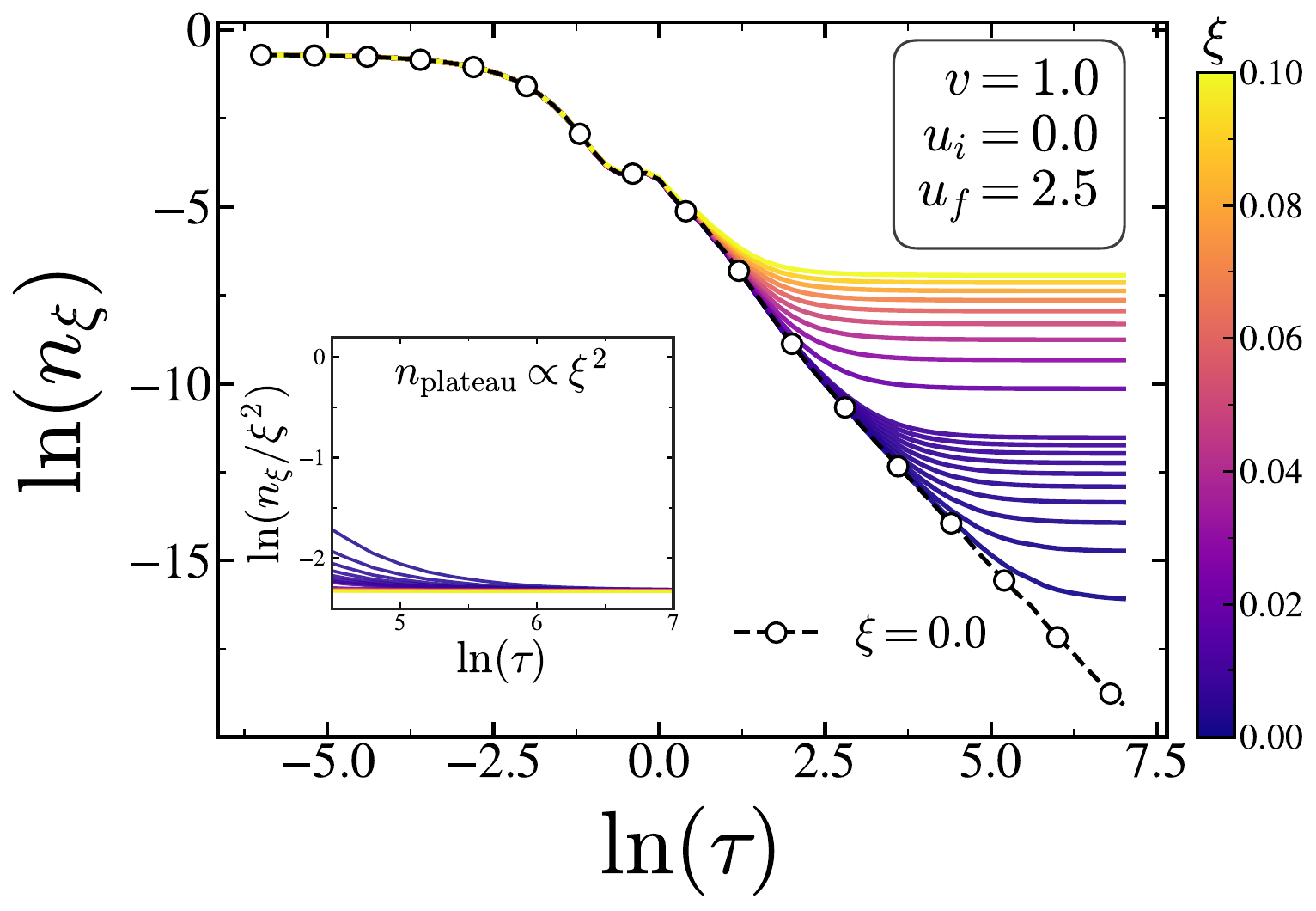}
			\label{fig:collapse_quench_u}
		\end{minipage}
		
		\begin{minipage}[t]{0.485\textwidth}
			\centering
			\setlength{\unitlength}{\textwidth}
			\begin{picture}(1, 0)
				\put(0.02, 0.02){\textbf{(c-1)}}
			\end{picture}
			\includegraphics[width=\textwidth]{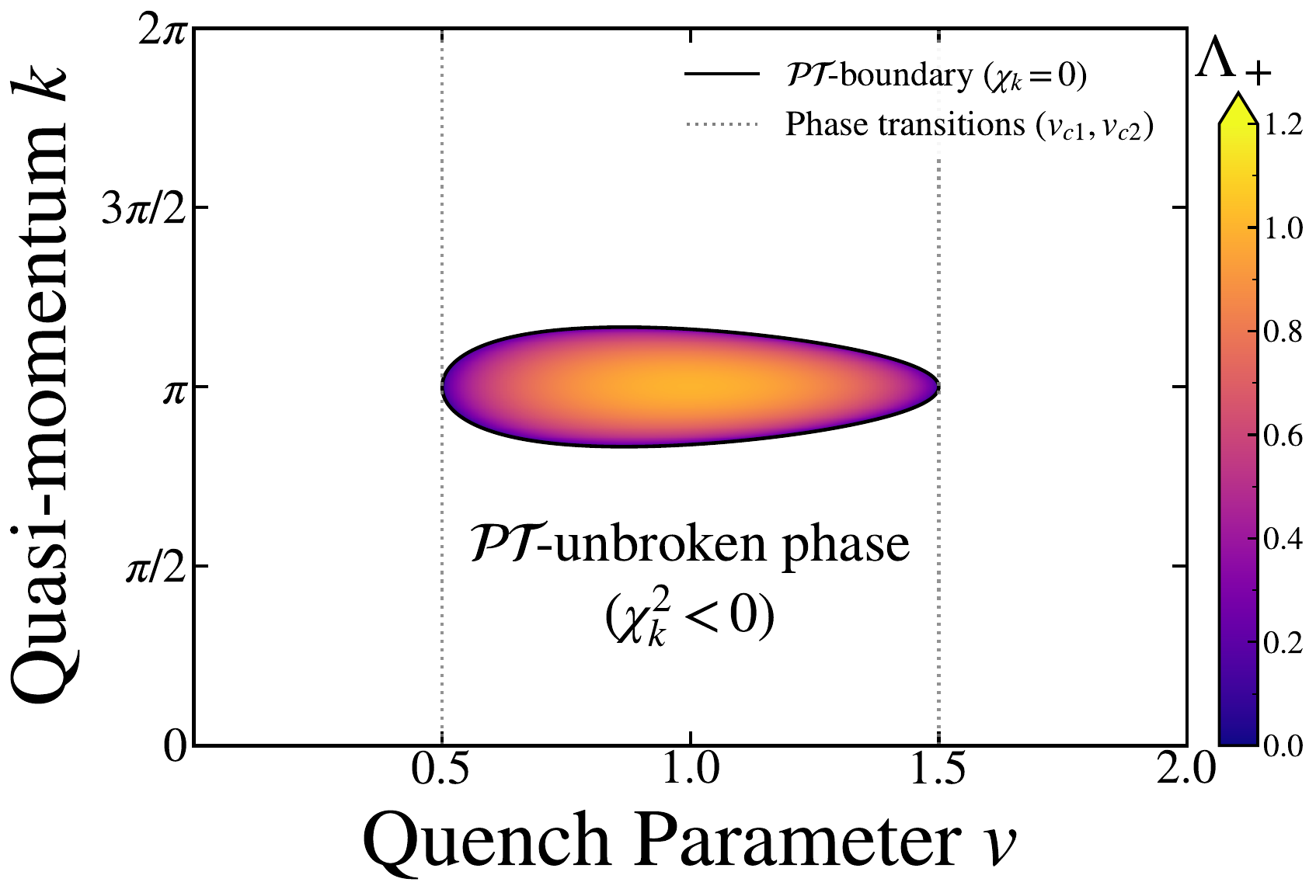}
			\label{fig:EV_quench_v}
		\end{minipage}\hfill
		\begin{minipage}[t]{0.485\textwidth}
			\centering
			\setlength{\unitlength}{\textwidth}
			\begin{picture}(1, 0)
				\put(0.02, 0.02){\textbf{(c-2)}}
			\end{picture}
			\includegraphics[width=\textwidth]{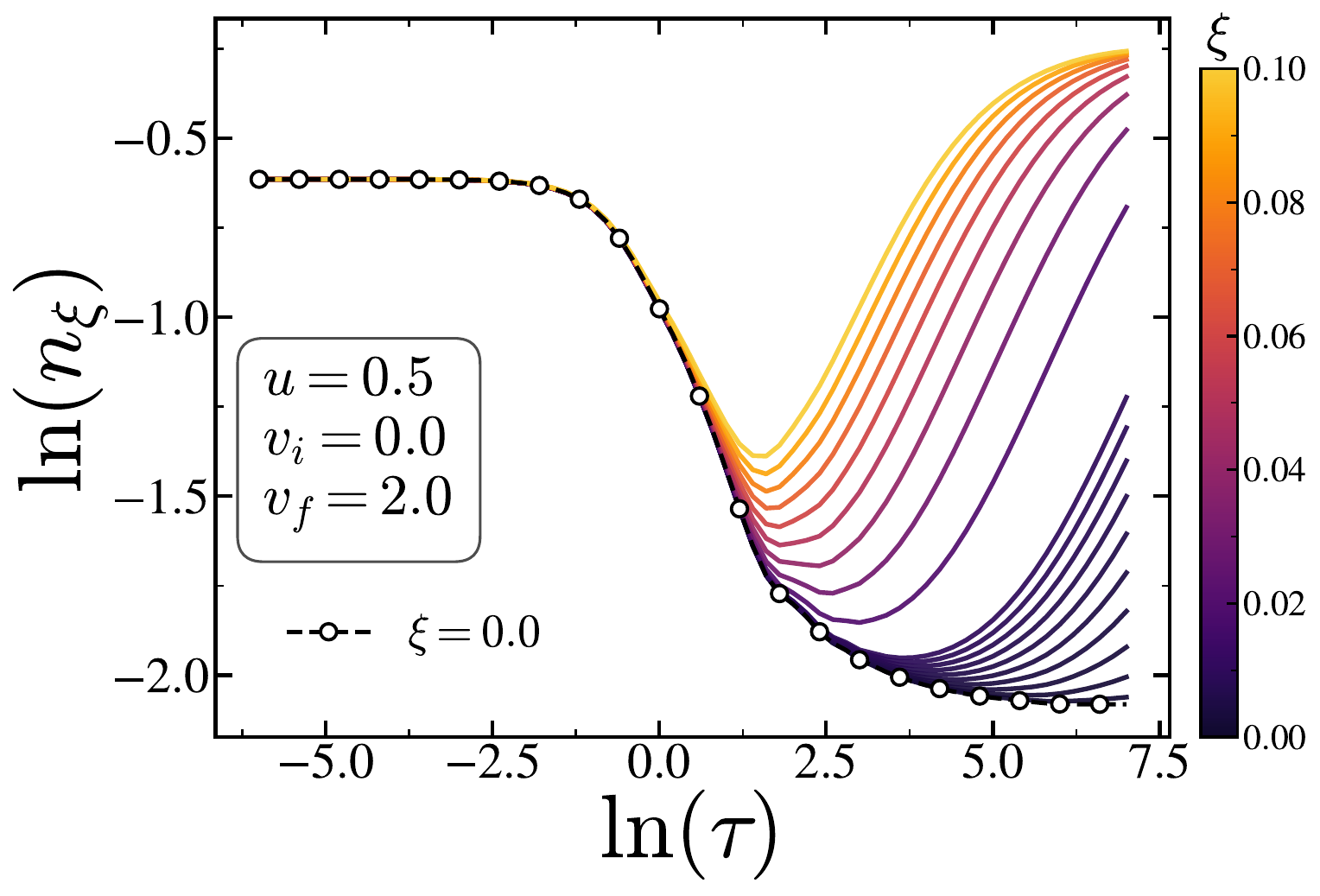}
			\label{fig:defect_quench_v}
		\end{minipage}
	\end{minipage}
	\caption{\justifying\textbf{Dynamical phase diagram and defect scaling in the noisy $\mathcal{PT}$-symmetric SSH model.} 
		( \textbf{a}) Non-Hermitian phase diagram in the $(u/w, v/w)$ plane depicting the $\mathcal{PT}$-symmetric trivial and topological phases, the $\mathcal{PT}$-broken regime, and the two quench trajectories: quench-$u$ (horizontal solid arrow) and quench-$v$ (vertical dashed arrow). 
		(\textbf{b-1}) Momentum-resolved dominant Liouvillian mode $\Lambda_+(k, u)$ along the quench-$u$ protocol across the global $\mathcal{PT}$-transition ($u=2$). 
		(\textbf{b-2}) Universal scaling and defect density $\ln(n)$ vs $\ln(\tau)$ for quench-$u$, showing the noise-dependent plateau regime with scaling collapse $n/\xi^2$ in the adiabatic limit (inset).
		(\textbf{c-1}) Momentum-resolved dominant Liouvillian mode $\Lambda_+(k, v)$ along the quench-$v$ path, highlighting the phase boundary ($\Lambda_+=0$). 
		(\textbf{c-2}) Defect density $\ln(n_\xi)$ as a function of ramp time $\ln(\tau)$ for quench-$v$, illustrating anti-Kibble–Zurek (AKZ) non-monotonic scaling and absence of plateaus.}
	
	\label{fig:SSH_summary}
\end{figure*}

To probe non-adiabatic defect dynamics, we consider a linear ramp protocol $u(t) = t/\tau$ over the duration $t \in [t_i, t_f]$ with ramp time $\tau$, interpolating from an initial unbroken configuration $u(t_i) = u_i$ to a final parameter $u(t_f) = u_f$ [see Fig.~\ref{fig:SSH_summary}(a), Quench--$u$]. For general translationally invariant systems that decouple into independent momentum modes, we define a momentum index $k$ within the underlying two-level model. Consequently, the dynamics of each $k$-mode are fully determined by the respective eigenvalues ($\Lambda_{j,k}$) and eigenvectors ($\mathbf{V}_{j,k}$) of the Liouvillian, i.e., $d\mathbf{R}_k/dt = \mathcal{L}_k(t) \mathbf{R}_k$. Substituting the model parameters into the general Liouvillian generator in Eq.~\eqref{eq:L_gen_compact}, in the extended basis $\mathbf{R}_k = (r_0, r_x, r_y, r_z)^T$, the instantaneous Liouvillian generator reads
\begin{equation}
	\mathcal{L}_k(t) = 
	\begin{pmatrix}
		2\xi^2 & 0 & 0 & 2u(t) \\
		0 & 0 & 0 & 2h_y \\
		0 & 0 & 0 & -2h_x \\
		2u(t) & -2h_y & 2h_x & 2\xi^2
	\end{pmatrix},
	\label{eq:Liouvillian_SSH}
\end{equation}
where $h_k^2 \equiv h_x^2 + h_y^2 = v^2 + w^2 + 2vw\cos k$. The instantaneous characteristic polynomial $\det[\lambda I - \mathcal{L}_k(t)] = \lambda \mathcal{P}(\lambda) = 0$ yields a trivial steady mode $\lambda_0 = 0$ alongside a non-trivial cubic sector:
\begin{equation}
	\mathcal{P}(\lambda) = \lambda^3 - 4\xi^2 \lambda^2 - 4\chi_k^2(t) \lambda - 8h_k^2 \xi^2 + \mathcal{O}(\xi^4) = 0,
	\label{eq:SSH_cubic_charpoly}
\end{equation}
where $\chi_k(t) \equiv \sqrt{u^2(t) - h_k^2}$. In the weak-noise regime ($\xi^2 \ll 1$), standard perturbation theory yields $\lambda = \lambda^{(0)} + \lambda^{(1)}\xi^2$, revealing the instantaneous dominant Liouvillian mode:
\begin{equation}
	\Lambda_{+,k}(t) \approx 2\chi_k(t) + \lambda_1(k, u(t))\xi^2, \quad \lambda_1(k, u) = \frac{2u^2 - h_k^2}{\chi_k^2}.
	\label{eq:lambda_leading_SSH}
\end{equation}
The underlying dynamical fate is critically governed by whether the ramp termination reaches the fully broken regime. For ramps terminating within the unbroken or partially broken phase ($u_f \le v+w = 2$), the condition of a positive leading Liouvillian rate spanning the entire Brillouin zone is violated [Fig.~\ref{fig:SSH_summary}(a)]. Consequently, no dynamical plateau emerges, and the ramp exhibits anti-Kibble--Zurek scaling dominated by non-adiabatic transitions near the exceptional lines.

In contrast, quenching across the global transition into the fully broken phase ($u_i=0 \to u_f=2.5 > 2$) ensures $\chi_k(t_f) > 0$ for all modes, establishing a macroscopic non-Hermitian attractor at late stages. Crucially, owing to the strong momentum anisotropy of the noise correction $\lambda_1(k, u)$, the resulting steady polarization retains an explicit $\xi$-dependence, realizing a noise-tunable plateau (case~(III)):
\begin{equation}
	n_{\mathrm{plat}} \propto \xi^2,
	\label{eq:plateau_SSH}
\end{equation}
which is confirmed by the clean scaling collapse $n_\xi / \xi^2$ in the adiabatic limit $\tau \to \infty$ [Fig.~\ref{fig:SSH_summary}(b)].

\subsection{Hopping-Noise Quench: Breakdown of Global Attractors and Anti-Kibble--Zurek Scaling.}To illustrate the criticality of the spectral gap condition, we examine an alternative configuration where stochastic fluctuations drive the intracell hopping amplitude, $v(t) \to v(t) + \eta(t)$ [Fig.~\ref{fig:SSH_summary}(a), Quench--$v$]. Here, the Hamiltonian is given by
\begin{equation}
	H_k(t) = \big(h_x(t) + \eta(t)\big)\sigma_x + h_y\sigma_y + iu\sigma_z,
\end{equation}
identifying a time-dependent coherent drive $\mathbf{d} = (h_x(t), h_y, 0)^T$, a static non-Hermitian drift $\boldsymbol{\gamma} = (0, 0, -u)^T$, and purely Hermitian noise along the $\hat{\mathbf{x}}$ direction ($\mathbf{m}_H = \hat{\mathbf{x}}$, $\xi_H^2 = \xi^2$, $\xi_{\mathrm{NH}} = 0$).

Linear ramps of the hopping parameter $v(t) = v_i + (v_f - v_i)t/\tau$ from $v_i = 0$ to $v_f = 2.0$ at fixed $u = 0.5$ fundamentally alter the Liouvillian spectrum. The instantaneous characteristic polynomial
\[
\det\!\left[\lambda I-\mathcal{L}_k(t)\right]
= \lambda\,q(\lambda)=0
\]
is obtained from the Liouvillian superoperator matrix
\begin{equation}
	\mathcal{L} = \begin{pmatrix}
		0 & 0 & 0 & 2u \\
		0 & 0 & 0 & 2h_y \\
		0 & 0 & -2\xi^2 & -2h_x \\
		2u & -2h_y & 2h_x & -2\xi^2
	\end{pmatrix}.
\end{equation}
This factorization yields a trivial steady mode $\lambda_0=0$, while the remaining modes are determined by the non-trivial cubic equation. Using $\chi_k^2(t) \equiv u^2-h_k^2(t)$, we write
\begin{equation}
	q(\lambda)
	= \lambda^3
	+4\xi^2\lambda^2
	-4\chi_k^2\lambda
	+8\xi^2\left(h_y^2-u^2\right)
	+\mathcal{O}(\xi^4)
	=0.
\end{equation}

In the weak-noise regime ($\xi^2 \ll 1$), standard perturbation theory gives $\lambda = \lambda^{(0)} + \lambda^{(1)}\xi^2$, where the unperturbed roots are $\lambda^{(0)} \in \{0,\,2\chi_k,\,-2\chi_k\}$. For the nonzero branches, the first-order correction is $\lambda^{(1)} = (2h_x^2 + h_y^2 - u^2)/\chi_k^2$. Thus, the dominant Liouvillian mode is
\begin{equation}
	\Lambda_+(t)
	\approx 2\chi_k(t)
	+ \frac{2h_x^2 + h_y^2 - u^2}{\chi_k^2}\,\xi^2.
\end{equation}
As shown in Fig.~\ref{fig:SSH_summary}(c-1), the $\mathcal{PT}$-broken regime ($\chi_k^2 > 0$) is confined to a closed momentum pocket around $k = \pi$ for $v \in [v_{c1}, v_{c2}]$ [$v_{c1,2} = w \mp u$], while the vast majority of the Brillouin zone remains $\mathcal{PT}$-symmetric ($\chi_k^2 < 0$). Consequently, the requisite condition of a strictly positive leading Liouvillian rate spanning all modes is globally violated. Deprived of a uniform non-Hermitian attractor, no steady-state plateau forms; instead, continuous non-Hermitian amplification over long ramp durations $\tau$ triggers pronounced anti-Kibble--Zurek (AKZ) behavior [Fig.~\ref{fig:SSH_summary}(c-2)], where defect generation scales inversely with adiabaticity and exhibits non-monotonic noise sensitivity.

\section{Algebraic Foundation of Krein-Charge Conservation as a Liouvillian Zero-Mode and the Necessity of the Spectral Gap}
\label{sec:SM_Krein_ZeroMode}

In this section, we provide a rigorous operator-algebraic derivation demonstrating that the single-mode pseudo-Hermitian Krein-charge conservation law emerges as a singular left zero-mode ($\Lambda_\Theta \equiv 0$) of the generalized Liouvillian superoperator $\mathcal{L}(\bm{k})$. We prove that this static algebraic invariance is structurally insufficient on its own to enforce dynamical plateau convergence, which fundamentally requires a strictly positive Liouvillian spectral gap $\mathrm{Re}(\Lambda_{+,\bm{k}}) > 0$. Furthermore, we demonstrate that when the noise-coupling geometry violates commutativity with the pseudo-Hermitian metric (as exemplified by the non-Hermitian SSH model), the conservation law fails identically ($dC/dt \neq 0$), whereas robust dynamical attractors persist under the Liouvillian framework.

\subsection{Mapping to the Extended Pauli-Bloch Liouvillian Generator}

Following the generalized master equation (Eq. (2)) in the main text, the time evolution of the noise-averaged density matrix $\rho_{\bm{k}}(t) = \langle \rho_\xi(\bm{k}, t) \rangle_\xi$ maps onto the linear dynamical system for the extended Pauli-Bloch state vector $\mathbf{R}_{\bm{k}}(t) \equiv (I_0, x_{\bm{k}}, y_{\bm{k}}, z_{\bm{k}})^T$:
\begin{equation}
	\frac{d\mathbf{R}_{\bm{k}}(t)}{dt} = \mathcal{L}(\bm{k}, t) \mathbf{R}_{\bm{k}}(t),
	\label{eq:SM_L_eom}
\end{equation}
where the $4 \times 4$ Liouvillian superoperator $\mathcal{L}(\bm{k}, t)$ is governed by the compact generator form defined in Eq.~\eqref{eq:L_gen_compact}. In the general parameterization, the off-diagonal cross-couplings are governed by:
\begin{equation}
	\mathcal{S}_{ij} \equiv 2\xi_H^2 m_{H,i} m_{H,j} + 2\xi_{\mathrm{NH}}^2 m_{\mathrm{NH},i} m_{\mathrm{NH},j} \quad (i \neq j).
	\label{eq:SM_S_ij_def}
\end{equation}
For the driven non-Hermitian imbalanced-pairing Kitaev model, using the Bogoliubov--de Gennes Hamiltonian in the Nambu basis given by Eq.~\eqref{eq:BdG_Matrix}, and substituting the decomposed parameters $\mathbf{d}_{\bm{k}}(t) = \left(\frac{1+g}{2}\Delta_{\bm{k}}, 0, -\mu_{\bm{k}}(t)\right)^T$, $\bm{\gamma}_{\bm{k}} = \left(0, -\frac{1-g}{2}\Delta_{\bm{k}}, 0\right)^T$, and $\mathbf{m}_H = (0,\, 0,\, 1)^T$ with $\xi_H = \xi$ and $\xi_{\mathrm{NH}} = 0$, the noise fluctuations are strictly aligned with the longitudinal axis. Consequently, all off-diagonal cross-couplings vanish identically:
\begin{equation}
	\mathcal{S}_{xy} = \mathcal{S}_{xz} = \mathcal{S}_{yz} = 0.
\end{equation}
The explicit Liouvillian generator in this regime reduces to:
\begin{equation}
	\mathcal{L}(\bm{k}, t) = 
	\begin{pmatrix}
		0 & 0 & (1-g)\Delta_{\bm{k}} & 0 \\
		0 & -2\xi^2 & 2 \mu_{\bm{k}}(t) & 0 \\
		(1-g)\Delta_{\bm{k}} & -2 \mu_{\bm{k}}(t) & -2\xi^2 & -(1+g)\Delta_{\bm{k}} \\
		0 & 0 & (1+g)\Delta_{\bm{k}} & 0
	\end{pmatrix}.
	\label{eq:SM_L_explicit}
\end{equation}

\subsection{Krein Metric Representation and the Dual Left Vector}

The deterministic Hamiltonian $H_{0,\bm{k}}$ is pseudo-Hermitian with respect to the Krein metric operator $\Theta = \mathrm{diag}(1, g^{-1})$, satisfying $H_{0,\bm{k}}^\dagger \Theta = \Theta H_{0,\bm{k}}$~\cite{Massarelli2022}. Rescaling by the constant $g$, we define $\tilde{\Theta} = g\Theta = \mathrm{diag}(g, 1) = \frac{1+g}{2} \mathbb{I} - \frac{1-g}{2} \sigma_z$. For any pure state $|\psi(\bm{k}, t)\rangle = (u_{\bm{k}}, v_{\bm{k}})^T$, the conserved Krein charge reads:
\begin{equation}
	C(\bm{k}, t) \equiv \mathrm{Tr}(\tilde{\Theta} \rho_{\bm{k}}) = g |u_{\bm{k}}|^2 + |v_{\bm{k}}|^2.
\end{equation}
In the extended Pauli-Bloch space $\mathbf{R}_{\bm{k}}$, this scalar invariant is expressed as a linear functional:
\begin{equation}
	C(\bm{k}, t) = \mathbf{W}_\Theta^T \mathbf{R}_{\bm{k}}(t),
\end{equation}
where the dual metric vector is explicitly defined as:
\begin{equation}
	\mathbf{W}_\Theta \equiv \begin{pmatrix} \frac{1+g}{2}, & 0, & 0, & -\frac{1-g}{2} \end{pmatrix}^T.
	\label{eq:SM_W_def}
\end{equation}

\subsection{Analytical Proof of the Liouvillian Zero-Mode and Metric-Preserving Cross-Couplings}

The exact stationarity of the Krein charge, $dC(\bm{k}, t)/dt = 0$, implies:
\begin{equation}
	\frac{d C(\bm{k}, t)}{dt} = \mathbf{W}_\Theta^T \frac{d\mathbf{R}_{\bm{k}}}{dt} = \mathbf{W}_\Theta^T \mathcal{L}(\bm{k}, t) \mathbf{R}_{\bm{k}}(t) = 0, \quad \forall \mathbf{R}_{\bm{k}}(t),
\end{equation}
which is fulfilled if and only if $\mathbf{W}_\Theta^T$ is an exact left null eigenvector ($\mathbf{W}_\Theta \in \ker(\mathcal{L}^T)$) of the Liouvillian generator:
\begin{equation}
	\mathbf{W}_\Theta^T \mathcal{L}(\bm{k}, t) = \mathbf{0}^T.
\end{equation}
Performing the direct algebraic multiplication using Eq.~\eqref{eq:SM_L_explicit} and evaluating component-wise, we obtain identically:
\begin{equation}
	\mathbf{W}_\Theta^T \mathcal{L}(\bm{k}, t) = \begin{pmatrix} 0 & 0 & 0 & 0 \end{pmatrix} = 0 \cdot \mathbf{W}_\Theta^T.
\end{equation}

\paragraph{Algebraic Equivalence with $\mathcal{S}_{ij} = 0$.}
To explicitly connect this result to the general cross-coupling framework of the main text, consider a set of noise operators $\mathcal{D}_\nu \equiv (\mathbf{m}_\nu \cdot \bm{\sigma})$, where $\nu$ denotes the specific stochastic channel. For the Krein charge to be conserved as a zero-mode, the Liouvillian generator $\mathcal{L}$ must act within the metric-invariant subspace. This condition is strictly satisfied if the noise operators commute with the metric operator, $[(\mathbf{m}_\nu \cdot \bm{\sigma}), \Theta] = 0$. Mathematically, this commutator vanishes if and only if the noise vectors are aligned with the longitudinal metric axis, i.e., $\mathbf{m}_\nu \parallel \hat{\mathbf{z}}$, which imposes $m_{\nu,x} = m_{\nu,y} = 0$ for all channels $\nu$. Substituting these metric-preserving conditions directly into the microscopic definition of the cross-coupling rates [Eq.~\eqref{eq:SM_S_ij_def}], we find:
\begin{equation}
	\mathcal{S}_{xy} = 0, \quad \mathcal{S}_{xz} = 0, \quad \mathcal{S}_{yz} = 0.
\end{equation}
Conversely, if transverse fluctuations exist ($m_{\nu,x} \neq 0$ or $m_{\nu,y} \neq 0$), metric commutativity is broken ($[(\mathbf{m}_\nu \cdot \bm{\sigma}), \Theta] \neq 0$), directly inducing non-vanishing off-diagonal rates ($\mathcal{S}_{xz} \neq 0$ or $\mathcal{S}_{yz} \neq 0$). In the generator $\mathcal{L}$, these non-zero $\mathcal{S}_{ij}$ terms immediately produce non-zero entries in $\mathbf{W}_\Theta^T \mathcal{L} \neq \mathbf{0}^T$, annihilating the left zero-mode. Thus, \emph{the vanishing of cross-couplings $\mathcal{S}_{ij} = 0$ is not a dynamical accident or fine-tuning, but the exact algebraic signature of noise preserving the metric symmetry of the non-Hermitian generator.}

\subsection{Insufficiency of Krein Invariance and the Indispensability of the Liouvillian Spectral Gap}

Although the left zero-mode fixes the static linear combination $(1+g)I_0(\bm{k}, t) - (1-g)z_{\bm{k}}(t) = 2C(\bm{k}, 0) = \text{const}$, the physically observed polarization $\mathcal{O}_z(\bm{k}, t) \equiv z_{\bm{k}}(t)/I_0(\bm{k}, t)$ satisfies:
\begin{equation}
	\mathcal{O}_z(\bm{k}, t) = \frac{1+g}{1-g} - \frac{2 C(\bm{k}, 0)}{(1-g) I_0(\bm{k}, t)}.
	\label{eq:SM_Oz_expansion}
\end{equation}
Equation~\eqref{eq:SM_Oz_expansion} proves that Krein conservation alone is \emph{insufficient} to guarantee a universal plateau: if the state norm $I_0(\bm{k}, t)$ remains bounded or decays, $\mathcal{O}_z(\bm{k}, t)$ retains a permanent memory of the initial state through $C(\bm{k}, 0)$.

A universal, history-independent plateau ($\mathcal{O}_z \to \frac{1+g}{1-g}$) requires the suppression of the memory term, $\lim_{t \to \infty} I_0(\bm{k}, t) \to \infty$. In spectral decomposition, the trajectory expands as $\mathbf{R}_{\bm{k}}(t) = \sum_{j} c_j e^{\int_0^t \Lambda_{j,\bm{k}}(t') dt'} \mathbf{V}_{j,\bm{k}}(t)$. Consequently, the asymptotic polarization is strictly governed by the \textbf{Liouvillian spectral gap}:
\begin{equation}
	\delta \Lambda_{\bm{k}} \equiv \mathrm{Re}(\Lambda_{+,\bm{k}}) - \max_{j \neq +} \mathrm{Re}(\Lambda_{j,\bm{k}}) > 0, \quad \text{with} \quad \mathrm{Re}(\Lambda_{+,\bm{k}}) > 0.
\end{equation}
This spectral gap provides the physical driving force:
\begin{enumerate}
	\item It dictates the exponential growth of the total ensemble norm $I_0(\bm{k}, t) \sim e^{\mathrm{Re}(\Lambda_{+,\bm{k}}) t}$, washing out initial-state memory.
	\item It establishes the finite characteristic plateau formation time:
	\begin{equation}
		t_{\mathrm{plat}} \sim \frac{1}{\delta \Lambda_{\bm{k}}}.
	\end{equation}
\end{enumerate}
Therefore, while Krein invariance defines a static invariant subspace, the Liouvillian spectral gap dictates the dynamical convergence to the non-Hermitian attractor.


\subsection{Breakdown of Krein Protection in the Non-Hermitian SSH Model}
\label{sec:ssh_dynamics_breakdown}

To establish the boundaries of symmetry-induced protection in stochastic systems, we contrast the asymptotic dynamics of the non-Hermitian Kitaev model with those of the non-Hermitian Su-Schrieffer-Heeger (SSH) model under dephasing noise. In the Kitaev model, the left zero-mode of the Liouvillian generator directly spans the longitudinal sector, constraining the state space and pinning the physical polarization to a robust, noise-independent attractor through the Liouvillian spectral gap.

In stark contrast, the SSH model reveals a fundamental limitation of algebraic conservation laws. While the system preserves an exact, noise-independent Krein charge associated with a zero-mode left eigenvector, this conserved quantity is confined entirely to the transverse sector of the generalized Bloch space. Because it lacks any longitudinal component, exact Krein invariance is intrinsically insufficient to constrain the physical polarization observable. 

Under stochastic dynamics, the unconstrained transverse fluctuations continuously feed into the longitudinal sector through an accumulated, history-dependent memory drift. Consequently, the longitudinal state vector fails to collapse onto a universal attractor. Instead of achieving exact polarization protection, the system exhibits a noise-tunable plateau in the slow-quench limit, where the asymptotic defect density scales quadratically with the noise strength [Eq.~(\ref{eq:plateau_SSH})]. This breakdown demonstrates a crucial physical principle: static conservation of a Krein charge does not guarantee the dynamical robustness of physical observables. True topological protection in stochastic non-Hermitian systems requires non-trivial alignment between the stochastic noise channels and the dominant longitudinal attractors of the Liouvillian spectrum.

\bibliography{References}

\begin{thebibliography}{115}%
\makeatletter
\providecommand \@ifxundefined [1]{%
 \@ifx{#1\undefined}
}%
\providecommand \@ifnum [1]{%
 \ifnum #1\expandafter \@firstoftwo
 \else \expandafter \@secondoftwo
 \fi
}%
\providecommand \@ifx [1]{%
 \ifx #1\expandafter \@firstoftwo
 \else \expandafter \@secondoftwo
 \fi
}%
\providecommand \natexlab [1]{#1}%
\providecommand \enquote  [1]{``#1''}%
\providecommand \bibnamefont  [1]{#1}%
\providecommand \bibfnamefont [1]{#1}%
\providecommand \citenamefont [1]{#1}%
\providecommand \href@noop [0]{\@secondoftwo}%
\providecommand \href [0]{\begingroup \@sanitize@url \@href}%
\providecommand \@href[1]{\@@startlink{#1}\@@href}%
\providecommand \@@href[1]{\endgroup#1\@@endlink}%
\providecommand \@sanitize@url [0]{\catcode `\\12\catcode `\$12\catcode
  `\&12\catcode `\#12\catcode `\^12\catcode `\_12\catcode `\%12\relax}%
\providecommand \@@startlink[1]{}%
\providecommand \@@endlink[0]{}%
\providecommand \url  [0]{\begingroup\@sanitize@url \@url }%
\providecommand \@url [1]{\endgroup\@href {#1}{\urlprefix }}%
\providecommand \urlprefix  [0]{URL }%
\providecommand \Eprint [0]{\href }%
\providecommand \doibase [0]{https://doi.org/}%
\providecommand \selectlanguage [0]{\@gobble}%
\providecommand \bibinfo  [0]{\@secondoftwo}%
\providecommand \bibfield  [0]{\@secondoftwo}%
\providecommand \translation [1]{[#1]}%
\providecommand \BibitemOpen [0]{}%
\providecommand \bibitemStop [0]{}%
\providecommand \bibitemNoStop [0]{.\EOS\space}%
\providecommand \EOS [0]{\spacefactor3000\relax}%
\providecommand \BibitemShut  [1]{\csname bibitem#1\endcsname}%
\let\auto@bib@innerbib\@empty
\bibitem [{\citenamefont {Kibble}(1976)}]{Kibble1976}%
  \BibitemOpen
  \bibfield  {author} {\bibinfo {author} {\bibfnamefont {T.~W.~B.}\
  \bibnamefont {Kibble}},\ }\bibfield  {title} {\bibinfo {title} {Topology of
  cosmic domains and strings},\ }\href
  {https://doi.org/10.1088/0305-4470/9/8/029} {\bibfield  {journal} {\bibinfo
  {journal} {Journal of Physics A: Mathematical and General}\ }\textbf
  {\bibinfo {volume} {9}},\ \bibinfo {pages} {1387} (\bibinfo {year}
  {1976})}\BibitemShut {NoStop}%
\bibitem [{\citenamefont {Zurek}(1985)}]{Zurek1985}%
  \BibitemOpen
  \bibfield  {author} {\bibinfo {author} {\bibfnamefont {W.~H.}\ \bibnamefont
  {Zurek}},\ }\bibfield  {title} {\bibinfo {title} {{Cosmological Experiments
  in Superfluid Helium?}},\ }\href {https://doi.org/10.1038/317505a0}
  {\bibfield  {journal} {\bibinfo  {journal} {Nature}\ }\textbf {\bibinfo
  {volume} {317}},\ \bibinfo {pages} {505} (\bibinfo {year}
  {1985})}\BibitemShut {NoStop}%
\bibitem [{\citenamefont {Dziarmaga}(2010)}]{Dziarmaga2010}%
  \BibitemOpen
  \bibfield  {author} {\bibinfo {author} {\bibfnamefont {J.}~\bibnamefont
  {Dziarmaga}},\ }\bibfield  {title} {\bibinfo {title} {Dynamics of a quantum
  phase transition and relaxation to a steady state},\ }\href
  {https://doi.org/10.1080/00018732.2010.514702} {\bibfield  {journal}
  {\bibinfo  {journal} {Advances in Physics}\ }\textbf {\bibinfo {volume}
  {59}},\ \bibinfo {pages} {1063–1189} (\bibinfo {year} {2010})}\BibitemShut
  {NoStop}%
\bibitem [{\citenamefont {Polkovnikov}\ \emph {et~al.}(2011)\citenamefont
  {Polkovnikov}, \citenamefont {Sengupta}, \citenamefont {Silva},\ and\
  \citenamefont {Vengalattore}}]{Polkovnikov2011}%
  \BibitemOpen
  \bibfield  {author} {\bibinfo {author} {\bibfnamefont {A.}~\bibnamefont
  {Polkovnikov}}, \bibinfo {author} {\bibfnamefont {K.}~\bibnamefont
  {Sengupta}}, \bibinfo {author} {\bibfnamefont {A.}~\bibnamefont {Silva}},\
  and\ \bibinfo {author} {\bibfnamefont {M.}~\bibnamefont {Vengalattore}},\
  }\bibfield  {title} {\bibinfo {title} {Colloquium: Nonequilibrium dynamics of
  closed interacting quantum systems},\ }\href
  {https://doi.org/10.1103/RevModPhys.83.863} {\bibfield  {journal} {\bibinfo
  {journal} {Rev. Mod. Phys.}\ }\textbf {\bibinfo {volume} {83}},\ \bibinfo
  {pages} {863} (\bibinfo {year} {2011})}\BibitemShut {NoStop}%
\bibitem [{\citenamefont {Sengupta}\ \emph {et~al.}(2008)\citenamefont
  {Sengupta}, \citenamefont {Sen},\ and\ \citenamefont
  {Mondal}}]{Sengupta2008}%
  \BibitemOpen
  \bibfield  {author} {\bibinfo {author} {\bibfnamefont {K.}~\bibnamefont
  {Sengupta}}, \bibinfo {author} {\bibfnamefont {D.}~\bibnamefont {Sen}},\ and\
  \bibinfo {author} {\bibfnamefont {S.}~\bibnamefont {Mondal}},\ }\bibfield
  {title} {\bibinfo {title} {Exact results for quench dynamics and defect
  production in a two-dimensional model},\ }\href
  {https://doi.org/10.1103/PhysRevLett.100.077204} {\bibfield  {journal}
  {\bibinfo  {journal} {Phys. Rev. Lett.}\ }\textbf {\bibinfo {volume} {100}},\
  \bibinfo {pages} {077204} (\bibinfo {year} {2008})}\BibitemShut {NoStop}%
\bibitem [{\citenamefont {Dutta}\ \emph {et~al.}(2015)\citenamefont {Dutta},
  \citenamefont {Aeppli}, \citenamefont {Chakrabarti}, \citenamefont
  {Divakaran}, \citenamefont {Rosenbaum},\ and\ \citenamefont
  {Sen}}]{Duttabook}%
  \BibitemOpen
  \bibfield  {author} {\bibinfo {author} {\bibfnamefont {A.}~\bibnamefont
  {Dutta}}, \bibinfo {author} {\bibfnamefont {G.}~\bibnamefont {Aeppli}},
  \bibinfo {author} {\bibfnamefont {B.~K.}\ \bibnamefont {Chakrabarti}},
  \bibinfo {author} {\bibfnamefont {U.}~\bibnamefont {Divakaran}}, \bibinfo
  {author} {\bibfnamefont {T.~F.}\ \bibnamefont {Rosenbaum}},\ and\ \bibinfo
  {author} {\bibfnamefont {D.}~\bibnamefont {Sen}},\ }\href@noop {} {\emph
  {\bibinfo {title} {Quantum Phase Transitions in Transverse Field Spin Models:
  From Statistical Physics to Quantum Information}}}\ (\bibinfo  {publisher}
  {Cambridge University Press},\ \bibinfo {year} {2015})\BibitemShut {NoStop}%
\bibitem [{\citenamefont {Sen}\ \emph {et~al.}(2008)\citenamefont {Sen},
  \citenamefont {Sengupta},\ and\ \citenamefont {Mondal}}]{Sen2008}%
  \BibitemOpen
  \bibfield  {author} {\bibinfo {author} {\bibfnamefont {D.}~\bibnamefont
  {Sen}}, \bibinfo {author} {\bibfnamefont {K.}~\bibnamefont {Sengupta}},\ and\
  \bibinfo {author} {\bibfnamefont {S.}~\bibnamefont {Mondal}},\ }\bibfield
  {title} {\bibinfo {title} {Defect production in nonlinear quench across a
  quantum critical point},\ }\href
  {https://doi.org/10.1103/PhysRevLett.101.016806} {\bibfield  {journal}
  {\bibinfo  {journal} {Phys. Rev. Lett.}\ }\textbf {\bibinfo {volume} {101}},\
  \bibinfo {pages} {016806} (\bibinfo {year} {2008})}\BibitemShut {NoStop}%
\bibitem [{\citenamefont {Mondal}\ \emph {et~al.}(2009)\citenamefont {Mondal},
  \citenamefont {Sengupta},\ and\ \citenamefont {Sen}}]{Shreyoshi2009}%
  \BibitemOpen
  \bibfield  {author} {\bibinfo {author} {\bibfnamefont {S.}~\bibnamefont
  {Mondal}}, \bibinfo {author} {\bibfnamefont {K.}~\bibnamefont {Sengupta}},\
  and\ \bibinfo {author} {\bibfnamefont {D.}~\bibnamefont {Sen}},\ }\bibfield
  {title} {\bibinfo {title} {Theory of defect production in nonlinear quench
  across a quantum critical point},\ }\href
  {https://doi.org/10.1103/PhysRevB.79.045128} {\bibfield  {journal} {\bibinfo
  {journal} {Phys. Rev. B}\ }\textbf {\bibinfo {volume} {79}},\ \bibinfo
  {pages} {045128} (\bibinfo {year} {2009})}\BibitemShut {NoStop}%
\bibitem [{\citenamefont {Das}\ \emph {et~al.}(2017)\citenamefont {Das},
  \citenamefont {Das}, \citenamefont {Galante}, \citenamefont {Myers},\ and\
  \citenamefont {Sengupta}}]{Das2017}%
  \BibitemOpen
  \bibfield  {author} {\bibinfo {author} {\bibfnamefont {D.}~\bibnamefont
  {Das}}, \bibinfo {author} {\bibfnamefont {S.~R.}\ \bibnamefont {Das}},
  \bibinfo {author} {\bibfnamefont {D.~A.}\ \bibnamefont {Galante}}, \bibinfo
  {author} {\bibfnamefont {R.~C.}\ \bibnamefont {Myers}},\ and\ \bibinfo
  {author} {\bibfnamefont {K.}~\bibnamefont {Sengupta}},\ }\bibfield  {title}
  {\bibinfo {title} {An exactly solvable quench protocol for integrable spin
  models},\ }\href {https://doi.org/10.1007/JHEP11(2017)157} {\bibfield
  {journal} {\bibinfo  {journal} {Journal of High Energy Physics}\ }\textbf
  {\bibinfo {volume} {2017}},\ \bibinfo {pages} {1} (\bibinfo {year}
  {2017})}\BibitemShut {NoStop}%
\bibitem [{\citenamefont {Pellegrini}\ \emph {et~al.}(2008)\citenamefont
  {Pellegrini}, \citenamefont {Montangero}, \citenamefont {Santoro},\ and\
  \citenamefont {Fazio}}]{Pellegrini2008}%
  \BibitemOpen
  \bibfield  {author} {\bibinfo {author} {\bibfnamefont {F.}~\bibnamefont
  {Pellegrini}}, \bibinfo {author} {\bibfnamefont {S.}~\bibnamefont
  {Montangero}}, \bibinfo {author} {\bibfnamefont {G.~E.}\ \bibnamefont
  {Santoro}},\ and\ \bibinfo {author} {\bibfnamefont {R.}~\bibnamefont
  {Fazio}},\ }\bibfield  {title} {\bibinfo {title} {Adiabatic quenches through
  an extended quantum critical region},\ }\href
  {https://doi.org/10.1103/PhysRevB.77.140404} {\bibfield  {journal} {\bibinfo
  {journal} {Phys. Rev. B}\ }\textbf {\bibinfo {volume} {77}},\ \bibinfo
  {pages} {140404(R)} (\bibinfo {year} {2008})}\BibitemShut {NoStop}%
\bibitem [{\citenamefont {Polkovnikov}(2005)}]{Polkovnikov2005}%
  \BibitemOpen
  \bibfield  {author} {\bibinfo {author} {\bibfnamefont {A.}~\bibnamefont
  {Polkovnikov}},\ }\bibfield  {title} {\bibinfo {title} {Universal adiabatic
  dynamics in the vicinity of a quantum critical point},\ }\href
  {https://doi.org/10.1103/PhysRevB.72.161201} {\bibfield  {journal} {\bibinfo
  {journal} {Phys. Rev. B}\ }\textbf {\bibinfo {volume} {72}},\ \bibinfo
  {pages} {161201} (\bibinfo {year} {2005})}\BibitemShut {NoStop}%
\bibitem [{\citenamefont {Kou}\ and\ \citenamefont {Li}(2023)}]{Kou2023}%
  \BibitemOpen
  \bibfield  {author} {\bibinfo {author} {\bibfnamefont {H.-C.}\ \bibnamefont
  {Kou}}\ and\ \bibinfo {author} {\bibfnamefont {P.}~\bibnamefont {Li}},\
  }\bibfield  {title} {\bibinfo {title} {Varying quench dynamics in the
  transverse ising chain: The kibble-zurek, saturated, and presaturated
  regimes},\ }\href {https://doi.org/10.1103/PhysRevB.108.214307} {\bibfield
  {journal} {\bibinfo  {journal} {Phys. Rev. B}\ }\textbf {\bibinfo {volume}
  {108}},\ \bibinfo {pages} {214307} (\bibinfo {year} {2023})}\BibitemShut
  {NoStop}%
\bibitem [{\citenamefont {Damski}(2005)}]{Damski2005}%
  \BibitemOpen
  \bibfield  {author} {\bibinfo {author} {\bibfnamefont {B.}~\bibnamefont
  {Damski}},\ }\bibfield  {title} {\bibinfo {title} {The simplest quantum model
  supporting the kibble-zurek mechanism of topological defect production:
  Landau-zener transitions from a new perspective},\ }\href
  {https://doi.org/10.1103/PhysRevLett.95.035701} {\bibfield  {journal}
  {\bibinfo  {journal} {Phys. Rev. Lett.}\ }\textbf {\bibinfo {volume} {95}},\
  \bibinfo {pages} {035701} (\bibinfo {year} {2005})}\BibitemShut {NoStop}%
\bibitem [{\citenamefont {Zamora}\ \emph {et~al.}(2020)\citenamefont {Zamora},
  \citenamefont {Dagvadorj}, \citenamefont {Comaron}, \citenamefont
  {Carusotto}, \citenamefont {Proukakis},\ and\ \citenamefont
  {Szyma\ifmmode~\acute{n}\else \'{n}\fi{}ska}}]{Zamora2020}%
  \BibitemOpen
  \bibfield  {author} {\bibinfo {author} {\bibfnamefont {A.}~\bibnamefont
  {Zamora}}, \bibinfo {author} {\bibfnamefont {G.}~\bibnamefont {Dagvadorj}},
  \bibinfo {author} {\bibfnamefont {P.}~\bibnamefont {Comaron}}, \bibinfo
  {author} {\bibfnamefont {I.}~\bibnamefont {Carusotto}}, \bibinfo {author}
  {\bibfnamefont {N.~P.}\ \bibnamefont {Proukakis}},\ and\ \bibinfo {author}
  {\bibfnamefont {M.~H.}\ \bibnamefont {Szyma\ifmmode~\acute{n}\else
  \'{n}\fi{}ska}},\ }\bibfield  {title} {\bibinfo {title} {Kibble-zurek
  mechanism in driven dissipative systems crossing a nonequilibrium phase
  transition},\ }\href {https://doi.org/10.1103/PhysRevLett.125.095301}
  {\bibfield  {journal} {\bibinfo  {journal} {Phys. Rev. Lett.}\ }\textbf
  {\bibinfo {volume} {125}},\ \bibinfo {pages} {095301} (\bibinfo {year}
  {2020})}\BibitemShut {NoStop}%
\bibitem [{\citenamefont {Jafari}\ \emph
  {et~al.}(2026{\natexlab{a}})\citenamefont {Jafari}, \citenamefont {Akbari},
  \citenamefont {Mardonov},\ and\ \citenamefont {Langari}}]{Jafari2026a}%
  \BibitemOpen
  \bibfield  {author} {\bibinfo {author} {\bibfnamefont {R.}~\bibnamefont
  {Jafari}}, \bibinfo {author} {\bibfnamefont {A.}~\bibnamefont {Akbari}},
  \bibinfo {author} {\bibfnamefont {S.}~\bibnamefont {Mardonov}},\ and\
  \bibinfo {author} {\bibfnamefont {A.}~\bibnamefont {Langari}},\ }\href
  {https://arxiv.org/abs/2609.00971} {\bibinfo {title} {Kibble--zurek mechanism
  and defect freezing in imbalanced-pairing kitaev models}} (\bibinfo {year}
  {2026}{\natexlab{a}}),\ \Eprint {https://arxiv.org/abs/2609.00971}
  {arXiv:2609.00971 [cond-mat.stat-mech]} \BibitemShut {NoStop}%
\bibitem [{\citenamefont {Puebla}\ \emph
  {et~al.}(2020{\natexlab{a}})\citenamefont {Puebla}, \citenamefont {Deffner},\
  and\ \citenamefont {Campbell}}]{Puebla2020b}%
  \BibitemOpen
  \bibfield  {author} {\bibinfo {author} {\bibfnamefont {R.}~\bibnamefont
  {Puebla}}, \bibinfo {author} {\bibfnamefont {S.}~\bibnamefont {Deffner}},\
  and\ \bibinfo {author} {\bibfnamefont {S.}~\bibnamefont {Campbell}},\
  }\bibfield  {title} {\bibinfo {title} {Kibble-zurek scaling in quantum speed
  limits for shortcuts to adiabaticity},\ }\href
  {https://doi.org/10.1103/PhysRevResearch.2.032020} {\bibfield  {journal}
  {\bibinfo  {journal} {Phys. Rev. Res.}\ }\textbf {\bibinfo {volume} {2}},\
  \bibinfo {pages} {032020} (\bibinfo {year} {2020}{\natexlab{a}})}\BibitemShut
  {NoStop}%
\bibitem [{\citenamefont {Roychowdhury}\ \emph {et~al.}(2021)\citenamefont
  {Roychowdhury}, \citenamefont {Moessner},\ and\ \citenamefont
  {Das}}]{Das2021}%
  \BibitemOpen
  \bibfield  {author} {\bibinfo {author} {\bibfnamefont {K.}~\bibnamefont
  {Roychowdhury}}, \bibinfo {author} {\bibfnamefont {R.}~\bibnamefont
  {Moessner}},\ and\ \bibinfo {author} {\bibfnamefont {A.}~\bibnamefont
  {Das}},\ }\bibfield  {title} {\bibinfo {title} {Dynamics and correlations at
  a quantum phase transition beyond kibble-zurek},\ }\href
  {https://doi.org/10.1103/PhysRevB.104.014406} {\bibfield  {journal} {\bibinfo
   {journal} {Phys. Rev. B}\ }\textbf {\bibinfo {volume} {104}},\ \bibinfo
  {pages} {014406} (\bibinfo {year} {2021})}\BibitemShut {NoStop}%
\bibitem [{\citenamefont {Dziarmaga}(2005)}]{Dziarmaga2005}%
  \BibitemOpen
  \bibfield  {author} {\bibinfo {author} {\bibfnamefont {J.}~\bibnamefont
  {Dziarmaga}},\ }\bibfield  {title} {\bibinfo {title} {Dynamics of a quantum
  phase transition: Exact solution of the quantum ising model},\ }\href
  {https://doi.org/10.1103/PhysRevLett.95.245701} {\bibfield  {journal}
  {\bibinfo  {journal} {Phys. Rev. Lett.}\ }\textbf {\bibinfo {volume} {95}},\
  \bibinfo {pages} {245701} (\bibinfo {year} {2005})}\BibitemShut {NoStop}%
\bibitem [{\citenamefont {Jafari}\ \emph
  {et~al.}(2026{\natexlab{b}})\citenamefont {Jafari}, \citenamefont {Akbari},
  \citenamefont {Mardonov},\ and\ \citenamefont {Yavartanoo}}]{Jafari2026b}%
  \BibitemOpen
  \bibfield  {author} {\bibinfo {author} {\bibfnamefont {R.}~\bibnamefont
  {Jafari}}, \bibinfo {author} {\bibfnamefont {A.}~\bibnamefont {Akbari}},
  \bibinfo {author} {\bibfnamefont {S.}~\bibnamefont {Mardonov}},\ and\
  \bibinfo {author} {\bibfnamefont {H.}~\bibnamefont {Yavartanoo}},\ }\href
  {https://arxiv.org/abs/2607.23956} {\bibinfo {title} {Dynamical quantum phase
  transitions in a pseudo-hermitian hamiltonian: The imbalanced-pairing kitaev
  model}} (\bibinfo {year} {2026}{\natexlab{b}}),\ \Eprint
  {https://arxiv.org/abs/2607.23956} {arXiv:2607.23956 [quant-ph]} \BibitemShut
  {NoStop}%
\bibitem [{\citenamefont {Zurek}\ \emph {et~al.}(2005)\citenamefont {Zurek},
  \citenamefont {Dorner},\ and\ \citenamefont {Zoller}}]{Zurek2005}%
  \BibitemOpen
  \bibfield  {author} {\bibinfo {author} {\bibfnamefont {W.~H.}\ \bibnamefont
  {Zurek}}, \bibinfo {author} {\bibfnamefont {U.}~\bibnamefont {Dorner}},\ and\
  \bibinfo {author} {\bibfnamefont {P.}~\bibnamefont {Zoller}},\ }\bibfield
  {title} {\bibinfo {title} {Dynamics of a quantum phase transition},\ }\href
  {https://doi.org/10.1103/PhysRevLett.95.105701} {\bibfield  {journal}
  {\bibinfo  {journal} {Phys. Rev. Lett.}\ }\textbf {\bibinfo {volume} {95}},\
  \bibinfo {pages} {105701} (\bibinfo {year} {2005})}\BibitemShut {NoStop}%
\bibitem [{\citenamefont {Kou}\ and\ \citenamefont {Li}(2022)}]{Kou2022}%
  \BibitemOpen
  \bibfield  {author} {\bibinfo {author} {\bibfnamefont {H.-C.}\ \bibnamefont
  {Kou}}\ and\ \bibinfo {author} {\bibfnamefont {P.}~\bibnamefont {Li}},\
  }\bibfield  {title} {\bibinfo {title} {Interferometry based on the quantum
  kibble-zurek mechanism},\ }\href
  {https://doi.org/10.1103/PhysRevB.106.184301} {\bibfield  {journal} {\bibinfo
   {journal} {Phys. Rev. B}\ }\textbf {\bibinfo {volume} {106}},\ \bibinfo
  {pages} {184301} (\bibinfo {year} {2022})}\BibitemShut {NoStop}%
\bibitem [{\citenamefont {Kou}\ \emph {et~al.}(2025)\citenamefont {Kou},
  \citenamefont {Zhang},\ and\ \citenamefont {Li}}]{Kou2025}%
  \BibitemOpen
  \bibfield  {author} {\bibinfo {author} {\bibfnamefont {H.-C.}\ \bibnamefont
  {Kou}}, \bibinfo {author} {\bibfnamefont {Z.-H.}\ \bibnamefont {Zhang}},\
  and\ \bibinfo {author} {\bibfnamefont {P.}~\bibnamefont {Li}},\ }\bibfield
  {title} {\bibinfo {title} {Kibble-zurek scaling immune to anti-kibble-zurek
  behavior in driven open systems at the limit of loss difference},\ }\href
  {https://doi.org/10.1103/PhysRevB.111.155152} {\bibfield  {journal} {\bibinfo
   {journal} {Phys. Rev. B}\ }\textbf {\bibinfo {volume} {111}},\ \bibinfo
  {pages} {155152} (\bibinfo {year} {2025})}\BibitemShut {NoStop}%
\bibitem [{\citenamefont {Bermudez}\ \emph {et~al.}(2009)\citenamefont
  {Bermudez}, \citenamefont {Patan\`e}, \citenamefont {Amico},\ and\
  \citenamefont {Martin-Delgado}}]{Bermudez2009}%
  \BibitemOpen
  \bibfield  {author} {\bibinfo {author} {\bibfnamefont {A.}~\bibnamefont
  {Bermudez}}, \bibinfo {author} {\bibfnamefont {D.}~\bibnamefont {Patan\`e}},
  \bibinfo {author} {\bibfnamefont {L.}~\bibnamefont {Amico}},\ and\ \bibinfo
  {author} {\bibfnamefont {M.~A.}\ \bibnamefont {Martin-Delgado}},\ }\bibfield
  {title} {\bibinfo {title} {Topology-induced anomalous defect production by
  crossing a quantum critical point},\ }\href
  {https://doi.org/10.1103/PhysRevLett.102.135702} {\bibfield  {journal}
  {\bibinfo  {journal} {Phys. Rev. Lett.}\ }\textbf {\bibinfo {volume} {102}},\
  \bibinfo {pages} {135702} (\bibinfo {year} {2009})}\BibitemShut {NoStop}%
\bibitem [{\citenamefont {Antunes}\ \emph {et~al.}(2006)\citenamefont
  {Antunes}, \citenamefont {Gandra},\ and\ \citenamefont
  {Rivers}}]{Antunes2006}%
  \BibitemOpen
  \bibfield  {author} {\bibinfo {author} {\bibfnamefont {N.~D.}\ \bibnamefont
  {Antunes}}, \bibinfo {author} {\bibfnamefont {P.}~\bibnamefont {Gandra}},\
  and\ \bibinfo {author} {\bibfnamefont {R.~J.}\ \bibnamefont {Rivers}},\
  }\bibfield  {title} {\bibinfo {title} {Is domain formation decided before or
  after the transition?},\ }\href {https://doi.org/10.1103/PhysRevD.73.125003}
  {\bibfield  {journal} {\bibinfo  {journal} {Phys. Rev. D}\ }\textbf {\bibinfo
  {volume} {73}},\ \bibinfo {pages} {125003} (\bibinfo {year}
  {2006})}\BibitemShut {NoStop}%
\bibitem [{\citenamefont {Lee}\ \emph {et~al.}(2015)\citenamefont {Lee},
  \citenamefont {Han},\ and\ \citenamefont {Choi}}]{Minchul2015}%
  \BibitemOpen
  \bibfield  {author} {\bibinfo {author} {\bibfnamefont {M.}~\bibnamefont
  {Lee}}, \bibinfo {author} {\bibfnamefont {S.}~\bibnamefont {Han}},\ and\
  \bibinfo {author} {\bibfnamefont {M.-S.}\ \bibnamefont {Choi}},\ }\bibfield
  {title} {\bibinfo {title} {Kibble-zurek mechanism in a topological phase
  transition},\ }\href {https://doi.org/10.1103/PhysRevB.92.035117} {\bibfield
  {journal} {\bibinfo  {journal} {Phys. Rev. B}\ }\textbf {\bibinfo {volume}
  {92}},\ \bibinfo {pages} {035117} (\bibinfo {year} {2015})}\BibitemShut
  {NoStop}%
\bibitem [{\citenamefont {Suzuki}(2010)}]{Suzuki2010}%
  \BibitemOpen
  \bibfield  {author} {\bibinfo {author} {\bibfnamefont {S.}~\bibnamefont
  {Suzuki}},\ }\bibinfo {title} {Quench dynamics of quantum and classical ising
  chains: From the viewpoint of the kibble--zurek mechanism},\ in\ \href
  {https://doi.org/10.1007/978-3-642-11470-0_5} {\emph {\bibinfo {booktitle}
  {Quantum Quenching, Annealing and Computation}}},\ \bibinfo {editor} {edited
  by\ \bibinfo {editor} {\bibfnamefont {A.~K.}\ \bibnamefont {Chandra}},
  \bibinfo {editor} {\bibfnamefont {A.}~\bibnamefont {Das}},\ and\ \bibinfo
  {editor} {\bibfnamefont {B.~K.}\ \bibnamefont {Chakrabarti}}}\ (\bibinfo
  {publisher} {Springer Berlin Heidelberg},\ \bibinfo {address} {Berlin,
  Heidelberg},\ \bibinfo {year} {2010})\ pp.\ \bibinfo {pages}
  {115--143}\BibitemShut {NoStop}%
\bibitem [{\citenamefont {Ladewig}\ \emph {et~al.}(2020)\citenamefont
  {Ladewig}, \citenamefont {Mathey},\ and\ \citenamefont
  {Diehl}}]{Ladewig2020}%
  \BibitemOpen
  \bibfield  {author} {\bibinfo {author} {\bibfnamefont {B.}~\bibnamefont
  {Ladewig}}, \bibinfo {author} {\bibfnamefont {S.}~\bibnamefont {Mathey}},\
  and\ \bibinfo {author} {\bibfnamefont {S.}~\bibnamefont {Diehl}},\ }\bibfield
   {title} {\bibinfo {title} {Kibble-zurek mechanism from different angles: The
  transverse xy model and subleading scalings},\ }\href
  {https://doi.org/10.1103/PhysRevB.102.104306} {\bibfield  {journal} {\bibinfo
   {journal} {Phys. Rev. B}\ }\textbf {\bibinfo {volume} {102}},\ \bibinfo
  {pages} {104306} (\bibinfo {year} {2020})}\BibitemShut {NoStop}%
\bibitem [{\citenamefont {Patan\`e}\ \emph {et~al.}(2008)\citenamefont
  {Patan\`e}, \citenamefont {Silva}, \citenamefont {Amico}, \citenamefont
  {Fazio},\ and\ \citenamefont {Santoro}}]{Dario2008}%
  \BibitemOpen
  \bibfield  {author} {\bibinfo {author} {\bibfnamefont {D.}~\bibnamefont
  {Patan\`e}}, \bibinfo {author} {\bibfnamefont {A.}~\bibnamefont {Silva}},
  \bibinfo {author} {\bibfnamefont {L.}~\bibnamefont {Amico}}, \bibinfo
  {author} {\bibfnamefont {R.}~\bibnamefont {Fazio}},\ and\ \bibinfo {author}
  {\bibfnamefont {G.~E.}\ \bibnamefont {Santoro}},\ }\bibfield  {title}
  {\bibinfo {title} {Adiabatic dynamics in open quantum critical many-body
  systems},\ }\href {https://doi.org/10.1103/PhysRevLett.101.175701} {\bibfield
   {journal} {\bibinfo  {journal} {Phys. Rev. Lett.}\ }\textbf {\bibinfo
  {volume} {101}},\ \bibinfo {pages} {175701} (\bibinfo {year}
  {2008})}\BibitemShut {NoStop}%
\bibitem [{\citenamefont {Patan\`e}\ \emph {et~al.}(2009)\citenamefont
  {Patan\`e}, \citenamefont {Amico}, \citenamefont {Silva}, \citenamefont
  {Fazio},\ and\ \citenamefont {Santoro}}]{Dario2009}%
  \BibitemOpen
  \bibfield  {author} {\bibinfo {author} {\bibfnamefont {D.}~\bibnamefont
  {Patan\`e}}, \bibinfo {author} {\bibfnamefont {L.}~\bibnamefont {Amico}},
  \bibinfo {author} {\bibfnamefont {A.}~\bibnamefont {Silva}}, \bibinfo
  {author} {\bibfnamefont {R.}~\bibnamefont {Fazio}},\ and\ \bibinfo {author}
  {\bibfnamefont {G.~E.}\ \bibnamefont {Santoro}},\ }\bibfield  {title}
  {\bibinfo {title} {Adiabatic dynamics of a quantum critical system coupled to
  an environment: Scaling and kinetic equation approaches},\ }\href
  {https://doi.org/10.1103/PhysRevB.80.024302} {\bibfield  {journal} {\bibinfo
  {journal} {Phys. Rev. B}\ }\textbf {\bibinfo {volume} {80}},\ \bibinfo
  {pages} {024302} (\bibinfo {year} {2009})}\BibitemShut {NoStop}%
\bibitem [{\citenamefont {Puebla}\ \emph
  {et~al.}(2020{\natexlab{b}})\citenamefont {Puebla}, \citenamefont {Smirne},
  \citenamefont {Huelga},\ and\ \citenamefont {Plenio}}]{Puebla2020}%
  \BibitemOpen
  \bibfield  {author} {\bibinfo {author} {\bibfnamefont {R.}~\bibnamefont
  {Puebla}}, \bibinfo {author} {\bibfnamefont {A.}~\bibnamefont {Smirne}},
  \bibinfo {author} {\bibfnamefont {S.~F.}\ \bibnamefont {Huelga}},\ and\
  \bibinfo {author} {\bibfnamefont {M.~B.}\ \bibnamefont {Plenio}},\ }\bibfield
   {title} {\bibinfo {title} {Universal anti-kibble-zurek scaling in fully
  connected systems},\ }\href {https://doi.org/10.1103/PhysRevLett.124.230602}
  {\bibfield  {journal} {\bibinfo  {journal} {Phys. Rev. Lett.}\ }\textbf
  {\bibinfo {volume} {124}},\ \bibinfo {pages} {230602} (\bibinfo {year}
  {2020}{\natexlab{b}})}\BibitemShut {NoStop}%
\bibitem [{\citenamefont {Balducci}\ \emph {et~al.}(2023)\citenamefont
  {Balducci}, \citenamefont {Beau}, \citenamefont {Yang}, \citenamefont
  {Gambassi},\ and\ \citenamefont {del Campo}}]{Balducci2023}%
  \BibitemOpen
  \bibfield  {author} {\bibinfo {author} {\bibfnamefont {F.}~\bibnamefont
  {Balducci}}, \bibinfo {author} {\bibfnamefont {M.}~\bibnamefont {Beau}},
  \bibinfo {author} {\bibfnamefont {J.}~\bibnamefont {Yang}}, \bibinfo {author}
  {\bibfnamefont {A.}~\bibnamefont {Gambassi}},\ and\ \bibinfo {author}
  {\bibfnamefont {A.}~\bibnamefont {del Campo}},\ }\bibfield  {title} {\bibinfo
  {title} {Large deviations beyond the kibble-zurek mechanism},\ }\href
  {https://doi.org/10.1103/PhysRevLett.131.230401} {\bibfield  {journal}
  {\bibinfo  {journal} {Phys. Rev. Lett.}\ }\textbf {\bibinfo {volume} {131}},\
  \bibinfo {pages} {230401} (\bibinfo {year} {2023})}\BibitemShut {NoStop}%
\bibitem [{\citenamefont {del Campo}\ \emph {et~al.}(2012)\citenamefont {del
  Campo}, \citenamefont {Rams},\ and\ \citenamefont {Zurek}}]{Adolfo2012}%
  \BibitemOpen
  \bibfield  {author} {\bibinfo {author} {\bibfnamefont {A.}~\bibnamefont {del
  Campo}}, \bibinfo {author} {\bibfnamefont {M.~M.}\ \bibnamefont {Rams}},\
  and\ \bibinfo {author} {\bibfnamefont {W.~H.}\ \bibnamefont {Zurek}},\
  }\bibfield  {title} {\bibinfo {title} {Assisted finite-rate adiabatic passage
  across a quantum critical point: Exact solution for the quantum ising
  model},\ }\href {https://doi.org/10.1103/PhysRevLett.109.115703} {\bibfield
  {journal} {\bibinfo  {journal} {Phys. Rev. Lett.}\ }\textbf {\bibinfo
  {volume} {109}},\ \bibinfo {pages} {115703} (\bibinfo {year}
  {2012})}\BibitemShut {NoStop}%
\bibitem [{\citenamefont {Jafari}\ and\ \citenamefont
  {Akbari}(2026)}]{Jafari2026bb}%
  \BibitemOpen
  \bibfield  {author} {\bibinfo {author} {\bibfnamefont {R.}~\bibnamefont
  {Jafari}}\ and\ \bibinfo {author} {\bibfnamefont {A.}~\bibnamefont
  {Akbari}},\ }\bibfield  {title} {\bibinfo {title} {Separation of the
  kibble-zurek mechanism from quantum criticality},\ }\href
  {https://doi.org/10.1103/9ts3-z9sk} {\bibfield  {journal} {\bibinfo
  {journal} {Phys. Rev. Res.}\ }\textbf {\bibinfo {volume} {8}},\ \bibinfo
  {pages} {L022027} (\bibinfo {year} {2026})}\BibitemShut {NoStop}%
\bibitem [{\citenamefont {De~Grandi}\ \emph {et~al.}(2010)\citenamefont
  {De~Grandi}, \citenamefont {Gritsev},\ and\ \citenamefont
  {Polkovnikov}}]{Grandi2010}%
  \BibitemOpen
  \bibfield  {author} {\bibinfo {author} {\bibfnamefont {C.}~\bibnamefont
  {De~Grandi}}, \bibinfo {author} {\bibfnamefont {V.}~\bibnamefont {Gritsev}},\
  and\ \bibinfo {author} {\bibfnamefont {A.}~\bibnamefont {Polkovnikov}},\
  }\bibfield  {title} {\bibinfo {title} {Quench dynamics near a quantum
  critical point},\ }\href {https://doi.org/10.1103/PhysRevB.81.012303}
  {\bibfield  {journal} {\bibinfo  {journal} {Phys. Rev. B}\ }\textbf {\bibinfo
  {volume} {81}},\ \bibinfo {pages} {012303} (\bibinfo {year}
  {2010})}\BibitemShut {NoStop}%
\bibitem [{\citenamefont {Polkovnikov}\ and\ \citenamefont
  {Gritsev}(2008)}]{Polkovnikov2008}%
  \BibitemOpen
  \bibfield  {author} {\bibinfo {author} {\bibfnamefont {A.}~\bibnamefont
  {Polkovnikov}}\ and\ \bibinfo {author} {\bibfnamefont {V.}~\bibnamefont
  {Gritsev}},\ }\bibfield  {title} {\bibinfo {title} {Breakdown of the
  adiabatic limit in low-dimensional gapless systems},\ }\href
  {https://doi.org/10.1038/nphys963} {\bibfield  {journal} {\bibinfo  {journal}
  {Nature Physics}\ }\textbf {\bibinfo {volume} {4}},\ \bibinfo {pages} {477}
  (\bibinfo {year} {2008})}\BibitemShut {NoStop}%
\bibitem [{\citenamefont {Jr.}\ and\ \citenamefont {Cosme}(2026)}]{Jara2026}%
  \BibitemOpen
  \bibfield  {author} {\bibinfo {author} {\bibfnamefont {R.~D.~J.}\
  \bibnamefont {Jr.}}\ and\ \bibinfo {author} {\bibfnamefont {J.~G.}\
  \bibnamefont {Cosme}},\ }\href
  {https://doi.org/https://doi.org/10.1103/jjqm-f8hm} {\bibinfo {title}
  {Universality of dissipative discrete time crystal formation}} (\bibinfo
  {year} {2026}),\ \Eprint {https://arxiv.org/abs/2507.18950} {arXiv:2507.18950
  [cond-mat.stat-mech]} \BibitemShut {NoStop}%
\bibitem [{\citenamefont {Liu}\ \emph {et~al.}(2018)\citenamefont {Liu},
  \citenamefont {Donadello}, \citenamefont {Lamporesi}, \citenamefont
  {Ferrari}, \citenamefont {Gou}, \citenamefont {Dalfovo},\ and\ \citenamefont
  {Proukakis}}]{Liu2018}%
  \BibitemOpen
  \bibfield  {author} {\bibinfo {author} {\bibfnamefont {I.-K.}\ \bibnamefont
  {Liu}}, \bibinfo {author} {\bibfnamefont {S.}~\bibnamefont {Donadello}},
  \bibinfo {author} {\bibfnamefont {G.}~\bibnamefont {Lamporesi}}, \bibinfo
  {author} {\bibfnamefont {G.}~\bibnamefont {Ferrari}}, \bibinfo {author}
  {\bibfnamefont {S.-C.}\ \bibnamefont {Gou}}, \bibinfo {author} {\bibfnamefont
  {F.}~\bibnamefont {Dalfovo}},\ and\ \bibinfo {author} {\bibfnamefont
  {N.}~\bibnamefont {Proukakis}},\ }\bibfield  {title} {\bibinfo {title}
  {Dynamical equilibration across a quenched phase transition in a trapped
  quantum gas},\ }\href {https://doi.org/10.1038/s42005-018-0023-6} {\bibfield
  {journal} {\bibinfo  {journal} {Communications Physics}\ }\textbf {\bibinfo
  {volume} {1}},\ \bibinfo {pages} {24} (\bibinfo {year} {2018})}\BibitemShut
  {NoStop}%
\bibitem [{\citenamefont {Bowick}\ \emph {et~al.}(1994)\citenamefont {Bowick},
  \citenamefont {Chandar}, \citenamefont {Schiff},\ and\ \citenamefont
  {Srivastava}}]{Mark1994}%
  \BibitemOpen
  \bibfield  {author} {\bibinfo {author} {\bibfnamefont {M.~J.}\ \bibnamefont
  {Bowick}}, \bibinfo {author} {\bibfnamefont {L.}~\bibnamefont {Chandar}},
  \bibinfo {author} {\bibfnamefont {E.~A.}\ \bibnamefont {Schiff}},\ and\
  \bibinfo {author} {\bibfnamefont {A.~M.}\ \bibnamefont {Srivastava}},\
  }\bibfield  {title} {\bibinfo {title} {The cosmological kibble mechanism in
  the laboratory: String formation in liquid crystals},\ }\href
  {https://doi.org/10.1126/science.263.5149.943} {\bibfield  {journal}
  {\bibinfo  {journal} {Science}\ }\textbf {\bibinfo {volume} {263}},\ \bibinfo
  {pages} {943} (\bibinfo {year} {1994})}\BibitemShut {NoStop}%
\bibitem [{\citenamefont {Maniv}\ \emph {et~al.}(2003)\citenamefont {Maniv},
  \citenamefont {Polturak},\ and\ \citenamefont {Koren}}]{Maniv2003}%
  \BibitemOpen
  \bibfield  {author} {\bibinfo {author} {\bibfnamefont {A.}~\bibnamefont
  {Maniv}}, \bibinfo {author} {\bibfnamefont {E.}~\bibnamefont {Polturak}},\
  and\ \bibinfo {author} {\bibfnamefont {G.}~\bibnamefont {Koren}},\ }\bibfield
   {title} {\bibinfo {title} {Observation of magnetic flux generated
  spontaneously during a rapid quench of superconducting films},\ }\href
  {https://doi.org/10.1103/PhysRevLett.91.197001} {\bibfield  {journal}
  {\bibinfo  {journal} {Phys. Rev. Lett.}\ }\textbf {\bibinfo {volume} {91}},\
  \bibinfo {pages} {197001} (\bibinfo {year} {2003})}\BibitemShut {NoStop}%
\bibitem [{\citenamefont {Corman}\ \emph {et~al.}(2014)\citenamefont {Corman},
  \citenamefont {Chomaz}, \citenamefont {Bienaim\'e}, \citenamefont
  {Desbuquois}, \citenamefont {Weitenberg}, \citenamefont {Nascimb\`ene},
  \citenamefont {Dalibard},\ and\ \citenamefont {Beugnon}}]{Corman2014}%
  \BibitemOpen
  \bibfield  {author} {\bibinfo {author} {\bibfnamefont {L.}~\bibnamefont
  {Corman}}, \bibinfo {author} {\bibfnamefont {L.}~\bibnamefont {Chomaz}},
  \bibinfo {author} {\bibfnamefont {T.}~\bibnamefont {Bienaim\'e}}, \bibinfo
  {author} {\bibfnamefont {R.}~\bibnamefont {Desbuquois}}, \bibinfo {author}
  {\bibfnamefont {C.}~\bibnamefont {Weitenberg}}, \bibinfo {author}
  {\bibfnamefont {S.}~\bibnamefont {Nascimb\`ene}}, \bibinfo {author}
  {\bibfnamefont {J.}~\bibnamefont {Dalibard}},\ and\ \bibinfo {author}
  {\bibfnamefont {J.}~\bibnamefont {Beugnon}},\ }\bibfield  {title} {\bibinfo
  {title} {Quench-induced supercurrents in an annular bose gas},\ }\href
  {https://doi.org/10.1103/PhysRevLett.113.135302} {\bibfield  {journal}
  {\bibinfo  {journal} {Phys. Rev. Lett.}\ }\textbf {\bibinfo {volume} {113}},\
  \bibinfo {pages} {135302} (\bibinfo {year} {2014})}\BibitemShut {NoStop}%
\bibitem [{\citenamefont {Navon}\ \emph {et~al.}(2015)\citenamefont {Navon},
  \citenamefont {Gaunt}, \citenamefont {Smith},\ and\ \citenamefont
  {Hadzibabic}}]{Navon2015}%
  \BibitemOpen
  \bibfield  {author} {\bibinfo {author} {\bibfnamefont {N.}~\bibnamefont
  {Navon}}, \bibinfo {author} {\bibfnamefont {A.~L.}\ \bibnamefont {Gaunt}},
  \bibinfo {author} {\bibfnamefont {R.~P.}\ \bibnamefont {Smith}},\ and\
  \bibinfo {author} {\bibfnamefont {Z.}~\bibnamefont {Hadzibabic}},\ }\bibfield
   {title} {\bibinfo {title} {Critical dynamics of spontaneous symmetry
  breaking in a homogeneous bose gas},\ }\href
  {https://doi.org/10.1126/science.1258676} {\bibfield  {journal} {\bibinfo
  {journal} {Science}\ }\textbf {\bibinfo {volume} {347}},\ \bibinfo {pages}
  {167} (\bibinfo {year} {2015})}\BibitemShut {NoStop}%
\bibitem [{\citenamefont {Deutschländer}\ \emph {et~al.}(2015)\citenamefont
  {Deutschländer}, \citenamefont {Dillmann}, \citenamefont {Maret},\ and\
  \citenamefont {Keim}}]{Sven2015}%
  \BibitemOpen
  \bibfield  {author} {\bibinfo {author} {\bibfnamefont {S.}~\bibnamefont
  {Deutschländer}}, \bibinfo {author} {\bibfnamefont {P.}~\bibnamefont
  {Dillmann}}, \bibinfo {author} {\bibfnamefont {G.}~\bibnamefont {Maret}},\
  and\ \bibinfo {author} {\bibfnamefont {P.}~\bibnamefont {Keim}},\ }\bibfield
  {title} {\bibinfo {title} {Kibble–zurek mechanism in colloidal
  monolayers},\ }\href {https://doi.org/10.1073/pnas.1500763112} {\bibfield
  {journal} {\bibinfo  {journal} {Proceedings of the National Academy of
  Sciences}\ }\textbf {\bibinfo {volume} {112}},\ \bibinfo {pages} {6925}
  (\bibinfo {year} {2015})}\BibitemShut {NoStop}%
\bibitem [{\citenamefont {Keesling}\ \emph {et~al.}(2019)\citenamefont
  {Keesling}, \citenamefont {Omran}, \citenamefont {Levine}, \citenamefont
  {Bernien}, \citenamefont {Pichler}, \citenamefont {Choi}, \citenamefont
  {Samajdar}, \citenamefont {Schwartz}, \citenamefont {Silvi}, \citenamefont
  {Sachdev} \emph {et~al.}}]{Keesling2019}%
  \BibitemOpen
  \bibfield  {author} {\bibinfo {author} {\bibfnamefont {A.}~\bibnamefont
  {Keesling}}, \bibinfo {author} {\bibfnamefont {A.}~\bibnamefont {Omran}},
  \bibinfo {author} {\bibfnamefont {H.}~\bibnamefont {Levine}}, \bibinfo
  {author} {\bibfnamefont {H.}~\bibnamefont {Bernien}}, \bibinfo {author}
  {\bibfnamefont {H.}~\bibnamefont {Pichler}}, \bibinfo {author} {\bibfnamefont
  {S.}~\bibnamefont {Choi}}, \bibinfo {author} {\bibfnamefont {R.}~\bibnamefont
  {Samajdar}}, \bibinfo {author} {\bibfnamefont {S.}~\bibnamefont {Schwartz}},
  \bibinfo {author} {\bibfnamefont {P.}~\bibnamefont {Silvi}}, \bibinfo
  {author} {\bibfnamefont {S.}~\bibnamefont {Sachdev}}, \emph {et~al.},\
  }\bibfield  {title} {\bibinfo {title} {Quantum kibble--zurek mechanism and
  critical dynamics on a programmable rydberg simulator},\ }\href
  {https://doi.org/10.1038/s41586-019-1070-1} {\bibfield  {journal} {\bibinfo
  {journal} {Nature}\ }\textbf {\bibinfo {volume} {568}},\ \bibinfo {pages}
  {207} (\bibinfo {year} {2019})}\BibitemShut {NoStop}%
\bibitem [{\citenamefont {Zhang}\ \emph {et~al.}(2017)\citenamefont {Zhang},
  \citenamefont {Cucchietti}, \citenamefont {Laflamme},\ and\ \citenamefont
  {Suter}}]{Zhang2017}%
  \BibitemOpen
  \bibfield  {author} {\bibinfo {author} {\bibfnamefont {J.}~\bibnamefont
  {Zhang}}, \bibinfo {author} {\bibfnamefont {F.~M.}\ \bibnamefont
  {Cucchietti}}, \bibinfo {author} {\bibfnamefont {R.}~\bibnamefont
  {Laflamme}},\ and\ \bibinfo {author} {\bibfnamefont {D.}~\bibnamefont
  {Suter}},\ }\bibfield  {title} {\bibinfo {title} {Defect production in
  non-equilibrium phase transitions: experimental investigation of the
  kibble–zurek mechanism in a two-qubit quantum simulator},\ }\href
  {https://doi.org/10.1088/1367-2630/aa6653} {\bibfield  {journal} {\bibinfo
  {journal} {New Journal of Physics}\ }\textbf {\bibinfo {volume} {19}},\
  \bibinfo {pages} {043001} (\bibinfo {year} {2017})}\BibitemShut {NoStop}%
\bibitem [{\citenamefont {Bando}\ \emph {et~al.}(2020)\citenamefont {Bando},
  \citenamefont {Susa}, \citenamefont {Oshiyama}, \citenamefont {Shibata},
  \citenamefont {Ohzeki}, \citenamefont {G\'omez-Ruiz}, \citenamefont {Lidar},
  \citenamefont {Suzuki}, \citenamefont {del Campo},\ and\ \citenamefont
  {Nishimori}}]{Bando2020}%
  \BibitemOpen
  \bibfield  {author} {\bibinfo {author} {\bibfnamefont {Y.}~\bibnamefont
  {Bando}}, \bibinfo {author} {\bibfnamefont {Y.}~\bibnamefont {Susa}},
  \bibinfo {author} {\bibfnamefont {H.}~\bibnamefont {Oshiyama}}, \bibinfo
  {author} {\bibfnamefont {N.}~\bibnamefont {Shibata}}, \bibinfo {author}
  {\bibfnamefont {M.}~\bibnamefont {Ohzeki}}, \bibinfo {author} {\bibfnamefont
  {F.~J.}\ \bibnamefont {G\'omez-Ruiz}}, \bibinfo {author} {\bibfnamefont
  {D.~A.}\ \bibnamefont {Lidar}}, \bibinfo {author} {\bibfnamefont
  {S.}~\bibnamefont {Suzuki}}, \bibinfo {author} {\bibfnamefont
  {A.}~\bibnamefont {del Campo}},\ and\ \bibinfo {author} {\bibfnamefont
  {H.}~\bibnamefont {Nishimori}},\ }\bibfield  {title} {\bibinfo {title}
  {Probing the universality of topological defect formation in a quantum
  annealer: Kibble-zurek mechanism and beyond},\ }\href
  {https://doi.org/10.1103/PhysRevResearch.2.033369} {\bibfield  {journal}
  {\bibinfo  {journal} {Phys. Rev. Res.}\ }\textbf {\bibinfo {volume} {2}},\
  \bibinfo {pages} {033369} (\bibinfo {year} {2020})}\BibitemShut {NoStop}%
\bibitem [{\citenamefont {Higuera-Quintero}\ \emph {et~al.}(2022)\citenamefont
  {Higuera-Quintero}, \citenamefont {Rodríguez}, \citenamefont {Quiroga},\
  and\ \citenamefont {Gómez-Ruiz}}]{Higuera2022}%
  \BibitemOpen
  \bibfield  {author} {\bibinfo {author} {\bibfnamefont {S.}~\bibnamefont
  {Higuera-Quintero}}, \bibinfo {author} {\bibfnamefont {F.~J.}\ \bibnamefont
  {Rodríguez}}, \bibinfo {author} {\bibfnamefont {L.}~\bibnamefont
  {Quiroga}},\ and\ \bibinfo {author} {\bibfnamefont {F.~J.}\ \bibnamefont
  {Gómez-Ruiz}},\ }\bibfield  {title} {\bibinfo {title} {Experimental
  validation of the kibble-zurek mechanism on a digital quantum computer},\
  }\bibfield  {journal} {\bibinfo  {journal} {Frontiers in Quantum Science and
  Technology}\ }\textbf {\bibinfo {volume} {Volume 1 - 2022}},\ \href
  {https://doi.org/10.3389/frqst.2022.1026025} {10.3389/frqst.2022.1026025}
  (\bibinfo {year} {2022})\BibitemShut {NoStop}%
\bibitem [{\citenamefont {Kolodrubetz}\ \emph {et~al.}(2012)\citenamefont
  {Kolodrubetz}, \citenamefont {Clark},\ and\ \citenamefont
  {Huse}}]{Kolodrubetz2012}%
  \BibitemOpen
  \bibfield  {author} {\bibinfo {author} {\bibfnamefont {M.}~\bibnamefont
  {Kolodrubetz}}, \bibinfo {author} {\bibfnamefont {B.~K.}\ \bibnamefont
  {Clark}},\ and\ \bibinfo {author} {\bibfnamefont {D.~A.}\ \bibnamefont
  {Huse}},\ }\bibfield  {title} {\bibinfo {title} {Nonequilibrium dynamic
  critical scaling of the quantum ising chain},\ }\href
  {https://doi.org/10.1103/PhysRevLett.109.015701} {\bibfield  {journal}
  {\bibinfo  {journal} {Phys. Rev. Lett.}\ }\textbf {\bibinfo {volume} {109}},\
  \bibinfo {pages} {015701} (\bibinfo {year} {2012})}\BibitemShut {NoStop}%
\bibitem [{\citenamefont {Chesler}\ \emph {et~al.}(2015)\citenamefont
  {Chesler}, \citenamefont {Garc\'{\i}a-Garc\'{\i}a},\ and\ \citenamefont
  {Liu}}]{Chesler2015}%
  \BibitemOpen
  \bibfield  {author} {\bibinfo {author} {\bibfnamefont {P.~M.}\ \bibnamefont
  {Chesler}}, \bibinfo {author} {\bibfnamefont {A.~M.}\ \bibnamefont
  {Garc\'{\i}a-Garc\'{\i}a}},\ and\ \bibinfo {author} {\bibfnamefont
  {H.}~\bibnamefont {Liu}},\ }\bibfield  {title} {\bibinfo {title} {Defect
  formation beyond kibble-zurek mechanism and holography},\ }\href
  {https://doi.org/10.1103/PhysRevX.5.021015} {\bibfield  {journal} {\bibinfo
  {journal} {Phys. Rev. X}\ }\textbf {\bibinfo {volume} {5}},\ \bibinfo {pages}
  {021015} (\bibinfo {year} {2015})}\BibitemShut {NoStop}%
\bibitem [{\citenamefont {Silvi}\ \emph {et~al.}(2016)\citenamefont {Silvi},
  \citenamefont {Morigi}, \citenamefont {Calarco},\ and\ \citenamefont
  {Montangero}}]{Silvi2016}%
  \BibitemOpen
  \bibfield  {author} {\bibinfo {author} {\bibfnamefont {P.}~\bibnamefont
  {Silvi}}, \bibinfo {author} {\bibfnamefont {G.}~\bibnamefont {Morigi}},
  \bibinfo {author} {\bibfnamefont {T.}~\bibnamefont {Calarco}},\ and\ \bibinfo
  {author} {\bibfnamefont {S.}~\bibnamefont {Montangero}},\ }\bibfield  {title}
  {\bibinfo {title} {Crossover from classical to quantum kibble-zurek
  scaling},\ }\href {https://doi.org/10.1103/PhysRevLett.116.225701} {\bibfield
   {journal} {\bibinfo  {journal} {Phys. Rev. Lett.}\ }\textbf {\bibinfo
  {volume} {116}},\ \bibinfo {pages} {225701} (\bibinfo {year}
  {2016})}\BibitemShut {NoStop}%
\bibitem [{\citenamefont {Deng}\ \emph {et~al.}(2026)\citenamefont {Deng},
  \citenamefont {Yang}, \citenamefont {Sun}, \citenamefont {Li},\ and\
  \citenamefont {Yu}}]{Deng2026}%
  \BibitemOpen
  \bibfield  {author} {\bibinfo {author} {\bibfnamefont {M.}~\bibnamefont
  {Deng}}, \bibinfo {author} {\bibfnamefont {S.}~\bibnamefont {Yang}}, \bibinfo
  {author} {\bibfnamefont {C.}~\bibnamefont {Sun}}, \bibinfo {author}
  {\bibfnamefont {F.}~\bibnamefont {Li}},\ and\ \bibinfo {author}
  {\bibfnamefont {X.-J.}\ \bibnamefont {Yu}},\ }\bibfield  {title} {\bibinfo
  {title} {Anomalous dynamical scaling at topological quantum criticality},\
  }\href {https://doi.org/10.1103/mqr4-wnny} {\bibfield  {journal} {\bibinfo
  {journal} {Phys. Rev. Lett.}\ }\textbf {\bibinfo {volume} {137}},\ \bibinfo
  {pages} {096605} (\bibinfo {year} {2026})}\BibitemShut {NoStop}%
\bibitem [{\citenamefont {De~Grandi}\ \emph {et~al.}(2011)\citenamefont
  {De~Grandi}, \citenamefont {Polkovnikov},\ and\ \citenamefont
  {Sandvik}}]{Grandi2011}%
  \BibitemOpen
  \bibfield  {author} {\bibinfo {author} {\bibfnamefont {C.}~\bibnamefont
  {De~Grandi}}, \bibinfo {author} {\bibfnamefont {A.}~\bibnamefont
  {Polkovnikov}},\ and\ \bibinfo {author} {\bibfnamefont {A.~W.}\ \bibnamefont
  {Sandvik}},\ }\bibfield  {title} {\bibinfo {title} {Universal nonequilibrium
  quantum dynamics in imaginary time},\ }\href
  {https://doi.org/10.1103/PhysRevB.84.224303} {\bibfield  {journal} {\bibinfo
  {journal} {Phys. Rev. B}\ }\textbf {\bibinfo {volume} {84}},\ \bibinfo
  {pages} {224303} (\bibinfo {year} {2011})}\BibitemShut {NoStop}%
\bibitem [{\citenamefont {Chandran}\ \emph {et~al.}(2012)\citenamefont
  {Chandran}, \citenamefont {Erez}, \citenamefont {Gubser},\ and\ \citenamefont
  {Sondhi}}]{Chandran2012}%
  \BibitemOpen
  \bibfield  {author} {\bibinfo {author} {\bibfnamefont {A.}~\bibnamefont
  {Chandran}}, \bibinfo {author} {\bibfnamefont {A.}~\bibnamefont {Erez}},
  \bibinfo {author} {\bibfnamefont {S.~S.}\ \bibnamefont {Gubser}},\ and\
  \bibinfo {author} {\bibfnamefont {S.~L.}\ \bibnamefont {Sondhi}},\ }\bibfield
   {title} {\bibinfo {title} {Kibble-zurek problem: Universality and the
  scaling limit},\ }\href {https://doi.org/10.1103/PhysRevB.86.064304}
  {\bibfield  {journal} {\bibinfo  {journal} {Phys. Rev. B}\ }\textbf {\bibinfo
  {volume} {86}},\ \bibinfo {pages} {064304} (\bibinfo {year}
  {2012})}\BibitemShut {NoStop}%
\bibitem [{\citenamefont {Hwang}\ \emph {et~al.}(2015)\citenamefont {Hwang},
  \citenamefont {Puebla},\ and\ \citenamefont {Plenio}}]{Hwang2015}%
  \BibitemOpen
  \bibfield  {author} {\bibinfo {author} {\bibfnamefont {M.-J.}\ \bibnamefont
  {Hwang}}, \bibinfo {author} {\bibfnamefont {R.}~\bibnamefont {Puebla}},\ and\
  \bibinfo {author} {\bibfnamefont {M.~B.}\ \bibnamefont {Plenio}},\ }\bibfield
   {title} {\bibinfo {title} {Quantum phase transition and universal dynamics
  in the rabi model},\ }\href {https://doi.org/10.1103/PhysRevLett.115.180404}
  {\bibfield  {journal} {\bibinfo  {journal} {Phys. Rev. Lett.}\ }\textbf
  {\bibinfo {volume} {115}},\ \bibinfo {pages} {180404} (\bibinfo {year}
  {2015})}\BibitemShut {NoStop}%
\bibitem [{\citenamefont {del Campo}(2018)}]{Adolfo2018}%
  \BibitemOpen
  \bibfield  {author} {\bibinfo {author} {\bibfnamefont {A.}~\bibnamefont {del
  Campo}},\ }\bibfield  {title} {\bibinfo {title} {Universal statistics of
  topological defects formed in a quantum phase transition},\ }\href
  {https://doi.org/10.1103/PhysRevLett.121.200601} {\bibfield  {journal}
  {\bibinfo  {journal} {Phys. Rev. Lett.}\ }\textbf {\bibinfo {volume} {121}},\
  \bibinfo {pages} {200601} (\bibinfo {year} {2018})}\BibitemShut {NoStop}%
\bibitem [{\citenamefont {Nikoghosyan}\ \emph {et~al.}(2016)\citenamefont
  {Nikoghosyan}, \citenamefont {Nigmatullin},\ and\ \citenamefont
  {Plenio}}]{Nikoghosyan2016}%
  \BibitemOpen
  \bibfield  {author} {\bibinfo {author} {\bibfnamefont {G.}~\bibnamefont
  {Nikoghosyan}}, \bibinfo {author} {\bibfnamefont {R.}~\bibnamefont
  {Nigmatullin}},\ and\ \bibinfo {author} {\bibfnamefont {M.~B.}\ \bibnamefont
  {Plenio}},\ }\bibfield  {title} {\bibinfo {title} {Universality in the
  dynamics of second-order phase transitions},\ }\href
  {https://doi.org/10.1103/PhysRevLett.116.080601} {\bibfield  {journal}
  {\bibinfo  {journal} {Phys. Rev. Lett.}\ }\textbf {\bibinfo {volume} {116}},\
  \bibinfo {pages} {080601} (\bibinfo {year} {2016})}\BibitemShut {NoStop}%
\bibitem [{\citenamefont {Deffner}(2017)}]{Deffner2017}%
  \BibitemOpen
  \bibfield  {author} {\bibinfo {author} {\bibfnamefont {S.}~\bibnamefont
  {Deffner}},\ }\bibfield  {title} {\bibinfo {title} {Kibble-zurek scaling of
  the irreversible entropy production},\ }\href
  {https://doi.org/10.1103/PhysRevE.96.052125} {\bibfield  {journal} {\bibinfo
  {journal} {Phys. Rev. E}\ }\textbf {\bibinfo {volume} {96}},\ \bibinfo
  {pages} {052125} (\bibinfo {year} {2017})}\BibitemShut {NoStop}%
\bibitem [{\citenamefont {Polkovnikov}\ and\ \citenamefont
  {Gritsev}(2010)}]{Polkovnikov2010}%
  \BibitemOpen
  \bibfield  {author} {\bibinfo {author} {\bibfnamefont {A.}~\bibnamefont
  {Polkovnikov}}\ and\ \bibinfo {author} {\bibfnamefont {V.}~\bibnamefont
  {Gritsev}},\ }\bibinfo {title} {Universal dynamics near quantum critical
  points},\ in\ \href {https://doi.org/https://doi.org/10.1201/b10273} {\emph
  {\bibinfo {booktitle} {Understanding quantum phase transitions}}},\ \bibinfo
  {editor} {edited by\ \bibinfo {editor} {\bibfnamefont {L.}~\bibnamefont
  {Carr}}}\ (\bibinfo  {publisher} {CRC Press},\ \bibinfo {year} {2010})\ pp.\
  \bibinfo {pages} {59--90}\BibitemShut {NoStop}%
\bibitem [{\citenamefont {Albash}\ and\ \citenamefont
  {Lidar}(2018)}]{Albash2018}%
  \BibitemOpen
  \bibfield  {author} {\bibinfo {author} {\bibfnamefont {T.}~\bibnamefont
  {Albash}}\ and\ \bibinfo {author} {\bibfnamefont {D.~A.}\ \bibnamefont
  {Lidar}},\ }\bibfield  {title} {\bibinfo {title} {Adiabatic quantum
  computation},\ }\href {https://doi.org/10.1103/RevModPhys.90.015002}
  {\bibfield  {journal} {\bibinfo  {journal} {Rev. Mod. Phys.}\ }\textbf
  {\bibinfo {volume} {90}},\ \bibinfo {pages} {015002} (\bibinfo {year}
  {2018})}\BibitemShut {NoStop}%
\bibitem [{\citenamefont {Dutta}\ \emph {et~al.}(2016)\citenamefont {Dutta},
  \citenamefont {Rahmani},\ and\ \citenamefont {del Campo}}]{Dutta2016}%
  \BibitemOpen
  \bibfield  {author} {\bibinfo {author} {\bibfnamefont {A.}~\bibnamefont
  {Dutta}}, \bibinfo {author} {\bibfnamefont {A.}~\bibnamefont {Rahmani}},\
  and\ \bibinfo {author} {\bibfnamefont {A.}~\bibnamefont {del Campo}},\
  }\bibfield  {title} {\bibinfo {title} {Anti-kibble-zurek behavior in crossing
  the quantum critical point of a thermally isolated system driven by a noisy
  control field},\ }\bibfield  {journal} {\bibinfo  {journal} {Physical Review
  Letters}\ }\textbf {\bibinfo {volume} {117}},\ \href
  {https://doi.org/10.1103/physrevlett.117.080402}
  {10.1103/physrevlett.117.080402} (\bibinfo {year} {2016})\BibitemShut
  {NoStop}%
\bibitem [{\citenamefont {Chenu}\ \emph {et~al.}(2017)\citenamefont {Chenu},
  \citenamefont {Beau}, \citenamefont {Cao},\ and\ \citenamefont {del
  Campo}}]{Chenu2017}%
  \BibitemOpen
  \bibfield  {author} {\bibinfo {author} {\bibfnamefont {A.}~\bibnamefont
  {Chenu}}, \bibinfo {author} {\bibfnamefont {M.}~\bibnamefont {Beau}},
  \bibinfo {author} {\bibfnamefont {J.}~\bibnamefont {Cao}},\ and\ \bibinfo
  {author} {\bibfnamefont {A.}~\bibnamefont {del Campo}},\ }\bibfield  {title}
  {\bibinfo {title} {Quantum simulation of generic many-body open system
  dynamics using classical noise},\ }\href
  {https://doi.org/10.1103/PhysRevLett.118.140403} {\bibfield  {journal}
  {\bibinfo  {journal} {Phys. Rev. Lett.}\ }\textbf {\bibinfo {volume} {118}},\
  \bibinfo {pages} {140403} (\bibinfo {year} {2017})}\BibitemShut {NoStop}%
\bibitem [{\citenamefont {Sadeghizade}\ \emph {et~al.}(2025)\citenamefont
  {Sadeghizade}, \citenamefont {Jafari},\ and\ \citenamefont
  {Langari}}]{Sadeghizade2025}%
  \BibitemOpen
  \bibfield  {author} {\bibinfo {author} {\bibfnamefont {S.}~\bibnamefont
  {Sadeghizade}}, \bibinfo {author} {\bibfnamefont {R.}~\bibnamefont
  {Jafari}},\ and\ \bibinfo {author} {\bibfnamefont {A.}~\bibnamefont
  {Langari}},\ }\bibfield  {title} {\bibinfo {title} {Anti-kibble-zurek
  behavior in the quantum xy spin-$\frac{1}{2}$ chain driven by correlated
  noisy magnetic field and anisotropy},\ }\href
  {https://doi.org/10.1103/PhysRevB.111.104310} {\bibfield  {journal} {\bibinfo
   {journal} {Phys. Rev. B}\ }\textbf {\bibinfo {volume} {111}},\ \bibinfo
  {pages} {104310} (\bibinfo {year} {2025})}\BibitemShut {NoStop}%
\bibitem [{\citenamefont {Singh}\ and\ \citenamefont
  {Gangadharaiah}(2021)}]{Singh2021}%
  \BibitemOpen
  \bibfield  {author} {\bibinfo {author} {\bibfnamefont {M.}~\bibnamefont
  {Singh}}\ and\ \bibinfo {author} {\bibfnamefont {S.}~\bibnamefont
  {Gangadharaiah}},\ }\bibfield  {title} {\bibinfo {title} {Driven quantum spin
  chain in the presence of noise: Anti-kibble-zurek behavior},\ }\href
  {https://doi.org/10.1103/PhysRevB.104.064313} {\bibfield  {journal} {\bibinfo
   {journal} {Phys. Rev. B}\ }\textbf {\bibinfo {volume} {104}},\ \bibinfo
  {pages} {064313} (\bibinfo {year} {2021})}\BibitemShut {NoStop}%
\bibitem [{\citenamefont {Gao}\ \emph {et~al.}(2017)\citenamefont {Gao},
  \citenamefont {Zhang}, \citenamefont {Yu},\ and\ \citenamefont
  {Zhu}}]{Gao2017}%
  \BibitemOpen
  \bibfield  {author} {\bibinfo {author} {\bibfnamefont {Z.-P.}\ \bibnamefont
  {Gao}}, \bibinfo {author} {\bibfnamefont {D.-W.}\ \bibnamefont {Zhang}},
  \bibinfo {author} {\bibfnamefont {Y.}~\bibnamefont {Yu}},\ and\ \bibinfo
  {author} {\bibfnamefont {S.-L.}\ \bibnamefont {Zhu}},\ }\bibfield  {title}
  {\bibinfo {title} {Anti-kibble-zurek behavior of a noisy transverse-field
  $\mathrm{XY}$ chain and its quantum simulation with two-level systems},\
  }\href {https://doi.org/10.1103/PhysRevB.95.224303} {\bibfield  {journal}
  {\bibinfo  {journal} {Phys. Rev. B}\ }\textbf {\bibinfo {volume} {95}},\
  \bibinfo {pages} {224303} (\bibinfo {year} {2017})}\BibitemShut {NoStop}%
\bibitem [{\citenamefont {Singh}\ \emph {et~al.}(2023)\citenamefont {Singh},
  \citenamefont {Dhara},\ and\ \citenamefont {Gangadharaiah}}]{Singh2023}%
  \BibitemOpen
  \bibfield  {author} {\bibinfo {author} {\bibfnamefont {M.}~\bibnamefont
  {Singh}}, \bibinfo {author} {\bibfnamefont {S.}~\bibnamefont {Dhara}},\ and\
  \bibinfo {author} {\bibfnamefont {S.}~\bibnamefont {Gangadharaiah}},\
  }\bibfield  {title} {\bibinfo {title} {Driven one-dimensional noisy kitaev
  chain},\ }\href {https://doi.org/10.1103/PhysRevB.107.014303} {\bibfield
  {journal} {\bibinfo  {journal} {Phys. Rev. B}\ }\textbf {\bibinfo {volume}
  {107}},\ \bibinfo {pages} {014303} (\bibinfo {year} {2023})}\BibitemShut
  {NoStop}%
\bibitem [{\citenamefont {Yang}\ and\ \citenamefont {Li}(2026)}]{Yang2026}%
  \BibitemOpen
  \bibfield  {author} {\bibinfo {author} {\bibfnamefont {M.}~\bibnamefont
  {Yang}}\ and\ \bibinfo {author} {\bibfnamefont {P.}~\bibnamefont {Li}},\
  }\href {https://arxiv.org/abs/2611.16237} {\bibinfo {title}
  {Anti-kibble-zurek behavior driven by dichotomous markov noise in a quantum
  ising chain: Generalized stochastic resonance in the defect density}}
  (\bibinfo {year} {2026}),\ \Eprint {https://arxiv.org/abs/2611.16237}
  {arXiv:2611.16237 [cond-mat.stat-mech]} \BibitemShut {NoStop}%
\bibitem [{\citenamefont {Gherardini}\ \emph {et~al.}(2024)\citenamefont
  {Gherardini}, \citenamefont {Buffoni},\ and\ \citenamefont
  {Defenu}}]{Gherardini2024}%
  \BibitemOpen
  \bibfield  {author} {\bibinfo {author} {\bibfnamefont {S.}~\bibnamefont
  {Gherardini}}, \bibinfo {author} {\bibfnamefont {L.}~\bibnamefont
  {Buffoni}},\ and\ \bibinfo {author} {\bibfnamefont {N.}~\bibnamefont
  {Defenu}},\ }\bibfield  {title} {\bibinfo {title} {Universal defects
  statistics with strong long-range interactions},\ }\href
  {https://doi.org/10.1103/PhysRevLett.133.113401} {\bibfield  {journal}
  {\bibinfo  {journal} {Phys. Rev. Lett.}\ }\textbf {\bibinfo {volume} {133}},\
  \bibinfo {pages} {113401} (\bibinfo {year} {2024})}\BibitemShut {NoStop}%
\bibitem [{\citenamefont {Li}\ \emph {et~al.}(2019)\citenamefont {Li},
  \citenamefont {Harter}, \citenamefont {Liu}, \citenamefont {de~Melo},
  \citenamefont {Joglekar},\ and\ \citenamefont {Luo}}]{Li2019}%
  \BibitemOpen
  \bibfield  {author} {\bibinfo {author} {\bibfnamefont {J.}~\bibnamefont
  {Li}}, \bibinfo {author} {\bibfnamefont {A.~K.}\ \bibnamefont {Harter}},
  \bibinfo {author} {\bibfnamefont {J.}~\bibnamefont {Liu}}, \bibinfo {author}
  {\bibfnamefont {L.}~\bibnamefont {de~Melo}}, \bibinfo {author} {\bibfnamefont
  {Y.~N.}\ \bibnamefont {Joglekar}},\ and\ \bibinfo {author} {\bibfnamefont
  {L.}~\bibnamefont {Luo}},\ }\bibfield  {title} {\bibinfo {title} {Observation
  of parity-time symmetry breaking transitions in a dissipative floquet system
  of ultracold atoms},\ }\bibfield  {journal} {\bibinfo  {journal} {Nature
  Communications}\ }\textbf {\bibinfo {volume} {10}},\ \href
  {https://doi.org/10.1038/s41467-019-08596-1} {10.1038/s41467-019-08596-1}
  (\bibinfo {year} {2019})\BibitemShut {NoStop}%
\bibitem [{\citenamefont {Ozawa}\ \emph {et~al.}(2019)\citenamefont {Ozawa},
  \citenamefont {Price}, \citenamefont {Amo}, \citenamefont {Goldman},
  \citenamefont {Hafezi}, \citenamefont {Lu}, \citenamefont {Rechtsman},
  \citenamefont {Schuster}, \citenamefont {Simon}, \citenamefont {Zilberberg},\
  and\ \citenamefont {Carusotto}}]{Ozawa2019}%
  \BibitemOpen
  \bibfield  {author} {\bibinfo {author} {\bibfnamefont {T.}~\bibnamefont
  {Ozawa}}, \bibinfo {author} {\bibfnamefont {H.~M.}\ \bibnamefont {Price}},
  \bibinfo {author} {\bibfnamefont {A.}~\bibnamefont {Amo}}, \bibinfo {author}
  {\bibfnamefont {N.}~\bibnamefont {Goldman}}, \bibinfo {author} {\bibfnamefont
  {M.}~\bibnamefont {Hafezi}}, \bibinfo {author} {\bibfnamefont
  {L.}~\bibnamefont {Lu}}, \bibinfo {author} {\bibfnamefont {M.~C.}\
  \bibnamefont {Rechtsman}}, \bibinfo {author} {\bibfnamefont {D.}~\bibnamefont
  {Schuster}}, \bibinfo {author} {\bibfnamefont {J.}~\bibnamefont {Simon}},
  \bibinfo {author} {\bibfnamefont {O.}~\bibnamefont {Zilberberg}},\ and\
  \bibinfo {author} {\bibfnamefont {I.}~\bibnamefont {Carusotto}},\ }\bibfield
  {title} {\bibinfo {title} {Topological photonics},\ }\href
  {https://doi.org/10.1103/RevModPhys.91.015006} {\bibfield  {journal}
  {\bibinfo  {journal} {Rev. Mod. Phys.}\ }\textbf {\bibinfo {volume} {91}},\
  \bibinfo {pages} {015006} (\bibinfo {year} {2019})}\BibitemShut {NoStop}%
\bibitem [{\citenamefont {Sim}\ \emph {et~al.}(2023)\citenamefont {Sim},
  \citenamefont {Defenu}, \citenamefont {Molignini},\ and\ \citenamefont
  {Chitra}}]{Sim2023}%
  \BibitemOpen
  \bibfield  {author} {\bibinfo {author} {\bibfnamefont {K.}~\bibnamefont
  {Sim}}, \bibinfo {author} {\bibfnamefont {N.}~\bibnamefont {Defenu}},
  \bibinfo {author} {\bibfnamefont {P.}~\bibnamefont {Molignini}},\ and\
  \bibinfo {author} {\bibfnamefont {R.}~\bibnamefont {Chitra}},\ }\bibfield
  {title} {\bibinfo {title} {Quantum metric unveils defect freezing in
  non-hermitian systems},\ }\href
  {https://doi.org/10.1103/PhysRevLett.131.156501} {\bibfield  {journal}
  {\bibinfo  {journal} {Phys. Rev. Lett.}\ }\textbf {\bibinfo {volume} {131}},\
  \bibinfo {pages} {156501} (\bibinfo {year} {2023})}\BibitemShut {NoStop}%
\bibitem [{\citenamefont {Najafzadeh}\ \emph {et~al.}(2026)\citenamefont
  {Najafzadeh}, \citenamefont {Sadeghizade}, \citenamefont {Jafari},\ and\
  \citenamefont {Langari}}]{Najafzadeh2026}%
  \BibitemOpen
  \bibfield  {author} {\bibinfo {author} {\bibfnamefont {H.}~\bibnamefont
  {Najafzadeh}}, \bibinfo {author} {\bibfnamefont {S.}~\bibnamefont
  {Sadeghizade}}, \bibinfo {author} {\bibfnamefont {R.}~\bibnamefont
  {Jafari}},\ and\ \bibinfo {author} {\bibfnamefont {A.}~\bibnamefont
  {Langari}},\ }\bibfield  {title} {\bibinfo {title} {Dissipation-induced
  deviations from kibble-zurek scaling in non-hermitian quantum annealing},\
  }\href {https://doi.org/10.1103/7vqp-cxmq} {\bibfield  {journal} {\bibinfo
  {journal} {Phys. Rev. B}\ }\textbf {\bibinfo {volume} {114}},\ \bibinfo
  {pages} {074303} (\bibinfo {year} {2026})}\BibitemShut {NoStop}%
\bibitem [{\citenamefont {D{\'o}ra}\ \emph {et~al.}(2019)\citenamefont
  {D{\'o}ra}, \citenamefont {Heyl},\ and\ \citenamefont {Moessner}}]{Dora2019}%
  \BibitemOpen
  \bibfield  {author} {\bibinfo {author} {\bibfnamefont {B.}~\bibnamefont
  {D{\'o}ra}}, \bibinfo {author} {\bibfnamefont {M.}~\bibnamefont {Heyl}},\
  and\ \bibinfo {author} {\bibfnamefont {R.}~\bibnamefont {Moessner}},\
  }\bibfield  {title} {\bibinfo {title} {The kibble-zurek mechanism at
  exceptional points},\ }\href {https://doi.org/10.1038/s41467-019-10048-9}
  {\bibfield  {journal} {\bibinfo  {journal} {Nature communications}\ }\textbf
  {\bibinfo {volume} {10}},\ \bibinfo {pages} {2254} (\bibinfo {year}
  {2019})}\BibitemShut {NoStop}%
\bibitem [{\citenamefont {Xiao}\ \emph {et~al.}(2021)\citenamefont {Xiao},
  \citenamefont {Qu}, \citenamefont {Wang}, \citenamefont {Li}, \citenamefont
  {Dai}, \citenamefont {D\'ora}, \citenamefont {Heyl}, \citenamefont
  {Moessner}, \citenamefont {Yi},\ and\ \citenamefont {Xue}}]{Xiao2021}%
  \BibitemOpen
  \bibfield  {author} {\bibinfo {author} {\bibfnamefont {L.}~\bibnamefont
  {Xiao}}, \bibinfo {author} {\bibfnamefont {D.}~\bibnamefont {Qu}}, \bibinfo
  {author} {\bibfnamefont {K.}~\bibnamefont {Wang}}, \bibinfo {author}
  {\bibfnamefont {H.-W.}\ \bibnamefont {Li}}, \bibinfo {author} {\bibfnamefont
  {J.-Y.}\ \bibnamefont {Dai}}, \bibinfo {author} {\bibfnamefont
  {B.}~\bibnamefont {D\'ora}}, \bibinfo {author} {\bibfnamefont
  {M.}~\bibnamefont {Heyl}}, \bibinfo {author} {\bibfnamefont {R.}~\bibnamefont
  {Moessner}}, \bibinfo {author} {\bibfnamefont {W.}~\bibnamefont {Yi}},\ and\
  \bibinfo {author} {\bibfnamefont {P.}~\bibnamefont {Xue}},\ }\bibfield
  {title} {\bibinfo {title} {Non-hermitian kibble-zurek mechanism with tunable
  complexity in single-photon interferometry},\ }\href
  {https://doi.org/10.1103/PRXQuantum.2.020313} {\bibfield  {journal} {\bibinfo
   {journal} {PRX Quantum}\ }\textbf {\bibinfo {volume} {2}},\ \bibinfo {pages}
  {020313} (\bibinfo {year} {2021})}\BibitemShut {NoStop}%
\bibitem [{\citenamefont {Deng}\ \emph
  {et~al.}(2025{\natexlab{a}})\citenamefont {Deng}, \citenamefont {Li},
  \citenamefont {Hu}, \citenamefont {Sun},\ and\ \citenamefont
  {Li}}]{Deng2025}%
  \BibitemOpen
  \bibfield  {author} {\bibinfo {author} {\bibfnamefont {M.}~\bibnamefont
  {Deng}}, \bibinfo {author} {\bibfnamefont {W.}~\bibnamefont {Li}}, \bibinfo
  {author} {\bibfnamefont {K.}~\bibnamefont {Hu}}, \bibinfo {author}
  {\bibfnamefont {C.}~\bibnamefont {Sun}},\ and\ \bibinfo {author}
  {\bibfnamefont {F.}~\bibnamefont {Li}},\ }\bibfield  {title} {\bibinfo
  {title} {Nonadiabatic dynamics and universal behaviors in non-hermitian
  systems under biorthogonal framework},\ }\href
  {https://doi.org/10.1103/h11w-kys5} {\bibfield  {journal} {\bibinfo
  {journal} {Phys. Rev. B}\ }\textbf {\bibinfo {volume} {112}},\ \bibinfo
  {pages} {L020306} (\bibinfo {year} {2025}{\natexlab{a}})}\BibitemShut
  {NoStop}%
\bibitem [{\citenamefont {Das~Agarwal}\ \emph {et~al.}(2024)\citenamefont
  {Das~Agarwal}, \citenamefont {Konar}, \citenamefont {Lakkaraju},\ and\
  \citenamefont {Sen(De)}}]{Agarwal2024}%
  \BibitemOpen
  \bibfield  {author} {\bibinfo {author} {\bibfnamefont {K.}~\bibnamefont
  {Das~Agarwal}}, \bibinfo {author} {\bibfnamefont {T.~K.}\ \bibnamefont
  {Konar}}, \bibinfo {author} {\bibfnamefont {L.~G.~C.}\ \bibnamefont
  {Lakkaraju}},\ and\ \bibinfo {author} {\bibfnamefont {A.}~\bibnamefont
  {Sen(De)}},\ }\bibfield  {title} {\bibinfo {title} {Detecting exceptional
  points through dynamics in non-hermitian systems},\ }\href
  {https://doi.org/10.1103/PhysRevA.110.012226} {\bibfield  {journal} {\bibinfo
   {journal} {Phys. Rev. A}\ }\textbf {\bibinfo {volume} {110}},\ \bibinfo
  {pages} {012226} (\bibinfo {year} {2024})}\BibitemShut {NoStop}%
\bibitem [{\citenamefont {Naji}\ \emph {et~al.}(2022)\citenamefont {Naji},
  \citenamefont {Jafari}, \citenamefont {Jafari},\ and\ \citenamefont
  {Akbari}}]{Naji2022}%
  \BibitemOpen
  \bibfield  {author} {\bibinfo {author} {\bibfnamefont {J.}~\bibnamefont
  {Naji}}, \bibinfo {author} {\bibfnamefont {M.}~\bibnamefont {Jafari}},
  \bibinfo {author} {\bibfnamefont {R.}~\bibnamefont {Jafari}},\ and\ \bibinfo
  {author} {\bibfnamefont {A.}~\bibnamefont {Akbari}},\ }\bibfield  {title}
  {\bibinfo {title} {Dissipative floquet dynamical quantum phase transition},\
  }\href {https://doi.org/10.1103/PhysRevA.105.022220} {\bibfield  {journal}
  {\bibinfo  {journal} {Phys. Rev. A}\ }\textbf {\bibinfo {volume} {105}},\
  \bibinfo {pages} {022220} (\bibinfo {year} {2022})}\BibitemShut {NoStop}%
\bibitem [{\citenamefont {Doppler}\ \emph {et~al.}(2016)\citenamefont
  {Doppler}, \citenamefont {Mailybaev}, \citenamefont {B{\"o}hm}, \citenamefont
  {Kuhl}, \citenamefont {Girschik}, \citenamefont {Libisch}, \citenamefont
  {Milburn}, \citenamefont {Rabl}, \citenamefont {Moiseyev},\ and\
  \citenamefont {Rotter}}]{Doppler2016}%
  \BibitemOpen
  \bibfield  {author} {\bibinfo {author} {\bibfnamefont {J.}~\bibnamefont
  {Doppler}}, \bibinfo {author} {\bibfnamefont {A.~A.}\ \bibnamefont
  {Mailybaev}}, \bibinfo {author} {\bibfnamefont {J.}~\bibnamefont {B{\"o}hm}},
  \bibinfo {author} {\bibfnamefont {U.}~\bibnamefont {Kuhl}}, \bibinfo {author}
  {\bibfnamefont {A.}~\bibnamefont {Girschik}}, \bibinfo {author}
  {\bibfnamefont {F.}~\bibnamefont {Libisch}}, \bibinfo {author} {\bibfnamefont
  {T.~J.}\ \bibnamefont {Milburn}}, \bibinfo {author} {\bibfnamefont
  {P.}~\bibnamefont {Rabl}}, \bibinfo {author} {\bibfnamefont {N.}~\bibnamefont
  {Moiseyev}},\ and\ \bibinfo {author} {\bibfnamefont {S.}~\bibnamefont
  {Rotter}},\ }\bibfield  {title} {\bibinfo {title} {Dynamically encircling an
  exceptional point for asymmetric mode switching},\ }\href
  {https://doi.org/10.1038/nature18605} {\bibfield  {journal} {\bibinfo
  {journal} {Nature}\ }\textbf {\bibinfo {volume} {537}},\ \bibinfo {pages}
  {76} (\bibinfo {year} {2016})}\BibitemShut {NoStop}%
\bibitem [{\citenamefont {Yao}\ and\ \citenamefont {Wang}(2018)}]{Yao2018}%
  \BibitemOpen
  \bibfield  {author} {\bibinfo {author} {\bibfnamefont {S.}~\bibnamefont
  {Yao}}\ and\ \bibinfo {author} {\bibfnamefont {Z.}~\bibnamefont {Wang}},\
  }\bibfield  {title} {\bibinfo {title} {Edge states and topological invariants
  of non-hermitian systems},\ }\href
  {https://doi.org/10.1103/PhysRevLett.121.086803} {\bibfield  {journal}
  {\bibinfo  {journal} {Phys. Rev. Lett.}\ }\textbf {\bibinfo {volume} {121}},\
  \bibinfo {pages} {086803} (\bibinfo {year} {2018})}\BibitemShut {NoStop}%
\bibitem [{\citenamefont {Bergholtz}\ \emph {et~al.}(2021)\citenamefont
  {Bergholtz}, \citenamefont {Budich},\ and\ \citenamefont
  {Kunst}}]{Bergholtz2021}%
  \BibitemOpen
  \bibfield  {author} {\bibinfo {author} {\bibfnamefont {E.~J.}\ \bibnamefont
  {Bergholtz}}, \bibinfo {author} {\bibfnamefont {J.~C.}\ \bibnamefont
  {Budich}},\ and\ \bibinfo {author} {\bibfnamefont {F.~K.}\ \bibnamefont
  {Kunst}},\ }\bibfield  {title} {\bibinfo {title} {Exceptional topology of
  non-hermitian systems},\ }\href
  {https://doi.org/10.1103/RevModPhys.93.015005} {\bibfield  {journal}
  {\bibinfo  {journal} {Rev. Mod. Phys.}\ }\textbf {\bibinfo {volume} {93}},\
  \bibinfo {pages} {015005} (\bibinfo {year} {2021})}\BibitemShut {NoStop}%
\bibitem [{\citenamefont {Ashida}\ \emph {et~al.}(2020)\citenamefont {Ashida},
  \citenamefont {Gong},\ and\ \citenamefont {Ueda}}]{Ashida2020}%
  \BibitemOpen
  \bibfield  {author} {\bibinfo {author} {\bibfnamefont {Y.}~\bibnamefont
  {Ashida}}, \bibinfo {author} {\bibfnamefont {Z.}~\bibnamefont {Gong}},\ and\
  \bibinfo {author} {\bibfnamefont {M.}~\bibnamefont {Ueda}},\ }\bibfield
  {title} {\bibinfo {title} {Non-hermitian physics},\ }\href
  {https://doi.org/10.1080/00018732.2021.1876991} {\bibfield  {journal}
  {\bibinfo  {journal} {Advances in Physics}\ }\textbf {\bibinfo {volume}
  {69}},\ \bibinfo {pages} {249–435} (\bibinfo {year} {2020})}\BibitemShut
  {NoStop}%
\bibitem [{\citenamefont {Hatano}\ and\ \citenamefont
  {Nelson}(1996)}]{Hatano1996}%
  \BibitemOpen
  \bibfield  {author} {\bibinfo {author} {\bibfnamefont {N.}~\bibnamefont
  {Hatano}}\ and\ \bibinfo {author} {\bibfnamefont {D.~R.}\ \bibnamefont
  {Nelson}},\ }\bibfield  {title} {\bibinfo {title} {Localization transitions
  in non-hermitian quantum mechanics},\ }\href
  {https://doi.org/10.1103/PhysRevLett.77.570} {\bibfield  {journal} {\bibinfo
  {journal} {Phys. Rev. Lett.}\ }\textbf {\bibinfo {volume} {77}},\ \bibinfo
  {pages} {570} (\bibinfo {year} {1996})}\BibitemShut {NoStop}%
\bibitem [{\citenamefont {Rudner}\ and\ \citenamefont
  {Levitov}(2009)}]{Rudner2009}%
  \BibitemOpen
  \bibfield  {author} {\bibinfo {author} {\bibfnamefont {M.~S.}\ \bibnamefont
  {Rudner}}\ and\ \bibinfo {author} {\bibfnamefont {L.~S.}\ \bibnamefont
  {Levitov}},\ }\bibfield  {title} {\bibinfo {title} {Topological transition in
  a non-hermitian quantum walk},\ }\href
  {https://doi.org/10.1103/PhysRevLett.102.065703} {\bibfield  {journal}
  {\bibinfo  {journal} {Phys. Rev. Lett.}\ }\textbf {\bibinfo {volume} {102}},\
  \bibinfo {pages} {065703} (\bibinfo {year} {2009})}\BibitemShut {NoStop}%
\bibitem [{\citenamefont {Hu}\ and\ \citenamefont {Hughes}(2011)}]{Hu2011}%
  \BibitemOpen
  \bibfield  {author} {\bibinfo {author} {\bibfnamefont {Y.~C.}\ \bibnamefont
  {Hu}}\ and\ \bibinfo {author} {\bibfnamefont {T.~L.}\ \bibnamefont
  {Hughes}},\ }\bibfield  {title} {\bibinfo {title} {Absence of topological
  insulator phases in non-hermitian $pt$-symmetric hamiltonians},\ }\href
  {https://doi.org/10.1103/PhysRevB.84.153101} {\bibfield  {journal} {\bibinfo
  {journal} {Phys. Rev. B}\ }\textbf {\bibinfo {volume} {84}},\ \bibinfo
  {pages} {153101} (\bibinfo {year} {2011})}\BibitemShut {NoStop}%
\bibitem [{\citenamefont {Borgnia}\ \emph {et~al.}(2020)\citenamefont
  {Borgnia}, \citenamefont {Kruchkov},\ and\ \citenamefont
  {Slager}}]{Borgnia2020}%
  \BibitemOpen
  \bibfield  {author} {\bibinfo {author} {\bibfnamefont {D.~S.}\ \bibnamefont
  {Borgnia}}, \bibinfo {author} {\bibfnamefont {A.~J.}\ \bibnamefont
  {Kruchkov}},\ and\ \bibinfo {author} {\bibfnamefont {R.-J.}\ \bibnamefont
  {Slager}},\ }\bibfield  {title} {\bibinfo {title} {Non-hermitian boundary
  modes and topology},\ }\href {https://doi.org/10.1103/PhysRevLett.124.056802}
  {\bibfield  {journal} {\bibinfo  {journal} {Phys. Rev. Lett.}\ }\textbf
  {\bibinfo {volume} {124}},\ \bibinfo {pages} {056802} (\bibinfo {year}
  {2020})}\BibitemShut {NoStop}%
\bibitem [{\citenamefont {Lee}\ \emph {et~al.}(2019)\citenamefont {Lee},
  \citenamefont {Ahn}, \citenamefont {Zhou},\ and\ \citenamefont
  {Vishwanath}}]{Lee2019}%
  \BibitemOpen
  \bibfield  {author} {\bibinfo {author} {\bibfnamefont {J.~Y.}\ \bibnamefont
  {Lee}}, \bibinfo {author} {\bibfnamefont {J.}~\bibnamefont {Ahn}}, \bibinfo
  {author} {\bibfnamefont {H.}~\bibnamefont {Zhou}},\ and\ \bibinfo {author}
  {\bibfnamefont {A.}~\bibnamefont {Vishwanath}},\ }\bibfield  {title}
  {\bibinfo {title} {Topological correspondence between hermitian and
  non-hermitian systems: Anomalous dynamics},\ }\href
  {https://doi.org/10.1103/PhysRevLett.123.206404} {\bibfield  {journal}
  {\bibinfo  {journal} {Phys. Rev. Lett.}\ }\textbf {\bibinfo {volume} {123}},\
  \bibinfo {pages} {206404} (\bibinfo {year} {2019})}\BibitemShut {NoStop}%
\bibitem [{\citenamefont {Chen}\ \emph {et~al.}(2020)\citenamefont {Chen},
  \citenamefont {Li}, \citenamefont {Fisher},\ and\ \citenamefont
  {Lucas}}]{Chen2020}%
  \BibitemOpen
  \bibfield  {author} {\bibinfo {author} {\bibfnamefont {X.}~\bibnamefont
  {Chen}}, \bibinfo {author} {\bibfnamefont {Y.}~\bibnamefont {Li}}, \bibinfo
  {author} {\bibfnamefont {M.~P.~A.}\ \bibnamefont {Fisher}},\ and\ \bibinfo
  {author} {\bibfnamefont {A.}~\bibnamefont {Lucas}},\ }\bibfield  {title}
  {\bibinfo {title} {Emergent conformal symmetry in nonunitary random dynamics
  of free fermions},\ }\href {https://doi.org/10.1103/PhysRevResearch.2.033017}
  {\bibfield  {journal} {\bibinfo  {journal} {Phys. Rev. Res.}\ }\textbf
  {\bibinfo {volume} {2}},\ \bibinfo {pages} {033017} (\bibinfo {year}
  {2020})}\BibitemShut {NoStop}%
\bibitem [{\citenamefont {Bender}(2007)}]{Bender2007}%
  \BibitemOpen
  \bibfield  {author} {\bibinfo {author} {\bibfnamefont {C.~M.}\ \bibnamefont
  {Bender}},\ }\bibfield  {title} {\bibinfo {title} {Making sense of
  non-hermitian hamiltonians},\ }\href
  {https://doi.org/10.1088/0034-4885/70/6/R03} {\bibfield  {journal} {\bibinfo
  {journal} {Reports on Progress in Physics}\ }\textbf {\bibinfo {volume}
  {70}},\ \bibinfo {pages} {947} (\bibinfo {year} {2007})}\BibitemShut
  {NoStop}%
\bibitem [{\citenamefont {Dvir}\ \emph {et~al.}(2023)\citenamefont {Dvir},
  \citenamefont {Wang}, \citenamefont {van Loo}, \citenamefont {Liu},
  \citenamefont {Mazur}, \citenamefont {Bordin}, \citenamefont {Ten~Haaf},
  \citenamefont {Wang}, \citenamefont {van Driel}, \citenamefont {Zatelli}
  \emph {et~al.}}]{Dvir2023}%
  \BibitemOpen
  \bibfield  {author} {\bibinfo {author} {\bibfnamefont {T.}~\bibnamefont
  {Dvir}}, \bibinfo {author} {\bibfnamefont {G.}~\bibnamefont {Wang}}, \bibinfo
  {author} {\bibfnamefont {N.}~\bibnamefont {van Loo}}, \bibinfo {author}
  {\bibfnamefont {C.-X.}\ \bibnamefont {Liu}}, \bibinfo {author} {\bibfnamefont
  {G.~P.}\ \bibnamefont {Mazur}}, \bibinfo {author} {\bibfnamefont
  {A.}~\bibnamefont {Bordin}}, \bibinfo {author} {\bibfnamefont {S.~L.}\
  \bibnamefont {Ten~Haaf}}, \bibinfo {author} {\bibfnamefont {J.-Y.}\
  \bibnamefont {Wang}}, \bibinfo {author} {\bibfnamefont {D.}~\bibnamefont {van
  Driel}}, \bibinfo {author} {\bibfnamefont {F.}~\bibnamefont {Zatelli}}, \emph
  {et~al.},\ }\bibfield  {title} {\bibinfo {title} {Realization of a minimal
  kitaev chain in coupled quantum dots},\ }\href
  {https://doi.org/10.1038/s41586-022-05585-1} {\bibfield  {journal} {\bibinfo
  {journal} {Nature}\ }\textbf {\bibinfo {volume} {614}},\ \bibinfo {pages}
  {445} (\bibinfo {year} {2023})}\BibitemShut {NoStop}%
\bibitem [{\citenamefont {Bender}\ and\ \citenamefont
  {Boettcher}(1998)}]{Bender1998}%
  \BibitemOpen
  \bibfield  {author} {\bibinfo {author} {\bibfnamefont {C.~M.}\ \bibnamefont
  {Bender}}\ and\ \bibinfo {author} {\bibfnamefont {S.}~\bibnamefont
  {Boettcher}},\ }\bibfield  {title} {\bibinfo {title} {Real spectra in
  non-hermitian hamiltonians having $\mathcal{P}\mathcal{T}$ symmetry},\ }\href
  {https://doi.org/10.1103/PhysRevLett.80.5243} {\bibfield  {journal} {\bibinfo
   {journal} {Phys. Rev. Lett.}\ }\textbf {\bibinfo {volume} {80}},\ \bibinfo
  {pages} {5243} (\bibinfo {year} {1998})}\BibitemShut {NoStop}%
\bibitem [{\citenamefont {El-Ganainy}\ \emph {et~al.}(2018)\citenamefont
  {El-Ganainy}, \citenamefont {Makris}, \citenamefont {Khajavikhan},
  \citenamefont {Musslimani}, \citenamefont {Rotter},\ and\ \citenamefont
  {Christodoulides}}]{ElGanainy2018}%
  \BibitemOpen
  \bibfield  {author} {\bibinfo {author} {\bibfnamefont {R.}~\bibnamefont
  {El-Ganainy}}, \bibinfo {author} {\bibfnamefont {K.~G.}\ \bibnamefont
  {Makris}}, \bibinfo {author} {\bibfnamefont {M.}~\bibnamefont {Khajavikhan}},
  \bibinfo {author} {\bibfnamefont {Z.~H.}\ \bibnamefont {Musslimani}},
  \bibinfo {author} {\bibfnamefont {S.}~\bibnamefont {Rotter}},\ and\ \bibinfo
  {author} {\bibfnamefont {D.~N.}\ \bibnamefont {Christodoulides}},\ }\bibfield
   {title} {\bibinfo {title} {{Non-Hermitian physics and PT symmetry}},\ }\href
  {https://doi.org/10.1038/nphys4323} {\bibfield  {journal} {\bibinfo
  {journal} {Nature Phys.}\ }\textbf {\bibinfo {volume} {14}},\ \bibinfo
  {pages} {11} (\bibinfo {year} {2018})}\BibitemShut {NoStop}%
\bibitem [{\citenamefont {Miri}\ and\ \citenamefont {Alù}(2019)}]{Miri2019}%
  \BibitemOpen
  \bibfield  {author} {\bibinfo {author} {\bibfnamefont {M.-A.}\ \bibnamefont
  {Miri}}\ and\ \bibinfo {author} {\bibfnamefont {A.}~\bibnamefont {Alù}},\
  }\bibfield  {title} {\bibinfo {title} {Exceptional points in optics and
  photonics},\ }\href {https://doi.org/10.1126/science.aar7709} {\bibfield
  {journal} {\bibinfo  {journal} {Science}\ }\textbf {\bibinfo {volume}
  {363}},\ \bibinfo {pages} {eaar7709} (\bibinfo {year} {2019})},\ \Eprint
  {https://arxiv.org/abs/https://www.science.org/doi/pdf/10.1126/science.aar7709}
  {https://www.science.org/doi/pdf/10.1126/science.aar7709} \BibitemShut
  {NoStop}%
\bibitem [{\citenamefont {Martinez-Azcona}\ \emph {et~al.}(2025)\citenamefont
  {Martinez-Azcona}, \citenamefont {Kundu}, \citenamefont {Saxena},
  \citenamefont {del Campo},\ and\ \citenamefont {Chenu}}]{Martinez2025}%
  \BibitemOpen
  \bibfield  {author} {\bibinfo {author} {\bibfnamefont {P.}~\bibnamefont
  {Martinez-Azcona}}, \bibinfo {author} {\bibfnamefont {A.}~\bibnamefont
  {Kundu}}, \bibinfo {author} {\bibfnamefont {A.}~\bibnamefont {Saxena}},
  \bibinfo {author} {\bibfnamefont {A.}~\bibnamefont {del Campo}},\ and\
  \bibinfo {author} {\bibfnamefont {A.}~\bibnamefont {Chenu}},\ }\bibfield
  {title} {\bibinfo {title} {Quantum dynamics with stochastic non-hermitian
  hamiltonians},\ }\href {https://doi.org/10.1103/5ksl-tjjm} {\bibfield
  {journal} {\bibinfo  {journal} {Phys. Rev. Lett.}\ }\textbf {\bibinfo
  {volume} {135}},\ \bibinfo {pages} {010402} (\bibinfo {year}
  {2025})}\BibitemShut {NoStop}%
\bibitem [{\citenamefont {Minganti}\ \emph {et~al.}(2019)\citenamefont
  {Minganti}, \citenamefont {Miranowicz}, \citenamefont {Chhajlany},\ and\
  \citenamefont {Nori}}]{Minganti2019}%
  \BibitemOpen
  \bibfield  {author} {\bibinfo {author} {\bibfnamefont {F.}~\bibnamefont
  {Minganti}}, \bibinfo {author} {\bibfnamefont {A.}~\bibnamefont
  {Miranowicz}}, \bibinfo {author} {\bibfnamefont {R.~W.}\ \bibnamefont
  {Chhajlany}},\ and\ \bibinfo {author} {\bibfnamefont {F.}~\bibnamefont
  {Nori}},\ }\bibfield  {title} {\bibinfo {title} {Quantum exceptional points
  of non-hermitian hamiltonians and liouvillians: The effects of quantum
  jumps},\ }\href {https://doi.org/10.1103/PhysRevA.100.062131} {\bibfield
  {journal} {\bibinfo  {journal} {Phys. Rev. A}\ }\textbf {\bibinfo {volume}
  {100}},\ \bibinfo {pages} {062131} (\bibinfo {year} {2019})}\BibitemShut
  {NoStop}%
\bibitem [{\citenamefont {Arkhipov}\ \emph {et~al.}(2020)\citenamefont
  {Arkhipov}, \citenamefont {Miranowicz}, \citenamefont {Minganti},\ and\
  \citenamefont {Nori}}]{Arkhipov2020}%
  \BibitemOpen
  \bibfield  {author} {\bibinfo {author} {\bibfnamefont {I.~I.}\ \bibnamefont
  {Arkhipov}}, \bibinfo {author} {\bibfnamefont {A.}~\bibnamefont
  {Miranowicz}}, \bibinfo {author} {\bibfnamefont {F.}~\bibnamefont
  {Minganti}},\ and\ \bibinfo {author} {\bibfnamefont {F.}~\bibnamefont
  {Nori}},\ }\bibfield  {title} {\bibinfo {title} {Liouvillian exceptional
  points of any order in dissipative linear bosonic systems: Coherence
  functions and switching between $\mathcal{PT}$ and anti-$\mathcal{PT}$
  symmetries},\ }\href {https://doi.org/10.1103/PhysRevA.102.033715} {\bibfield
   {journal} {\bibinfo  {journal} {Phys. Rev. A}\ }\textbf {\bibinfo {volume}
  {102}},\ \bibinfo {pages} {033715} (\bibinfo {year} {2020})}\BibitemShut
  {NoStop}%
\bibitem [{\citenamefont {Cattaneo}\ \emph {et~al.}(2020)\citenamefont
  {Cattaneo}, \citenamefont {Giorgi}, \citenamefont {Maniscalco},\ and\
  \citenamefont {Zambrini}}]{Cattaneo2020}%
  \BibitemOpen
  \bibfield  {author} {\bibinfo {author} {\bibfnamefont {M.}~\bibnamefont
  {Cattaneo}}, \bibinfo {author} {\bibfnamefont {G.~L.}\ \bibnamefont
  {Giorgi}}, \bibinfo {author} {\bibfnamefont {S.}~\bibnamefont {Maniscalco}},\
  and\ \bibinfo {author} {\bibfnamefont {R.}~\bibnamefont {Zambrini}},\
  }\bibfield  {title} {\bibinfo {title} {Symmetry and block structure of the
  liouvillian superoperator in partial secular approximation},\ }\href
  {https://doi.org/10.1103/PhysRevA.101.042108} {\bibfield  {journal} {\bibinfo
   {journal} {Phys. Rev. A}\ }\textbf {\bibinfo {volume} {101}},\ \bibinfo
  {pages} {042108} (\bibinfo {year} {2020})}\BibitemShut {NoStop}%
\bibitem [{\citenamefont {Gorini}\ \emph {et~al.}(1976)\citenamefont {Gorini},
  \citenamefont {Kossakowski},\ and\ \citenamefont {Sudarshan}}]{Gorini1976}%
  \BibitemOpen
  \bibfield  {author} {\bibinfo {author} {\bibfnamefont {V.}~\bibnamefont
  {Gorini}}, \bibinfo {author} {\bibfnamefont {A.}~\bibnamefont
  {Kossakowski}},\ and\ \bibinfo {author} {\bibfnamefont {E.~C.~G.}\
  \bibnamefont {Sudarshan}},\ }\bibfield  {title} {\bibinfo {title} {Completely
  positive dynamical semigroups of n‐level systems},\ }\href
  {https://doi.org/10.1063/1.522979} {\bibfield  {journal} {\bibinfo  {journal}
  {Journal of Mathematical Physics}\ }\textbf {\bibinfo {volume} {17}},\
  \bibinfo {pages} {821} (\bibinfo {year} {1976})}\BibitemShut {NoStop}%
\bibitem [{\citenamefont {Lindblad}(1976)}]{Lindblad1976}%
  \BibitemOpen
  \bibfield  {author} {\bibinfo {author} {\bibfnamefont {G.}~\bibnamefont
  {Lindblad}},\ }\bibfield  {title} {\bibinfo {title} {On the generators of
  quantum dynamical semigroups},\ }\href {https://doi.org/10.1007/BF01608499}
  {\bibfield  {journal} {\bibinfo  {journal} {Communications in mathematical
  physics}\ }\textbf {\bibinfo {volume} {48}},\ \bibinfo {pages} {119}
  (\bibinfo {year} {1976})}\BibitemShut {NoStop}%
\bibitem [{\citenamefont {Novikov}(1965{\natexlab{a}})}]{Novikov}%
  \BibitemOpen
  \bibfield  {author} {\bibinfo {author} {\bibfnamefont {E.~A.}\ \bibnamefont
  {Novikov}},\ }\bibfield  {title} {\bibinfo {title} {Functionals and the
  random-force method in turbulence theory},\ }\href
  {http://www.jetp.ras.ru/cgi-bin/e/index/e/20/5/p1290?a=list} {\bibfield
  {journal} {\bibinfo  {journal} {Sov. Phys. JETP}\ }\textbf {\bibinfo {volume}
  {20}},\ \bibinfo {pages} {1290} (\bibinfo {year}
  {1965}{\natexlab{a}})}\BibitemShut {NoStop}%
\bibitem [{Sup()}]{SuppMat}%
  \BibitemOpen
  \href@noop {} {}\bibinfo {note} {See Supplemental Material for Noise-Immune
  Attractors in Stochastically Driven Non-Hermitian Systems}\BibitemShut
  {NoStop}%
\bibitem [{\citenamefont {G{\"u}nther}\ \emph {et~al.}(2005)\citenamefont
  {G{\"u}nther}, \citenamefont {Stefani},\ and\ \citenamefont
  {Znojil}}]{Gunther2005}%
  \BibitemOpen
  \bibfield  {author} {\bibinfo {author} {\bibfnamefont {U.}~\bibnamefont
  {G{\"u}nther}}, \bibinfo {author} {\bibfnamefont {F.}~\bibnamefont
  {Stefani}},\ and\ \bibinfo {author} {\bibfnamefont {M.}~\bibnamefont
  {Znojil}},\ }\bibfield  {title} {\bibinfo {title} {Mhd $\alpha$2-dynamo,
  squire equation and pt-symmetric interpolation between square well and
  harmonic oscillator},\ }\href {https://doi.org/10.1063/1.1915293} {\bibfield
  {journal} {\bibinfo  {journal} {Journal of Mathematical Physics}\ }\textbf
  {\bibinfo {volume} {46}},\ \bibinfo {pages} {063504} (\bibinfo {year}
  {2005})}\BibitemShut {NoStop}%
\bibitem [{\citenamefont {Mostafazadeh}(2006)}]{Mostafazadeh2006}%
  \BibitemOpen
  \bibfield  {author} {\bibinfo {author} {\bibfnamefont {A.}~\bibnamefont
  {Mostafazadeh}},\ }\bibfield  {title} {\bibinfo {title} {Krein-space
  formulation of $\mathcal{PT}$symmetry, $\mathcal{CPT}$-inner products, and
  pseudo-hermiticity},\ }\href {https://doi.org/10.1007/s10582-006-0388-8}
  {\bibfield  {journal} {\bibinfo  {journal} {Czechoslovak Journal of Physics}\
  }\textbf {\bibinfo {volume} {56}},\ \bibinfo {pages} {919} (\bibinfo {year}
  {2006})}\BibitemShut {NoStop}%
\bibitem [{\citenamefont {G{\"u}nther}\ and\ \citenamefont
  {Samsonov}(2008)}]{Gunther2008}%
  \BibitemOpen
  \bibfield  {author} {\bibinfo {author} {\bibfnamefont {U.}~\bibnamefont
  {G{\"u}nther}}\ and\ \bibinfo {author} {\bibfnamefont {B.~F.}\ \bibnamefont
  {Samsonov}},\ }\bibfield  {title} {\bibinfo {title}
  {$\mathcal{P}\mathcal{T}$-symmetric brachistochrone problem, lorentz boosts,
  and nonunitary operator equivalence classes},\ }\href
  {https://doi.org/10.1103/PhysRevA.78.042115} {\bibfield  {journal} {\bibinfo
  {journal} {Phys. Rev. A}\ }\textbf {\bibinfo {volume} {78}},\ \bibinfo
  {pages} {042115} (\bibinfo {year} {2008})}\BibitemShut {NoStop}%
\bibitem [{\citenamefont {Massarelli}\ \emph {et~al.}(2022)\citenamefont
  {Massarelli}, \citenamefont {Khait},\ and\ \citenamefont
  {Paramekanti}}]{Massarelli2022}%
  \BibitemOpen
  \bibfield  {author} {\bibinfo {author} {\bibfnamefont {G.}~\bibnamefont
  {Massarelli}}, \bibinfo {author} {\bibfnamefont {I.}~\bibnamefont {Khait}},\
  and\ \bibinfo {author} {\bibfnamefont {A.}~\bibnamefont {Paramekanti}},\
  }\bibfield  {title} {\bibinfo {title} {Krein-unitary schrieffer-wolff
  transformation and band touchings in bosonic bogoliubov--de gennes and other
  krein-hermitian hamiltonians},\ }\href
  {https://doi.org/10.1103/PhysRevB.106.144434} {\bibfield  {journal} {\bibinfo
   {journal} {Phys. Rev. B}\ }\textbf {\bibinfo {volume} {106}},\ \bibinfo
  {pages} {144434} (\bibinfo {year} {2022})}\BibitemShut {NoStop}%
\bibitem [{\citenamefont {Li}\ \emph {et~al.}(2018)\citenamefont {Li},
  \citenamefont {Zhang}, \citenamefont {Zhang},\ and\ \citenamefont
  {Song}}]{Li2018}%
  \BibitemOpen
  \bibfield  {author} {\bibinfo {author} {\bibfnamefont {C.}~\bibnamefont
  {Li}}, \bibinfo {author} {\bibfnamefont {X.~Z.}\ \bibnamefont {Zhang}},
  \bibinfo {author} {\bibfnamefont {G.}~\bibnamefont {Zhang}},\ and\ \bibinfo
  {author} {\bibfnamefont {Z.}~\bibnamefont {Song}},\ }\bibfield  {title}
  {\bibinfo {title} {Topological phases in a kitaev chain with imbalanced
  pairing},\ }\href {https://doi.org/10.1103/PhysRevB.97.115436} {\bibfield
  {journal} {\bibinfo  {journal} {Phys. Rev. B}\ }\textbf {\bibinfo {volume}
  {97}},\ \bibinfo {pages} {115436} (\bibinfo {year} {2018})}\BibitemShut
  {NoStop}%
\bibitem [{\citenamefont {Wang}\ \emph {et~al.}(2017)\citenamefont {Wang},
  \citenamefont {Lin}, \citenamefont {Zhang},\ and\ \citenamefont
  {Song}}]{Wang2017DKC}%
  \BibitemOpen
  \bibfield  {author} {\bibinfo {author} {\bibfnamefont {P.}~\bibnamefont
  {Wang}}, \bibinfo {author} {\bibfnamefont {S.}~\bibnamefont {Lin}}, \bibinfo
  {author} {\bibfnamefont {G.}~\bibnamefont {Zhang}},\ and\ \bibinfo {author}
  {\bibfnamefont {Z.}~\bibnamefont {Song}},\ }\bibfield  {title} {\bibinfo
  {title} {Topological gapless phase in kitaev model on square lattice},\
  }\href {https://doi.org/10.1038/s41598-017-17334-w} {\bibfield  {journal}
  {\bibinfo  {journal} {Scientific Reports}\ }\textbf {\bibinfo {volume} {7}},\
  \bibinfo {pages} {17179} (\bibinfo {year} {2017})}\BibitemShut {NoStop}%
\bibitem [{\citenamefont {Li}\ \emph {et~al.}(2006)\citenamefont {Li},
  \citenamefont {Ding}, \citenamefont {Yu}, \citenamefont {Roscilde},\ and\
  \citenamefont {Haas}}]{Li2006}%
  \BibitemOpen
  \bibfield  {author} {\bibinfo {author} {\bibfnamefont {W.}~\bibnamefont
  {Li}}, \bibinfo {author} {\bibfnamefont {L.}~\bibnamefont {Ding}}, \bibinfo
  {author} {\bibfnamefont {R.}~\bibnamefont {Yu}}, \bibinfo {author}
  {\bibfnamefont {T.}~\bibnamefont {Roscilde}},\ and\ \bibinfo {author}
  {\bibfnamefont {S.}~\bibnamefont {Haas}},\ }\bibfield  {title} {\bibinfo
  {title} {Scaling behavior of entanglement in two- and three-dimensional
  free-fermion systems},\ }\href {https://doi.org/10.1103/PhysRevB.74.073103}
  {\bibfield  {journal} {\bibinfo  {journal} {Phys. Rev. B}\ }\textbf {\bibinfo
  {volume} {74}},\ \bibinfo {pages} {073103} (\bibinfo {year}
  {2006})}\BibitemShut {NoStop}%
\bibitem [{\citenamefont {Lieu}(2018{\natexlab{a}})}]{Lieu2018}%
  \BibitemOpen
  \bibfield  {author} {\bibinfo {author} {\bibfnamefont {S.}~\bibnamefont
  {Lieu}},\ }\bibfield  {title} {\bibinfo {title} {Topological phases in the
  non-hermitian su-schrieffer-heeger model},\ }\href
  {https://doi.org/10.1103/PhysRevB.97.045106} {\bibfield  {journal} {\bibinfo
  {journal} {Phys. Rev. B}\ }\textbf {\bibinfo {volume} {97}},\ \bibinfo
  {pages} {045106} (\bibinfo {year} {2018}{\natexlab{a}})}\BibitemShut
  {NoStop}%
\bibitem [{\citenamefont {Szameit}\ and\ \citenamefont
  {Nolte}(2010)}]{Szameit2010}%
  \BibitemOpen
  \bibfield  {author} {\bibinfo {author} {\bibfnamefont {A.}~\bibnamefont
  {Szameit}}\ and\ \bibinfo {author} {\bibfnamefont {S.}~\bibnamefont
  {Nolte}},\ }\bibfield  {title} {\bibinfo {title} {Discrete optics in
  femtosecond-laser-written photonic structures},\ }\href
  {https://doi.org/10.1088/0953-4075/43/16/163001} {\bibfield  {journal}
  {\bibinfo  {journal} {Journal of Physics B: Atomic, Molecular and Optical
  Physics}\ }\textbf {\bibinfo {volume} {43}},\ \bibinfo {pages} {163001}
  (\bibinfo {year} {2010})}\BibitemShut {NoStop}%
\bibitem [{\citenamefont {Cerjan}\ \emph {et~al.}(2019)\citenamefont {Cerjan},
  \citenamefont {Huang}, \citenamefont {Wang}, \citenamefont {Chen},
  \citenamefont {Chong},\ and\ \citenamefont {Rechtsman}}]{Cerjan2019}%
  \BibitemOpen
  \bibfield  {author} {\bibinfo {author} {\bibfnamefont {A.}~\bibnamefont
  {Cerjan}}, \bibinfo {author} {\bibfnamefont {S.}~\bibnamefont {Huang}},
  \bibinfo {author} {\bibfnamefont {M.}~\bibnamefont {Wang}}, \bibinfo {author}
  {\bibfnamefont {K.~P.}\ \bibnamefont {Chen}}, \bibinfo {author}
  {\bibfnamefont {Y.}~\bibnamefont {Chong}},\ and\ \bibinfo {author}
  {\bibfnamefont {M.~C.}\ \bibnamefont {Rechtsman}},\ }\bibfield  {title}
  {\bibinfo {title} {Experimental realization of a weyl exceptional ring},\
  }\href {https://doi.org/10.1038/s41566-019-0453-z} {\bibfield  {journal}
  {\bibinfo  {journal} {Nature Photonics}\ }\textbf {\bibinfo {volume} {13}},\
  \bibinfo {pages} {623–628} (\bibinfo {year} {2019})}\BibitemShut {NoStop}%
\bibitem [{\citenamefont {Slootman}\ \emph {et~al.}(2024)\citenamefont
  {Slootman}, \citenamefont {Cherifi}, \citenamefont {Eek}, \citenamefont
  {Arouca}, \citenamefont {Bergholtz}, \citenamefont {Bourennane},\ and\
  \citenamefont {Smith}}]{Slootman2024}%
  \BibitemOpen
  \bibfield  {author} {\bibinfo {author} {\bibfnamefont {E.}~\bibnamefont
  {Slootman}}, \bibinfo {author} {\bibfnamefont {W.}~\bibnamefont {Cherifi}},
  \bibinfo {author} {\bibfnamefont {L.}~\bibnamefont {Eek}}, \bibinfo {author}
  {\bibfnamefont {R.}~\bibnamefont {Arouca}}, \bibinfo {author} {\bibfnamefont
  {E.~J.}\ \bibnamefont {Bergholtz}}, \bibinfo {author} {\bibfnamefont
  {M.}~\bibnamefont {Bourennane}},\ and\ \bibinfo {author} {\bibfnamefont
  {C.~M.}\ \bibnamefont {Smith}},\ }\bibfield  {title} {\bibinfo {title}
  {Breaking and resurgence of symmetry in the non-hermitian
  su-schrieffer-heeger model in photonic waveguides},\ }\href
  {https://doi.org/10.1103/PhysRevResearch.6.023140} {\bibfield  {journal}
  {\bibinfo  {journal} {Phys. Rev. Res.}\ }\textbf {\bibinfo {volume} {6}},\
  \bibinfo {pages} {023140} (\bibinfo {year} {2024})}\BibitemShut {NoStop}%
\bibitem [{\citenamefont {Najafzadeh}\ \emph {et~al.}()\citenamefont
  {Najafzadeh}, \citenamefont {Jafari},\ and\ \citenamefont {Langari}}]{main}%
  \BibitemOpen
  \bibfield  {author} {\bibinfo {author} {\bibfnamefont {H.}~\bibnamefont
  {Najafzadeh}}, \bibinfo {author} {\bibfnamefont {R.}~\bibnamefont {Jafari}},\
  and\ \bibinfo {author} {\bibfnamefont {A.}~\bibnamefont {Langari}},\
  }\bibfield  {title} {\bibinfo {title} {Noise-immune attractors in
  stochastically driven non-hermitian systems},\ }\href@noop {} {\
  }\BibitemShut {NoStop}%
\bibitem [{\citenamefont {Novikov}(1965{\natexlab{b}})}]{NovikovSM}%
  \BibitemOpen
  \bibfield  {author} {\bibinfo {author} {\bibfnamefont {E.~A.}\ \bibnamefont
  {Novikov}},\ }\bibfield  {title} {\bibinfo {title} {Functionals and the
  random-force method in turbulence theory},\ }\href
  {http://www.jetp.ras.ru/cgi-bin/e/index/e/20/5/p1290?a=list} {\bibfield
  {journal} {\bibinfo  {journal} {Sov. Phys. JETP}\ }\textbf {\bibinfo {volume}
  {20}},\ \bibinfo {pages} {1290} (\bibinfo {year}
  {1965}{\natexlab{b}})}\BibitemShut {NoStop}%
\bibitem [{\citenamefont {Deng}\ \emph
  {et~al.}(2025{\natexlab{b}})\citenamefont {Deng}, \citenamefont {Li},
  \citenamefont {Hu}, \citenamefont {Sun},\ and\ \citenamefont
  {Li}}]{Deng2025SM}%
  \BibitemOpen
  \bibfield  {author} {\bibinfo {author} {\bibfnamefont {M.}~\bibnamefont
  {Deng}}, \bibinfo {author} {\bibfnamefont {W.}~\bibnamefont {Li}}, \bibinfo
  {author} {\bibfnamefont {K.}~\bibnamefont {Hu}}, \bibinfo {author}
  {\bibfnamefont {C.}~\bibnamefont {Sun}},\ and\ \bibinfo {author}
  {\bibfnamefont {F.}~\bibnamefont {Li}},\ }\bibfield  {title} {\bibinfo
  {title} {Nonadiabatic dynamics and universal behaviors in non-hermitian
  systems under biorthogonal framework},\ }\href
  {https://doi.org/10.1103/h11w-kys5} {\bibfield  {journal} {\bibinfo
  {journal} {Phys. Rev. B}\ }\textbf {\bibinfo {volume} {112}},\ \bibinfo
  {pages} {L020306} (\bibinfo {year} {2025}{\natexlab{b}})}\BibitemShut
  {NoStop}%
\bibitem [{\citenamefont {Jafari}\ \emph
  {et~al.}(2026{\natexlab{c}})\citenamefont {Jafari}, \citenamefont {Akbari},
  \citenamefont {Mardonov},\ and\ \citenamefont {Langari}}]{Jafari2026aSM}%
  \BibitemOpen
  \bibfield  {author} {\bibinfo {author} {\bibfnamefont {R.}~\bibnamefont
  {Jafari}}, \bibinfo {author} {\bibfnamefont {A.}~\bibnamefont {Akbari}},
  \bibinfo {author} {\bibfnamefont {S.}~\bibnamefont {Mardonov}},\ and\
  \bibinfo {author} {\bibfnamefont {A.}~\bibnamefont {Langari}},\ }\href
  {https://arxiv.org/abs/2609.00971} {\bibinfo {title} {Kibble--zurek mechanism
  and defect freezing in imbalanced-pairing kitaev models}} (\bibinfo {year}
  {2026}{\natexlab{c}}),\ \Eprint {https://arxiv.org/abs/2609.00971}
  {arXiv:2609.00971 [cond-mat.stat-mech]} \BibitemShut {NoStop}%
\bibitem [{\citenamefont {Jafari}\ \emph
  {et~al.}(2026{\natexlab{d}})\citenamefont {Jafari}, \citenamefont {Akbari},
  \citenamefont {Mardonov},\ and\ \citenamefont {Yavartanoo}}]{Jafari2026bSM}%
  \BibitemOpen
  \bibfield  {author} {\bibinfo {author} {\bibfnamefont {R.}~\bibnamefont
  {Jafari}}, \bibinfo {author} {\bibfnamefont {A.}~\bibnamefont {Akbari}},
  \bibinfo {author} {\bibfnamefont {S.}~\bibnamefont {Mardonov}},\ and\
  \bibinfo {author} {\bibfnamefont {H.}~\bibnamefont {Yavartanoo}},\ }\href
  {https://arxiv.org/abs/2607.23956} {\bibinfo {title} {Dynamical quantum phase
  transitions in a pseudo-hermitian hamiltonian: The imbalanced-pairing kitaev
  model}} (\bibinfo {year} {2026}{\natexlab{d}}),\ \Eprint
  {https://arxiv.org/abs/2607.23956} {arXiv:2607.23956 [quant-ph]} \BibitemShut
  {NoStop}%
\bibitem [{\citenamefont {Lieu}(2018{\natexlab{b}})}]{Lieu2018SM}%
  \BibitemOpen
  \bibfield  {author} {\bibinfo {author} {\bibfnamefont {S.}~\bibnamefont
  {Lieu}},\ }\bibfield  {title} {\bibinfo {title} {Topological phases in the
  non-hermitian su-schrieffer-heeger model},\ }\href
  {https://doi.org/10.1103/PhysRevB.97.045106} {\bibfield  {journal} {\bibinfo
  {journal} {Phys. Rev. B}\ }\textbf {\bibinfo {volume} {97}},\ \bibinfo
  {pages} {045106} (\bibinfo {year} {2018}{\natexlab{b}})}\BibitemShut
  {NoStop}%
\end{thebibliography}%

\end{document}